\documentclass[%
 reprint,
 superscriptaddress,
 amsmath,amssymb,
 aps,
]{revtex4-2}

\usepackage{graphicx}
\usepackage{dcolumn}
\usepackage{bm}
\usepackage{aas_macros}
\usepackage{subcaption} 
\usepackage{float}
\usepackage{hyperref}
\usepackage[dvipsnames]{xcolor}
\usepackage{comment}
\usepackage{booktabs}  

\newcommand{\rev}[1]{{ #1}}

\begin{document}


\title{Model-Independent Measurement of Baryon Gas Fractions through Galaxy-Galaxy Lensing and the Kinematic Sunyaev-Zel'dovich Effect}

\author{R. Henry Liu}
\email{rh\_liu@berkeley.edu}
\affiliation{Department of Physics, University of California, Berkeley, CA 94720, USA}
\affiliation{Lawrence Berkeley National Laboratory, 1 Cyclotron Road, Berkeley, CA 94720, USA}
\affiliation{Berkeley Center for Cosmological Physics, Department of Physics,
University of California, Berkeley, CA 94720, USA}
\author{Uro\v{s} Seljak}
\affiliation{Berkeley Center for Cosmological Physics, Department of Physics,
University of California, Berkeley, CA 94720, USA}
\affiliation{Lawrence Berkeley National Laboratory, 1 Cyclotron Road, Berkeley, CA 94720, USA}
\author{Simone Ferraro}
\affiliation{Lawrence Berkeley National Laboratory, 1 Cyclotron Road, Berkeley, CA 94720, USA}
\affiliation{Berkeley Center for Cosmological Physics, Department of Physics,
University of California, Berkeley, CA 94720, USA}
\author{Boryana Hadzhiyska}
\affiliation{Institute of Astronomy, University of Cambridge, Madingley Road, Cambridge CB3 0HA, UK}
\affiliation{Kavli Institute for Cosmology Cambridge, Madingley Road, Cambridge, CB3 0HA, UK}
\author{Frank J. Qu}
\affiliation{Kavli Institute for Particle Astrophysics and Cosmology,
Stanford University, 452 Lomita Mall, Stanford, CA, 94305, USA}
\affiliation{Department of Physics, Stanford University, 382 Via Pueblo Mall, Stanford, CA, 94305, USA}
\affiliation{SLAC National Accelerator Laboratory, 2575 Sand Hill Road, Menlo Park, California 94025, USA}
\author{Bernardita \surname{Ried Guachalla}}
\affiliation{Kavli Institute for Particle Astrophysics and Cosmology, Stanford University, 452 Lomita Mall, Stanford, CA, 94305, USA}
\affiliation{Department of Physics, Stanford University, 382 Via Pueblo Mall, Stanford, CA, 94305, USA}
\affiliation{SLAC National Accelerator Laboratory, 2575 Sand Hill Road, Menlo Park, California 94025, USA}
\author{Emmanuel Schaan}
\affiliation{Kavli Institute for Particle Astrophysics and Cosmology, Stanford University, 452 Lomita Mall, Stanford, CA, 94305, USA}
\affiliation{SLAC National Accelerator Laboratory, 2575 Sand Hill Road, Menlo Park, California 94025, USA}
\author{Jessica Nicole Aguilar}
\affiliation{Lawrence Berkeley National Laboratory, 1 Cyclotron Road, Berkeley, CA 94720, USA}
\author{Steven Ahlen}
\affiliation{Department of Physics, Boston University, 590 Commonwealth Avenue, Boston, MA 02215 USA}
\author{Anton \surname{Baleato Lizancos}}
\affiliation{Department of Physics, University of California, Berkeley, CA 94720, USA}
\affiliation{Lawrence Berkeley National Laboratory, 1 Cyclotron Road, Berkeley, CA 94720, USA}
\author{Florian Beutler}
\affiliation{Institute for Astronomy, University of Edinburgh, Royal Observatory, Blackford Hill, Edinburgh EH9 3HJ, UK}
\author{Davide Bianchi}
\affiliation{Dipartimento di Fisica ``Aldo Pontremoli'', Universit\`a degli Studi di Milano, Via Celoria 16, I-20133 Milano, Italy}
\affiliation{INAF-Osservatorio Astronomico di Brera, Via Brera 28, 20122 Milano, Italy}
\author{David Brooks}
\affiliation{Department of Physics \& Astronomy, University College London, Gower Street, London, WC1E 6BT, UK}
\author{Aurelio \surname{Carnero Rosell}}
\affiliation{Departamento de Astrof\'{\i}sica, Universidad de La Laguna (ULL), E-38206, La Laguna, Tenerife, Spain}
\affiliation{Instituto de Astrof\'{\i}sica de Canarias, C/ V\'{\i}a L\'{a}ctea, s/n, E-38205 La Laguna, Tenerife, Spain}
\author{Francisco Javier Castander}
\affiliation{Institut d'Estudis Espacials de Catalunya (IEEC), c/ Esteve Terradas 1, Edifici RDIT, Campus PMT-UPC, 08860 Castelldefels, Spain}
\affiliation{Institute of Space Sciences, ICE-CSIC, Campus UAB, Carrer de Can Magrans s/n, 08913 Bellaterra, Barcelona, Spain}
\author{Todd Claybaugh}
\affiliation{Lawrence Berkeley National Laboratory, 1 Cyclotron Road, Berkeley, CA 94720, USA}
\author{Andrei Cuceu}
\affiliation{Lawrence Berkeley National Laboratory, 1 Cyclotron Road, Berkeley, CA 94720, USA}
\author{Axel \surname{de la Macorra}}
\affiliation{Instituto de F\'{\i}sica, Universidad Nacional Aut\'{o}noma de M\'{e}xico,  Circuito de la Investigaci\'{o}n Cient\'{\i}fica, Ciudad Universitaria, Cd. de M\'{e}xico  C.~P.~04510,  M\'{e}xico}
\author{Jaime E. Forero-Romero}
\affiliation{Departamento de F\'isica, Universidad de los Andes, Cra. 1 No. 18A-10, Edificio Ip, CP 111711, Bogot\'a, Colombia}
\affiliation{Observatorio Astron\'omico, Universidad de los Andes, Cra. 1 No. 18A-10, Edificio H, CP 111711 Bogot\'a, Colombia}
\author{Enrique Gazta\~{n}aga}
\affiliation{Institut d'Estudis Espacials de Catalunya (IEEC), c/ Esteve Terradas 1, Edifici RDIT, Campus PMT-UPC, 08860 Castelldefels, Spain}
\affiliation{Institute of Cosmology and Gravitation, University of Portsmouth, Dennis Sciama Building, Portsmouth, PO1 3FX, UK}
\affiliation{Institute of Space Sciences, ICE-CSIC, Campus UAB, Carrer de Can Magrans s/n, 08913 Bellaterra, Barcelona, Spain}
\author{Satya \surname{Gontcho A Gontcho}}
\affiliation{University of Virginia, Department of Astronomy, Charlottesville, VA 22904, USA}
\author{Gaston Gutierrez}
\affiliation{Fermi National Accelerator Laboratory, PO Box 500, Batavia, IL 60510, USA}
\author{Klaus Honscheid}
\affiliation{Center for Cosmology and AstroParticle Physics, The Ohio State University, 191 West Woodruff Avenue, Columbus, OH 43210, USA}
\affiliation{Department of Physics, The Ohio State University, 191 West Woodruff Avenue, Columbus, OH 43210, USA}
\affiliation{The Ohio State University, Columbus, 43210 OH, USA}
\author{Dragan Huterer}
\affiliation{Department of Physics, University of Michigan, 450 Church Street, Ann Arbor, MI 48109, USA}
\affiliation{University of Michigan, 500 S. State Street, Ann Arbor, MI 48109, USA}
\author{Mustapha Ishak}
\affiliation{Department of Physics, The University of Texas at Dallas, 800 W. Campbell Rd., Richardson, TX 75080, USA}
\author{Stephanie Juneau}
\affiliation{NSF NOIRLab, 950 N. Cherry Ave., Tucson, AZ 85719, USA}
\author{Tanveer Karim}
\affiliation{Department of Astronomy \& Astrophysics, University of Toronto, Toronto, ON M5S 3H4, Canada}
\author{David Kirkby}
\affiliation{Department of Physics and Astronomy, University of California, Irvine, 92697, USA}
\author{Anthony Kremin}
\affiliation{Lawrence Berkeley National Laboratory, 1 Cyclotron Road, Berkeley, CA 94720, USA}
\author{Ofer Lahav}
\affiliation{Department of Physics \& Astronomy, University College London, Gower Street, London, WC1E 6BT, UK}
\author{Martin Landriau}
\affiliation{Lawrence Berkeley National Laboratory, 1 Cyclotron Road, Berkeley, CA 94720, USA}
\author{Laurent \surname{Le Guillou}}
\affiliation{Sorbonne Universit\'{e}, CNRS/IN2P3, Laboratoire de Physique Nucl\'{e}aire et de Hautes Energies (LPNHE), FR-75005 Paris, France}
\author{Martine Lokken}
\affiliation{Institut de F\'{i}sica d’Altes Energies (IFAE), The Barcelona Institute of Science and Technology, Edifici Cn, Campus UAB, 08193, Bellaterra (Barcelona), Spain}
\author{Marc Manera}
\affiliation{Departament de F\'{i}sica, Serra H\'{u}nter, Universitat Aut\`{o}noma de Barcelona, 08193 Bellaterra (Barcelona), Spain}
\affiliation{Institut de F\'{i}sica d’Altes Energies (IFAE), The Barcelona Institute of Science and Technology, Edifici Cn, Campus UAB, 08193, Bellaterra (Barcelona), Spain}
\author{Aaron Meisner}
\affiliation{NSF NOIRLab, 950 N. Cherry Ave., Tucson, AZ 85719, USA}
\author{Ramon Miquel}
\affiliation{Instituci\'{o} Catalana de Recerca i Estudis Avan\c{c}ats, Passeig de Llu\'{\i}s Companys, 23, 08010 Barcelona, Spain}
\affiliation{Institut de F\'{i}sica d’Altes Energies (IFAE), The Barcelona Institute of Science and Technology, Edifici Cn, Campus UAB, 08193, Bellaterra (Barcelona), Spain}
\author{John Moustakas}
\affiliation{Department of Physics and Astronomy, Siena University, 515 Loudon Road, Loudonville, NY 12211, USA}
\author{Seshadri Nadathur}
\affiliation{Institute of Cosmology and Gravitation, University of Portsmouth, Dennis Sciama Building, Portsmouth, PO1 3FX, UK}
\author{Jeffrey A. Newman}
\affiliation{Department of Physics \& Astronomy and Pittsburgh Particle Physics, Astrophysics, and Cosmology Center (PITT PACC), University of Pittsburgh, 3941 O'Hara Street, Pittsburgh, PA 15260, USA}
\author{Hernan Enrique Noriega}
\affiliation{Instituto de Ciencias F\'{\i}sicas, Universidad Nacional Aut\'onoma de M\'exico, Av. Universidad s/n, Cuernavaca, Morelos, C.~P.~62210, M\'exico}
\affiliation{Instituto de F\'{\i}sica, Universidad Nacional Aut\'{o}noma de M\'{e}xico,  Circuito de la Investigaci\'{o}n Cient\'{\i}fica, Ciudad Universitaria, Cd. de M\'{e}xico  C.~P.~04510,  M\'{e}xico}
\author{Will Percival}
\affiliation{Department of Physics and Astronomy, University of Waterloo, 200 University Ave W, Waterloo, ON N2L 3G1, Canada}
\affiliation{Perimeter Institute for Theoretical Physics, 31 Caroline St. North, Waterloo, ON N2L 2Y5, Canada}
\affiliation{Waterloo Centre for Astrophysics, University of Waterloo, 200 University Ave W, Waterloo, ON N2L 3G1, Canada}
\author{Ignasi P\'{e}rez-R\`{a}fols}
\affiliation{Departament de F\'isica, EEBE, Universitat Polit\`ecnica de Catalunya, c/Eduard Maristany 10, 08930 Barcelona, Spain}
\author{Francisco Prada}
\affiliation{Instituto de Astrof\'{i}sica de Andaluc\'{i}a (CSIC), Glorieta de la Astronom\'{i}a, s/n, E-18008 Granada, Spain}
\author{Corentin Ravoux}
\affiliation{Universit\'{e} Clermont-Auvergne, CNRS, LPCA, 63000 Clermont-Ferrand, France}
\author{Graziano Rossi}
\affiliation{Department of Physics and Astronomy, Sejong University, 209 Neungdong-ro, Gwangjin-gu, Seoul 05006, Republic of Korea}
\author{Lado Samushia}
\affiliation{Abastumani Astrophysical Observatory, Tbilisi, GE-0179, Georgia}
\affiliation{Department of Physics, Kansas State University, 116 Cardwell Hall, Manhattan, KS 66506, USA}
\author{Eusebio Sanchez}
\affiliation{CIEMAT, Avenida Complutense 40, E-28040 Madrid, Spain}
\author{Christoph Saulder}
\affiliation{Max Planck Institute for Extraterrestrial Physics, Gie\ss enbachstra\ss e 1, 85748 Garching, Germany}
\author{David Schlegel}
\affiliation{Lawrence Berkeley National Laboratory, 1 Cyclotron Road, Berkeley, CA 94720, USA}
\author{Michael Schubnell}
\affiliation{Department of Physics, University of Michigan, 450 Church Street, Ann Arbor, MI 48109, USA}
\affiliation{University of Michigan, 500 S. State Street, Ann Arbor, MI 48109, USA}
\author{Hee-Jong Seo}
\affiliation{Department of Physics \& Astronomy, Ohio University, 139 University Terrace, Athens, OH 45701, USA}
\author{Ma\l{}gorzata Siudek}
\affiliation{Institute of Space Sciences, ICE-CSIC, Campus UAB, Carrer de Can Magrans s/n, 08913 Bellaterra, Barcelona, Spain}
\affiliation{Instituto de Astrof\'{\i}sica de Canarias, C/ V\'{\i}a L\'{a}ctea, s/n, E-38205 La Laguna, Tenerife, Spain}
\author{Gregory Tarl\'{e}}
\affiliation{University of Michigan, 500 S. State Street, Ann Arbor, MI 48109, USA}
\author{Benjamin Alan Weaver}
\affiliation{NSF NOIRLab, 950 N. Cherry Ave., Tucson, AZ 85719, USA}
\author{Rongpu Zhou}
\affiliation{Lawrence Berkeley National Laboratory, 1 Cyclotron Road, Berkeley, CA 94720, USA}


\date{\today}

\begin{abstract}
Baryon feedback is a leading source of systematic uncertainty for cosmology from weak lensing, but measurements of the gas distribution around galaxies have largely relied on parametric profile models or simulation-calibrated frameworks. We present model independent measurements of the radial gas fraction profile around galaxies, combining galaxy-galaxy lensing and kinematic Sunyaev-Zel'dovich (kSZ) effect. We introduce a method that applies the same radial $\Delta\Sigma$ aperture filter to both the galaxy-galaxy lensing shear field and the velocity-weighted kSZ temperature maps; their ratio directly yields $\Delta\Sigma$ filtered gas fraction $f_{\rm gas}(R)$, the ratio of ionized gas to total matter as a function of projected radius. We apply this approach to DESI DR2 Bright Galaxy Survey (BGS, $\bar{z} \approx 0.26$) and Luminous Red Galaxy (LRG, $0.4 < z < 1.1$) samples, using ACT DR6 component-separated CMB maps for kSZ, and the HSC Year 3 shear catalog for weak lensing. After correcting for ACT beam suppression using a simulation-calibrated compensation factor, we detect baryon depletion relative to the cosmic mean baryon fraction at ${\rm SNR} = 17.2$ (BGS) and ${\rm SNR} = 13.8$ (LRG bin~1). Comparison with six hydrodynamical simulations from the Illustris, IllustrisTNG, SIMBA, and FLAMINGO suites shows that the fiducial FLAMINGO run most closely tracks the observed radial gas distribution in both samples, with Illustris-1 and TNG300-1 bracketing the data from below and above at small radii. 
We find a mild discrepancy between the kSZ measurements in the South and North Galactic Caps in one LRG redshift bin, which may indicate residual systematics.
We show that the ratio is robust to splits by stellar mass and to the removal of close pairs.
These results establish $\Delta\Sigma$ filtering of kSZ versus weak lensing signal as a model-independent probe of the baryon distribution around galaxies. 
\end{abstract}

\maketitle


\section{Introduction}
\label{sec:intro}

Weak gravitational lensing surveys have entered a precision regime in which baryonic feedback processes are a leading source of systematic uncertainty.
Energy injection from active galactic nuclei (AGN) and supernovae redistributes gas from the inner regions of halos to their outskirts and beyond, suppressing the matter power spectrum on scales $k \gtrsim 0.5\;h\,\mathrm{Mpc}^{-1}$ by up to $50\%$ relative to the dark-matter-only prediction~\cite{Chisari2018, Dalal2023}.
Unmodeled baryonic suppression biases cosmological parameter inference, such as the amplitude of matter fluctuations $S_8$, from current and upcoming lensing analyses.

Observational constraints on baryonic feedback have been accumulated across multiple complementary probes. X-ray emission from hot intracluster gas has mapped gas depletion in the centers of massive groups and clusters \cite{Vikhlinin2006_Xray, Akino2022, Popesso2026_eROSITA, Lau2025_Xray, Eckert2026, Seppi2026, Khalil2026}, while stacked thermal SZ measurements have extended this to lower halo masses \cite{Greco2015_tSZ, Pandey2022_tSZ, Liu2025_tSZ}, and to somewhat larger scales. 
\rev{However, both X-ray and tSZ signals are dominated by dense, hot gas, and therefore lose sensitivity to the cooler, lower-pressure gas that feedback displaces to and beyond the virial radius \cite{DeGraaff2019, Tanimura2019, Tanimura2020}.}

Weak gravitational lensing constrains the total projected matter distribution, but struggles to separate baryonic feedback from cosmological parameters on its own \cite{Mandelbaum2006, Amon2022_S8, Terasawa2025_HSC}. The kSZ effect has emerged as a direct tracer of the free-electron momentum field around galaxies to scales larger than probed previously \cite{ACTPol:2015teu, SchaanFerraro2021, Amodeo2021, Hadzhiyska2024a, RiedGuachalla2025, Hadzhiyska2025_missing_baryons}, and fast radio burst dispersion measures now offer a complementary line-of-sight probe of the warm-hot baryon budget \cite{McQuinn2014_FRBs, Macquart2020_FRBs, Medlock2025_FRBs,Reischke2025_FRBs, Leung2025_FRBs, Wayland2026_FRBs,Sharma2026_Backlight,Sharma2026_Baryons, Ebina2026}.
However, because these measurements probe either the gas or the total matter content but rarely both simultaneously, translating them into gas fractions has generally required parametric profile models coupled to 
galaxy clustering  \cite{Hadzhiyska2024a} or 
weak lensing \cite{Siegel2025} halo mass calibrations. These approaches 
are however not direct, \rev{and rely on modelling assumptions and are limited because of }
uncertainties with regard to satellite 
fractions and halo mass calibrations. 



Of these methods, weak galaxy-galaxy lensing (GGL) provides a direct measurement of the projected total matter distribution around galaxies.
The tangential shear of background source galaxies induced by a foreground lens sample yields the excess surface mass density $\Delta\Sigma(R)$ as a function of projected separation $R$, tracing the combined dark matter and baryonic mass profile~\cite{Sheldon2004, Mandelbaum2006, Leauthaud2017, Heydenreich2025}.
GGL is therefore sensitive to the total gravitational potential, but is unable on its own to distinguish the baryonic contribution from the dark matter component.

Meanwhile, the kinematic Sunyaev-Zel'dovich (kSZ) effect~\cite{Sunyaev1980a} offers a complementary probe of the free-electron distribution.
When CMB photons inverse-Compton scatter off electrons in the ionized gas surrounding galaxies, the resulting frequency-independent temperature shift is proportional to the line-of-sight peculiar momentum of the gas.
First detected in pairwise cluster motions \cite{Hand2012}, the kSZ signal can also be extracted by stacking CMB temperature maps at the positions of spectroscopic galaxies weighted by reconstructed radial velocities~\cite{SchaanFerraro2016, SchaanFerraro2021, Amodeo2021}.
Recent analyses combining ACT with photometric and spectroscopic galaxy samples from the Dark Energy Spectroscopic Instrument (DESI) have detected the kSZ signal at high significance across a range of halo masses and redshifts~\cite{Hadzhiyska2024a, RiedGuachalla2025,Qu2026_kSZ,Hadzhiyska2026_kSZ}.

To connect these diverse observational probes to the underlying baryon distribution, cosmological hydrodynamical simulations such as IllustrisTNG~\cite{Nelson2019_IllustrisTNG}, SIMBA~\cite{Dave2019_SIMBA}, and FLAMINGO~\cite{Schaye2023_FLAMINGO} self-consistently evolve gas, stars, and black holes, but must rely on sub-grid prescriptions for unresolved processes such as AGN jets, winds, and supernova-driven outflows.
An alternative class of methods avoid the expense of full hydrodynamical simulations by applying analytic baryon correction models (BCMs) to gravity-only $N$-body outputs \cite{Schneider2020}.
While both approaches have been successful at reproducing many observed galaxy and halo properties, the predicted matter power spectrum suppression varies significantly across simulation codes, feedback parameter choices, and modeling frameworks.
This model dependence propagates directly into the derived cosmological constraints, motivating the development of observational probes that can measure the impact of feedback without relying on a specific simulation or modeling prescription.

Several recent studies have combined kSZ measurements with gravitational lensing to constrain the baryon and gas content of halos.
Hadzhiyska et al.~\cite{Hadzhiyska2025_missing_baryons} combined kSZ measurements with CMB lensing to measure the gas fraction in galaxy groups, while Bigwood et al.~\cite{Bigwood2024_ksz_wl} jointly analyzed kSZ profiles and cosmic shear from the Dark Energy Survey (DES) to constrain baryon feedback models.
Siegel et al.~\cite{Siegel2025, Siegel2025_feedback} combined X-ray, kSZ, and weak lensing measurements to constrain the matter power spectrum suppression, and McCarthy et al.~\cite{McCarthy2025_FLAMINGO} and Bigwood et al.~\cite{Bigwood2025} used the FLAMINGO and other hydrodynamical simulations to interpret joint kSZ and GGL observations as benchmarks for AGN feedback models.
These analyses represent important advances, but generally rely on parametric profile models or simulation-calibrated frameworks to connect the observables to the underlying baryon distribution.

In this work, we introduce a model-independent approach to measuring the radial gas fraction around galaxies.
\rev{Here, we define the term model-independent in a specific sense: no model of the baryon distribution enters the construction of the observable. No functional form is assumed for either profile and no fit is performed; the total matter is measured directly by lensing around the same galaxies, so no halo mass calibration, stellar-to-halo mass relation, or satellite fraction is needed; and the identical filter is applied to both signals, so the ratio is formed bin by bin at fixed angular scale. Prior gas fraction measurements have relied on at least one such intermediary, whether a parametric gas profile fit to the kSZ signal \cite{Amodeo2021, Bigwood2024_ksz_wl}, or a halo mass inferred from clustering or from a halo model fit to a lensing profile \cite{Hadzhiyska2024a, Hadzhiyska2025_missing_baryons},
or a simulation-based forward model of the observable \cite{McCarthy2025_FLAMINGO, Siegel2025}. 
Each is well motivated, but each imports a prior on the gas distribution into a measurement whose purpose is to constrain it. }

We apply the same $\Delta\Sigma$ aperture filter to both the GGL shear signal and the kSZ signal, projecting each observable into identical radial bins. While these 
filters are compensated, meaning that 
they integrate to zero, in practice their 
radial profiles are quite similar to the 3d profile of gas and total matter. 
Their radial bin ratio yields a compensated filter profile of the gas fraction $f_{\rm gas}(R)$. 
To our knowledge, this is the first measurement of the gas fraction profile in which an identical $\Delta\Sigma$ aperture filter is applied to both observables, eliminating the need for profile modeling or simulation based calibrations.

We apply this method to spectroscopic lens samples drawn from the second data release of DESI~\cite{DESI2016a, DESI2022.KP1.Instr}, including the Bright Galaxy Survey (BGS)~\cite{Hahn2023_DESIBGS} at low redshift and Luminous Red Galaxies (LRGs)~\cite{Zhou2023_DESILRG} spanning $0.4 < z < 1.1$.
The kSZ signal is extracted from ACT DR6 component-separated CMB temperature maps~\cite{Coulton2024}, using velocity reconstructions presented previously in papers by Qu et al.~\cite{Qu2026_kSZ} and Hadzhiyska et al.~\cite{Hadzhiyska2026_kSZ}.
The weak lensing shear measurements come from the third-year shape catalog of the Hyper Suprime-Cam Subaru Strategic Program (HSC Y3)~\cite{HSC2022.DR3, Li2022_HSCY3_ShearCatalog}.
The combination of DESI's dense spectroscopic coverage, ACT's high-resolution CMB maps, and HSC's deep imaging enables precise, matched measurements of both the gas and total matter profiles around the same galaxy populations.

The remainder of this paper is organized as follows.
In Sec.~\ref{sec:data}, we describe the DESI, ACT, and HSC datasets.
In Sec.~\ref{sec:methods}, we present the theory and methodology for the GGL and kSZ measurements and their combination.
In Sec.~\ref{sec:results}, we present our main results.
In Sec.~\ref{sec:simulations}, we compare our measurements to predictions from hydrodynamical simulations.
We conclude in Sec.~\ref{sec:conclusion}.

\section{Data}
\label{sec:data}

\begin{figure*}[htbp!]
    \centering
    \includegraphics[width=0.9\textwidth]{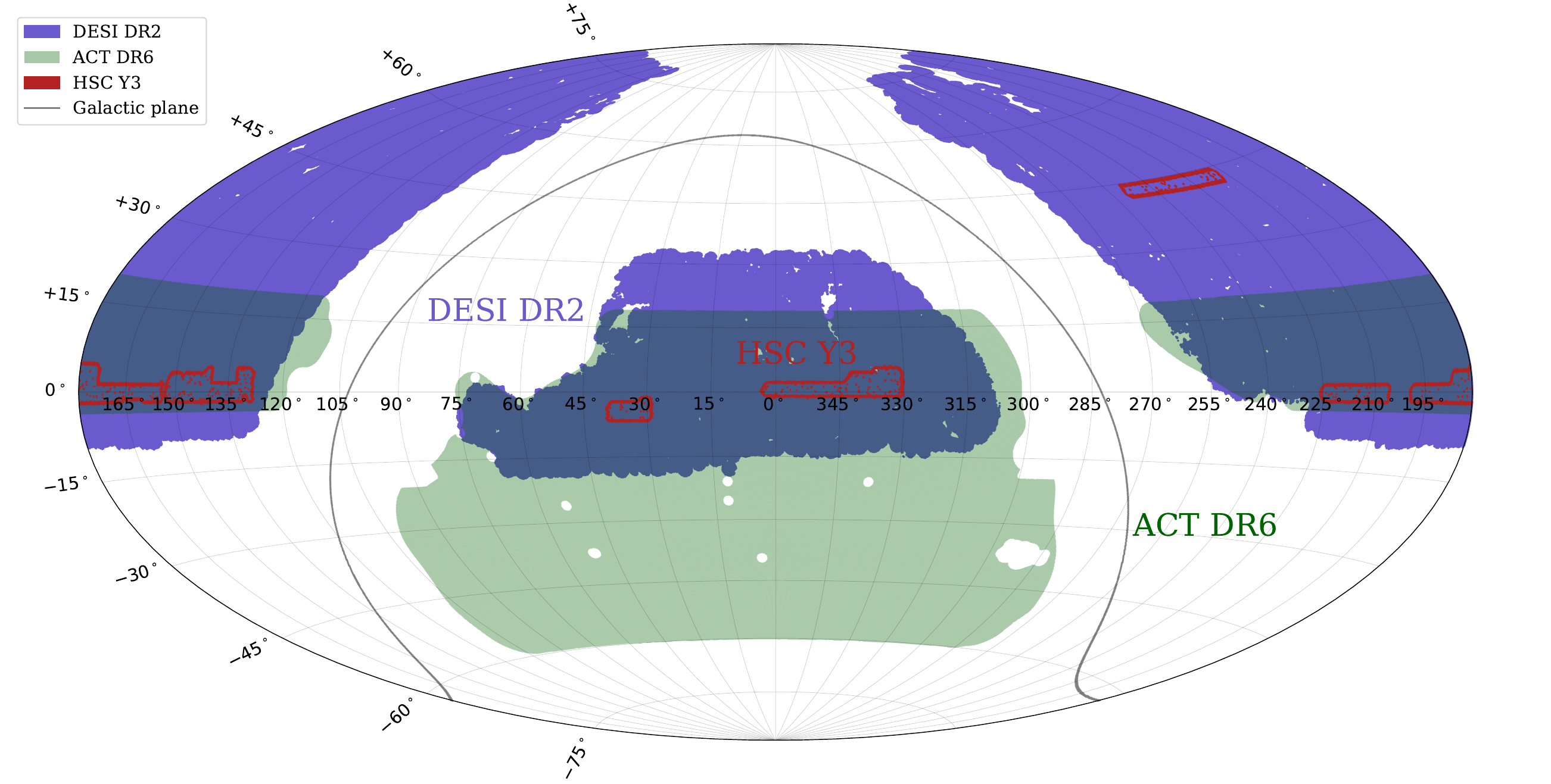}
    \caption{Sky footprints of the DESI DR2 (blue), ACT DR6 (green), and HSC Y3 (red). The DESI--ACT overlap is used for the kSZ measurements and the DESI--HSC overlap for the weak lensing measurements (see also Table~\ref{tab:surveyArea}).}
    \label{fig:footprints}
\end{figure*}

\subsection{Dark Energy Spectroscopic Instrument}

The Dark Energy Spectroscopic Instrument (DESI) is a robotic, fiber-fed, highly multiplexed spectroscopic surveyor that operates on the Mayall 4-meter telescope at Kitt Peak National Observatory \cite{DESI2022.KP1.Instr}. DESI, which can obtain simultaneous spectra of almost 5,000 objects over a 3° field \cite{DESI2016b, Silber2023, Miller2024}, is conducting an eight-year survey of about $17{,}000\,\mathrm{deg}^2$ of the sky. The full survey will lead to 63 million spectroscopically-confirmed galaxies and quasars spanning the spectroscopic redshifts $0.1 < z < 1.6$, substantially more compared to the initial forecasts of 39 million \cite{DESI2016a}. The sheer scale of the DESI survey necessitates multiple supporting software pipelines and products \cite{Guy2023,Schlafly2023}.

DESI target selection is based on the DESI Legacy Imaging Surveys \cite{Zou2017, Dey2019}, which provide deep optical and infrared imaging over the DESI footprint. The target selection process makes use of machine learning methods and imaging quality cuts to identify and classify various classes of objects \cite{Myers2023}, including Bright Galaxy Survey (BGS) targets \cite{Hahn2023_DESIBGS}, Luminous Red Galaxies (LRGs) \cite{Zhou2023_DESILRG}, Emission Line Galaxies (ELGs) \cite{Raichoor2023_DESIELG}, and quasars \cite{Chaussidon2023_DESIQSOs}. Each class of targets is selected based on specific photometric criteria designed to optimize the scientific return of the survey.
This present work makes use of the DESI BGS-BRIGHT and LRG samples from the DESI DR2 catalogues. The BGS-BRIGHT sample included in this work has a magnitude cut of $r < 20.2$, and consists of 4.85 million galaxies with a mean redshift of $0.26$. The LRG sample contains 4.4 million galaxies with a median redshift of $0.74$.

For the DESI BGS tracer, we additionally follow Hadzhiyska et al. \cite{Hadzhiyska2026_kSZ} and match to the DESI DR2 value-added catalogue for physical properties derived from the CIGALE code \cite{Siudek2024_CIGALE, Siudek2025_CIGALE}. The CIGALE code fits galaxy spectral energy distributions (SEDs) from the optical to the infrared, using stellar population synthesis, delayed star formation histories with optional bursts, nebular emission, dust attenuation and re-emission, and AGN templates. The fits incorporate photometry from the Wide-field Infrared Survey Explorer (WISE) and yield posteriors for stellar mass, star formation rate, and AGN fraction. The DESI DR2 Value Added Catalogue (VAC) contains $\sim$ 26 million galaxies across all primary DESI target types. This value-added stellar mass catalogue is used in Section~\ref{sec:mass_bins} to illustrate the mass dependence of our measurements.


We make use of the DESI DR2 catalogue, with the combined BGS catalogue as well as the LRG catalogue divided into four redshift bins, for both the galaxy-galaxy lensing measurements and the kSZ measurements, ensuring that both profiles are derived from the same underlying galaxy population.


\begin{table}
    \centering
    \setlength{\tabcolsep}{10pt}  
    \begin{tabular}{l|cc}
    \hline\hline
     Sample & $\overline{z}$ & $N_{\rm objects}$\\
    \hline
     BGS       &   0.26 & 2,224,546  \\
     LRG bin 1 &   0.51 &   597,109  \\
     LRG bin 2 &   0.71 &   918,354  \\
     LRG bin 3 &   0.87 &   731,638  \\
     LRG bin 4 &   1.01 &   297,823  \\
    \hline\hline
    \end{tabular}
    \caption{Mean redshift and number of objects for the DESI DR2 BGS and LRG samples within the ACT DR6 footprint, after sample cuts. The LRG sample is divided into four redshift bins, with the mean redshift and number of galaxies of each bin listed here.}
    \label{tab:samples}
\end{table}


\subsection{Atacama Cosmology Telescope}

For the CMB component of our measurements, we make use of harmonic-space Internal Linear Combination (ILC) maps from the Atacama Cosmology Telescope (ACT) Data Release 6 \cite{Coulton2024}. ACT was a six-meter telescope located in the Atacama Desert in Chile, which observed the CMB with high sensitivity and angular resolution between 2007 and 2022. ACT surveyed roughly $18{,}000\,\mathrm{deg}^2$ of the sky (before masking) in three frequency bands: f90 (77-112 GHz), f150 (124-172 GHz), and f220 (182-277 GHz). 

The ACT maps used in this work are the result of a component separation process that combines data from all three frequency bands to produce a clean CMB map with minimized foreground contamination. The ILC method works by linearly combining the multi-frequency maps in harmonic space, weighting them in such a way as to minimize the variance of the resulting map while preserving the CMB signal. This process effectively removes contributions from foreground sources such as galactic dust and extragalactic point sources, which can contaminate the kSZ signal.

The maps are from the night-time portion of the ACT DR6 data (version dr6.01), provided at 0.5 arcmin pixel resolution and smoothed with a Gaussian beam of FWHM = 1.6 arcmin. To further mitigate contamination from massive galaxy clusters whose thermal Sunyaev-Zel'dovich emission can bias the kSZ signal, we apply the ACT DR6 cluster mask \cite{Hilton2021}, which places holes of 6 arcminute radius around 4,922 standard clusters and 10 arcminute radius around 436 extended clusters. Following the approach of \cite{SchaanFerraro2021}, when extracting kSZ profiles, galaxies are additionally excluded if any masked pixel falls within their stacking aperture, or if their filtered temperature at any aperture exceeds a threshold corresponding to the $5\sigma$ equivalent for the sample size; these per-galaxy cuts are described in Section~\ref{sec:ksz_methods}. After applying all masks, the effective ACT DR6 sky area used for our kSZ measurements is $13{,}320\,\mathrm{deg}^2$ (Table~\ref{tab:surveyArea}), consistent with the footprint used in the companion analyses of Qu et al.~\cite{Qu2026_kSZ} and Hadzhiyska et al.~\cite{Hadzhiyska2026_kSZ}.



\subsection{Hyper Suprime-Cam}

The Hyper Suprime-Cam Subaru Strategic Program (HSC) is a wide-field optical imaging camera mounted on the Subaru Telescope, located at the Mauna Kea Observatory in Hawaii. HSC has a field of view of 1.5 degrees in diameter and is equipped with a 870-megapixel camera, making it one of the most powerful instruments for wide-field imaging in the optical regime \cite{HSC2018}. The HSC survey has been designed to cover a large area of the sky with deep imaging in multiple bands, making it an ideal dataset for weak lensing studies.

For the lensing component of this work, we make use of the shear catalogue from the HSC Survey Data Release 3 (DR3) \cite{HSC2022.DR3,Li2022_HSCY3_ShearCatalog}, which provides shape measurements for millions of galaxies across the sky. The HSC Y3 shear catalogue includes measurements of galaxy shapes, photometric redshifts, and other relevant properties that are essential for performing galaxy-galaxy lensing analyses.

However, rather than making use of the official HSC photometric redshift calibrations \cite{Rau2023}, which results in wider priors in lensing measurements \cite{Dalal2023}, we make use of a newer re-calibration of the HSC Photometric redshifts using the Dark Energy Spectroscopic Instrument (DESI) spectroscopic redshifts \cite{ChoppinDeJanvry2025}. This re-calibration significantly improves the precision of the photometric redshifts for the HSC shear catalogue, which in turn leads to more precise galaxy-galaxy lensing measurements.

We choose to make use of the HSC Y3 shear catalogue for our lensing measurements rather than catalogues from the Dark Energy Survey (DES) and the Kilo-Degree Survey (KiDS) due to the superior depth of the HSC. HSC catalogues extend further in redshift than DES and KiDS, allowing us to extend our measurements to the entirety of DESI LRGs. Though as we will show in Sec. \ref{sec:results}, the highest redshift samples are still dominated by noise in the lensing and SZ measurements.


\subsection{Footprints and Overlap}

The footprints of all 3 surveys used in this analysis are shown in Figure~\ref{fig:footprints}. Table \ref{tab:surveyArea} summarizes the sky areas covered by each survey and their overlaps. The overlapping area between DESI DR2 and ACT DR6 is $6{,}709\,\mathrm{deg}^2$, which is used for the kSZ measurements. The overlapping area between DESI DR2 and HSC Y3 is $546.3\,\mathrm{deg}^2$, which is used for the galaxy-galaxy lensing measurements. We note that the sky areas of ACT DR6 represent the masked area after removing regions with high galactic dust contamination, detected tSZ clusters and point sources, which is the relevant area for our kSZ measurements. 
\rev{The kSZ and lensing profiles are therefore measured over different and largely disjoint sky areas. The DESI BGS and LRG samples are selected with uniform photometric criteria across the footprint \cite{Hahn2023_DESIBGS,Zhou2023_DESILRG}. We verify in Appendix~\ref{app:consistency_tests} that the lens population inside the HSC footprint is statistically indistinguishable from the full sample in the properties that set the signal, so that the ratio of Eq.~\ref{eq:fgas_ratio} is well defined.}

\begin{table}
    \centering
    \setlength{\tabcolsep}{10pt}  
    \begin{tabular}{l|c}
        \hline \hline
        Survey & Area\\
        \hline
        DESI DR2 &  $14{,}282\,\mathrm{deg}^2$\\
        ACT DR6 &  $13{,}320\,\mathrm{deg}^2$\\
        HSC Y3 &  $548.8\,\mathrm{deg}^2$\\
        DESI DR2 $\cap$ ACT DR6 &  $6{,}709\,\mathrm{deg}^2$\\
        DESI DR2 $\cap$ HSC Y3 &  $546.3\,\mathrm{deg}^2$\\
        \hline \hline
    \end{tabular}    
    \caption{Sky areas of the DESI DR2, ACT DR6 (masked), and HSC Y3 surveys used in this work, as well as the overlapping areas between DESI and ACT, and between DESI and HSC. We note that the sky areas of ACT DR6 represent the masked area after removing regions with high galactic dust contamination, tSZ clusters and point sources, which is the relevant area for our kSZ measurements.}
    \label{tab:surveyArea}
\end{table}

\section{Theory and Methodology}
\label{sec:methods}


Measurements of galaxy-galaxy lensing profiles and kSZ profiles are computed using traditionally different observables. As a result, direct comparison of these two profiles is not straightforward; applying a common $\Delta\Sigma$ aperture filter to both (Sections~\ref{sec:lensing_methods}--\ref{sec:combine_measurements}) is the key step that makes it possible.

Throughout this work we assume a flat $\Lambda$CDM cosmology consistent with the Planck 2018 TT,TE,EE+lowE+lensing+BAO results \citep{Planck2018_cosmological_parameters}: $h = 0.6766$, $\Omega_m = 0.3097$, $\Omega_b = 0.0490$, yielding a cosmic baryon fraction $f_b = \Omega_b / \Omega_m = 0.158$. All simulated-profile comparisons use the cosmological parameters native to each simulation (Section~\ref{sec:simulations}).
\rev{Each simulation retains its own cosmological parameters, and each simulated $f_{\rm gas}$ profile is normalised by its own $\Omega_b / \Omega_m$ value. We expect the residual differences to be subdominant to the variations in feedback prescription, since $f_{\rm gas}$ is a ratio of two profiles stacked around the same objects at fixed comoving number density, so that common amplitude factors cancel and cosmology enters principally through a small shift in the abundance-matched halo mass.}

\subsection{Galaxy-Galaxy Lensing}
\label{sec:lensing_methods}

Galaxy-galaxy lensing (GGL) measures the coherent distortion of background source galaxy shapes induced by the gravitational potential of foreground lens galaxies. This weak lensing signal provides a direct probe of the total matter distribution around the lenses, including both luminous and dark matter, making it complementary to other tracers that are more sensitive to baryonic components.

The fundamental observable in GGL is the tangential shear, $\gamma_t$, which quantifies the preferential alignment of background galaxy ellipticities in the direction tangential to the lens-source separation vector. In the weak lensing regime, where lensing distortions are small, the observed ellipticity of a source galaxy is approximately equal to the reduced shear $g = \gamma/(1-\kappa)$, where $\gamma$ is the shear and $\kappa$ is the convergence. For weak lensing where $\kappa \ll 1$, we can approximate the measured tangential ellipticity as directly proportional to the tangential shear: $\gamma_t \approx g_t$.

The tangential shear is directly related to the excess surface density profile through the lensing efficiency:
\begin{equation}
    \Delta\Sigma(R) = \Sigma_{\rm crit} \langle \gamma_t(R) \rangle,
\end{equation}
where the angle brackets denote an average over all lens-source pairs at projected separation $R$, and $\Sigma_{\rm crit}$ is the critical surface density for lensing, defined as:
\begin{equation}
    \Sigma_{\rm crit} = \frac{c^2}{4\pi G} \frac{D_s}{D_l D_{ls}} (1+z_l)^{-2}.
\end{equation}
Here, $D_s$, $D_l$, and $D_{ls}$ are the angular diameter distances to the source, to the lens, and between the lens and source, respectively. The factor $(1+z_l)^{-2}$ converts $\Sigma_{\rm crit}$ from physical to comoving surface density units, consistent with the lensing pipeline convention.

The excess surface density $\Delta\Sigma(R)$ itself has a direct physical interpretation in terms of the projected matter distribution:
\begin{equation}
\label{eq:dsigma_lensing}
    \Delta\Sigma(R) = \bar{\Sigma}(<R) - \Sigma(R),
\end{equation}
where $\bar{\Sigma}(<R)$ is the mean surface density within projected radius $R$, and $\Sigma(R)$ is the azimuthally-averaged surface density at radius $R$. This filter is insensitive to the constant
mass density contribution, which cancels out in $\Delta\Sigma(R)$.

In practice, we measure $\Delta\Sigma(R)$ by stacking the tangential shear profiles around a large sample of lens galaxies. For each lens-source pair, we compute the tangential component of the source's ellipticity relative to the lens position, weight it by the inverse critical surface density and additional factors accounting for measurement uncertainties, and average over all pairs in radial bins. This stacking procedure enhances the signal-to-noise ratio and provides an average profile representative of the lens sample.

We employ the publicly available code \texttt{dsigma} \citep{dsigma2022}, first used in \cite{Lange2024}, to compute the galaxy-galaxy lensing profiles. The code takes as input the positions and spectroscopic redshifts of lens galaxies from DESI, the positions, photometric redshifts, and shape measurements of source galaxies from HSC, and computes the weighted tangential shear profile $\gamma_t(R)$ in radial bins. It then converts this to the excess surface density profile $\Delta\Sigma(R)$ by applying the appropriate critical surface density for each lens-source pair based on their geometric configuration. 

\subsubsection{Lens and Source Selection}

Lens galaxies are drawn from the DESI DR2 spectroscopic catalogues and required to have valid systematic weights ($w_{\rm sys} > 0$) and finite spectroscopic redshifts and sky positions. The BGS-BRIGHT sample is used as a single redshift bin spanning the full sample redshift range, with no additional cuts beyond the BGS-BRIGHT magnitude selection ($r < 20.2$, Section~\ref{sec:data}). The LRG sample is divided into four spectroscopic redshift bins: $0.4 < z < 0.6$, $0.6 < z < 0.8$, $0.8 < z < 0.95$, and $0.95 < z < 1.1$.

Source galaxies are drawn from the HSC Y3 shear catalogue \citep{HSC2022.DR3}, using shapes measured with the re-Gaussianization PSF correction algorithm \citep{Hirata2003} in the $i$ band. 
We use three tomographic source bins defined by the DNNz photometric redshift \citep{ChoppinDeJanvry2025}: bin~2 ($0.6 \leq z_p < 0.9$), bin~3 ($0.9 \leq z_p < 1.2$), and bin~4 ($1.2 \leq z_p < 1.5$). The bin assignment for each source is taken from the \texttt{hsc\_y3\_zbin} column of the HSC Y3 catalogue. The calibrated redshift distribution $n(z)$ for each bin, derived from DESI clustering redshifts following Choppin de Janvry et al.~\cite{ChoppinDeJanvry2025}, is used to compute an effective critical surface density integrated over the source population \citep{Heydenreich2025, dsigma2022}:
\begin{equation}
    \Sigma_{\rm crit,eff}^{-1}(z_l) = \int dz_s \, n(z_s) \, \Sigma_{\rm crit}^{-1}(z_l, z_s),
    \label{eq:sigma_crit_eff}
\end{equation}
where $n(z_s)$ is the normalised source redshift distribution for the tomographic bin. This single effective value is applied to all lens-source pairs in a given (lens bin, source bin) combination, avoiding the need for per-pair photometric redshift estimates. All HSC source bins sufficiently behind a given lens redshift bin are used independently, and their $\Delta\Sigma$ profiles are shown as separate measurements in the results. HSC bin~2 ($0.6 \leq z_p < 0.9$) is not used for LRG bin~2 ($0.6 < z < 0.8$), since the two bins overlap substantially in redshift and HSC bin~2 sources would contribute little lensing signal.

For each lens-source pair, we impose a minimum redshift separation $\Delta z = \bar{z}_s - z_l > 0.1$, where $\bar{z}_s$ is the mean redshift of the source tomographic bin derived from $n(z)$, to suppress contamination from galaxies physically associated with the lens. For HSC bin~2 paired with LRG bin~1, this threshold is relaxed to $\Delta z > 0.0$, since the lower edge of that source bin ($z_p = 0.6$) coincides with the upper edge of the lens bin, and a stricter cut would discard a large fraction of valid pairs.

In addition to the boost factor discussed below, two systematic corrections are applied to the $\Delta\Sigma$ profiles. First, we perform random subtraction: $\Delta\Sigma$ is measured around a catalogue of DESI random points drawn from the same angular mask as the lens sample and subtracted from the lens signal, removing additive systematic contributions from survey geometry and large-scale structure along the line of sight \citep{Heydenreich2025}. Second, shear responsivity and per-source multiplicative shear bias corrections are applied using the HSC Y3 shear calibration, where the multiplicative bias $m$ is stored per source in the HSC catalogue. No photo-z dilution correction is required, as the full tomographic $n(z)$ is used in $\Sigma_{\rm crit}$ rather than individual photometric redshift point estimates. We also do not apply a lens magnification bias correction. \rev{This additive term grows with redshift and matters most at large $r_p$. For the BGS and LRG bin 1 samples on which our quantitative conclusions rest, and on the $\lesssim 6$~cMpc scales used here, it is well below our statistical uncertainty on $f_{\rm gas}^{\rm obs}$; we do not use the higher-redshift LRG bins for quantitative inference.}

\subsubsection{Radial Binning and Covariance}

We measure $\Delta\Sigma(R)$ in 9 linearly spaced angular bins spanning $1$--$6$ arcminutes, converted to comoving projected separations at the mean spectroscopic redshift of each lens bin. All projected separations, surface densities, and the $\Sigma_{\rm crit}$ computation are performed in comoving coordinates throughout, consistent with the comoving-frame kSZ $\Delta\Sigma$ filter used for direct comparison. This angular range is chosen to match the scale of the kSZ $\Delta\Sigma$ filter, enabling a direct radial bin-by-bin comparison between the two profiles. Depending on the lens redshift, the resulting comoving scales span approximately $0.3$--$6\,{\rm cMpc}$.

Statistical uncertainties are estimated from a jackknife covariance matrix computed over 100 spatially contiguous fields, defined by $k$-means clustering of the lens positions on the sky. The full covariance matrix is retained and propagated through to the kSZ-to-lensing ratio.

\subsubsection{The Boost Factor}

Galaxy-galaxy lensing measurements on small scales are affected by source-lens clustering: the physical clustering of source galaxies around the lens population causes an excess of source galaxies in the lens environment relative to the survey-wide background distribution. Since galaxies physically associated with the lens contribute little to no lensing signal (their redshifts are similar to those of the lenses), this excess dilutes the measured $\Delta\Sigma$ profile. A multiplicative boost factor correction is commonly applied to account for this effect \citep{Simet2015, Leauthaud2017}.

The boost factor $b(R)$ at projected separation $R$ is estimated by comparing the weighted number of source galaxies around true lens positions to that around random lens positions,
\begin{equation}
    b(R) = \frac{\sum_r w_{\rm sys}}{\sum_l w_{\rm sys}} \cdot \frac{\sum_{ls} w_{\rm sys}\, w_{ls}}{\sum_{rs} w_{\rm sys}\, w_{rs}},
\end{equation}
where the sums run over all lens--source ($ls$) and random--source ($rs$) pairs within each radial bin, $w_{\rm sys}$ are systematic weights, and $w_{ls}$, $w_{rs}$ are the lens-source and random-source pair weights, respectively \citep{Heydenreich2025}. When $b > 1$, the uncorrected profile is diluted and the corrected profile is $\Delta\Sigma = b\,\Delta\Sigma_{\rm raw}$.

However, the boost factor estimator is itself susceptible to several additional biases of uncertain magnitude that depend on the specific telescope, instrument, and data reduction pipeline, including blending of foreground and lens light into source photometry \citep{Simet2015}, contamination of the source catalog by lens-associated galaxies \citep{Leauthaud2017}, and source magnification. Following \cite{Heydenreich2025}, we therefore do not apply a boost factor correction to our lensing profiles. 
Furthermore, Lange et al.~\cite{Lange2024} demonstrated that boost factor corrections induce only percent-level changes in the lensing data vector and that differences between surveys are negligible.

\subsection{The kSZ Effect}
\label{sec:ksz_methods}

The kinematic Sunyaev-Zel'dovich (kSZ) effect is a secondary anisotropy in the Cosmic Microwave Background (CMB). It is caused by the inverse Compton scattering of CMB photons off free electrons that have a non-zero bulk velocity relative to the CMB rest frame. The free electrons are typically found in the hot gas within galaxy clusters and groups, as well as in the intergalactic medium.

The kSZ effect manifests as a small temperature fluctuation in the CMB, which can be expressed as:
\begin{equation}
\label{eq:T_ksz}
    \frac{\Delta T_{\rm kSZ}(\hat{\mathbf{n}})}{T_{\rm CMB}} = 
    - \frac{1}{c} \int \frac{d\chi}{1+z} n_e(\chi \hat{\mathbf{n}}, z) \,\sigma_T\, e^{-\tau} \,(\mathbf{v} \cdot \hat{\mathbf{n}}),
\end{equation}
where $c$ is the speed of light, $\chi$ is the comoving distance along the line of sight, $z$ is the redshift corresponding to that distance, $n_e$ is the free electron number density, $\sigma_T$ is the Thomson scattering cross-section, $\mathbf{v} \cdot \hat{\mathbf{n}}$ is the line-of-sight component of the electron peculiar velocity and $\tau$ is the optical depth to Thomson scattering, defined as:
\begin{equation}
    \tau(\chi) = \int_0^\chi d\chi' \frac{n_e(\chi') \sigma_T}{1+z'}.
\end{equation}
In the case of groups and galaxy clusters, the optical depth $\tau$ is typically small, and the exponential factor $e^{-\tau}$ can be approximated as unity. Therefore, the kSZ temperature fluctuation can be simplified to:
\begin{equation}
    \frac{\Delta T_{\rm kSZ}(\hat{\mathbf{n}})}{T_{\rm CMB}} \approx - \frac{1}{c} \int \frac{d\chi}{1+z} n_e(\chi \hat{\mathbf{n}}, z) \,\sigma_T\, (\mathbf{v} \cdot \hat{\mathbf{n}})
\end{equation}
For an individual halo $i$, the kSZ temperature fluctuation can be expressed as:
\begin{equation}
\label{eq:T_ksz_individual}
    \Delta T_{\rm kSZ}^i \approx T_{\rm CMB} \tau_i \frac{v_{\rm los,i}^{\rm halo}}{c},
\end{equation}
where $\tau_i$ is the optical depth of the halo, and $v_{\rm los,i}^{\rm halo}$ is the line-of-sight velocity of the halo. A key challenge in measuring the kSZ effect then is to obtain reliable estimates of the line-of-sight velocity of the halo. This process will be described in more detail in the following section.

\subsubsection{Velocity Reconstruction}
\label{sec:vel_recon}

The kSZ effect for each galaxy (Eq.~\ref{eq:T_ksz_individual}) requires a per-galaxy estimate of the line-of-sight peculiar velocity $v_{\rm los}$. 
We make use of the same BAO reconstruction formalism as in \citep{RiedGuachalla2024, Hadzhiyska2024b}, using the same velocity-reconstructed catalogues in \citep{Qu2026_kSZ} and \citep{Hadzhiyska2026_kSZ} in this work to ensure consistency.

The linearized continuity equation relates the velocity field $\mathbf{v}$ of galaxies to the galaxy overdensity $\delta_g$. Including the redshift space distortions (RSD), this continuity equation is expressed as:
\begin{equation}
    \nabla \cdot \mathbf{v} + \frac{f}{b} \nabla \cdot [(\mathbf{v} \cdot \mathbf{\hat{n}})\mathbf{\hat{n}}] = -a Hf\frac{\delta_g}{b},
\end{equation}
where $a$ is the scale factor, $H$ the Hubble parameter, $f\equiv \rm{d} \ln{D(a)} /\rm{d} \ln{a}$ the linear growth factor, and $b$ is the large scale galaxy bias. The second term on the left hand side represents the correction induced by RSD, with $v^{\rm LOS} = \mathbf{v} \cdot \mathbf{\hat{n}} \approx v^{\rm LOS}_e $ the electron bulk velocity of the halo.


The DESI DR2 catalogues provide both pre-reconstruction and post-reconstruction galaxy positions.  BAO reconstruction displaces each galaxy by an estimate of its Lagrangian displacement $\boldsymbol{\Psi}$, computed from the galaxy density field smoothed on a scale $R_s = 15\,h^{-1}\,\mathrm{Mpc}$ using the Zel'dovich approximation.  Under the symmetric reconstruction (RecSym) convention, the line-of-sight component of this displacement for galaxy $i$ is
\begin{equation}
    \Psi_{{\rm los},i} = \frac{\chi_{{\rm pre},i} - \chi_{{\rm post},i}}{1 + f(z_i)},
    \label{eq:psi_los}
\end{equation}
where $\chi_{\rm pre}$ and $\chi_{\rm post}$ are the comoving distances to the pre- and post-reconstruction positions, and the factor $(1+f)^{-1}$ corrects for the contribution of redshift-space distortions to the displacement, with $f(z) = d\ln D / d\ln a$ the linear growth rate.  The corresponding line-of-sight peculiar velocity follows from the linearised continuity equation,
\begin{equation}
    \hat{v}_{{\rm los},i} = a(z_i)\,H(z_i)\,f(z_i)\,\Psi_{{\rm los},i},
    \label{eq:v_recon}
\end{equation}
where $a$ is the scale factor and $H(z)$ is the Hubble parameter.  The reconstruction is performed with the MultiGrid solver in the \texttt{pyrecon} package.

For the LRG sample we make use of the official DESI velocity reconstruction, with the full reconstruction procedure is described in Qu et al.~\cite{Qu2026_kSZ}; for the BGS sample we make use of the custom DESI BGS velocity reconstruction as described and used in Hadzhiyska et al.~\cite{Hadzhiyska2026_kSZ}. The methodology and its performance on $N$-body simulations are characterised in Ried Guachalla et al.~\cite{RiedGuachalla2024} and Hadzhiyska et al.~\cite{Hadzhiyska2024b}.

\subsubsection{Velocity Reconstruction Factor}
\label{sec:vel_recon_factor}

The kSZ estimator of Hadzhiyska et al.~\cite{Hadzhiyska2024a} stacks CMB temperature stamps weighted by the estimated line-of-sight velocity $\hat{v}_i$ of each foreground galaxy, reconstructed from the surrounding density field. The stacked signal scales as $r_V \, \tau \, v_{\rm los} T_{\rm CMB} / c$, where $r_V$ is the cross-correlation coefficient between the estimated and true halo line-of-sight velocities, while $v_{\rm los}$ is the RMS line-of-sight velocity. 
This quality factor $r_V < 1$ reflects imperfections in the velocity reconstruction: on large scales this is due to 
contributions from modes
larger than the survey, 
while on small scales this 
is due to the breakdown 
of linearized continuity 
equation caused by nonlinear 
effects. 
\rev{We note however that the scale dependence of the reconstruction fidelity is integrated into the calibrated value~\cite{RiedGuachalla2024,Hadzhiyska2024b}.}
These corrections must be accounted for in the kSZ measurement; the measured stacked kSZ profiles were divided by $r_V$ to obtain the corrected kSZ stacked measurements. We adopt $r_V = 0.65$ for the spectroscopic LRG sample following Qu et al.~\cite{Qu2026_kSZ} and $r_V = 0.64$ for the BGS sample following Hadzhiyska et al.~\cite{Hadzhiyska2026_kSZ}.


\subsubsection{Stacking}

We measure kSZ profiles by stacking velocity-weighted postage stamps of the ACT DR6 ILC temperature map.  For each galaxy $i$ that overlaps the ACT footprint, a small cutout is extracted at its sky position at $0.25\,\mathrm{arcmin}$ resolution using a cylindrical equal-area (CEA) projection.  Objects whose cutouts intersect a masked or point-source-contaminated pixel are excluded.  Statistical outliers (objects whose filtered temperature magnitude exceeds a threshold corresponding to the $5\sigma$ equivalent for the sample size) are also rejected.

We zero the mean velocity of each stacking sample before stacking, to ensure that the resulting kSZ profile is not biased by tSZ or CIB contamination. The filtered temperature $T_i(R)$ at angular radius $R$ is obtained by applying the $\Delta\Sigma$ aperture filter to each stamp (described below).  The kSZ profile is then estimated using a velocity-weighted, uniform mean estimator:
\begin{equation}
    \hat{T}_{\rm kSZ}(R) =
        \frac{\hat{\sigma}_v}{r_V \sum_i \hat{v}_i^2}
        \sum_i \hat{v}_i\,T_i(R),
    \label{eq:ksz_estimator}
\end{equation}
where $\hat{v}_i = -\hat{v}_{{\rm los},i}/c$ is the dimensionless reconstructed line-of-sight velocity, $\hat{\sigma}_v = \mathrm{std}(\hat{v}_i)$ is the standard deviation of the reconstructed velocities across the sample, and $r_V$ is the velocity reconstruction cross-correlation coefficient (Section~\ref{sec:vel_recon_factor}).  
For correct $r_V$, Eq. \ref{eq:ksz_estimator} returns an unbiased stacked profile despite the imperfect velocity reconstruction. 

\subsubsection{The $\Delta\Sigma$ filter.}

\begin{figure}
    \centering
    \includegraphics[width=0.9\columnwidth]{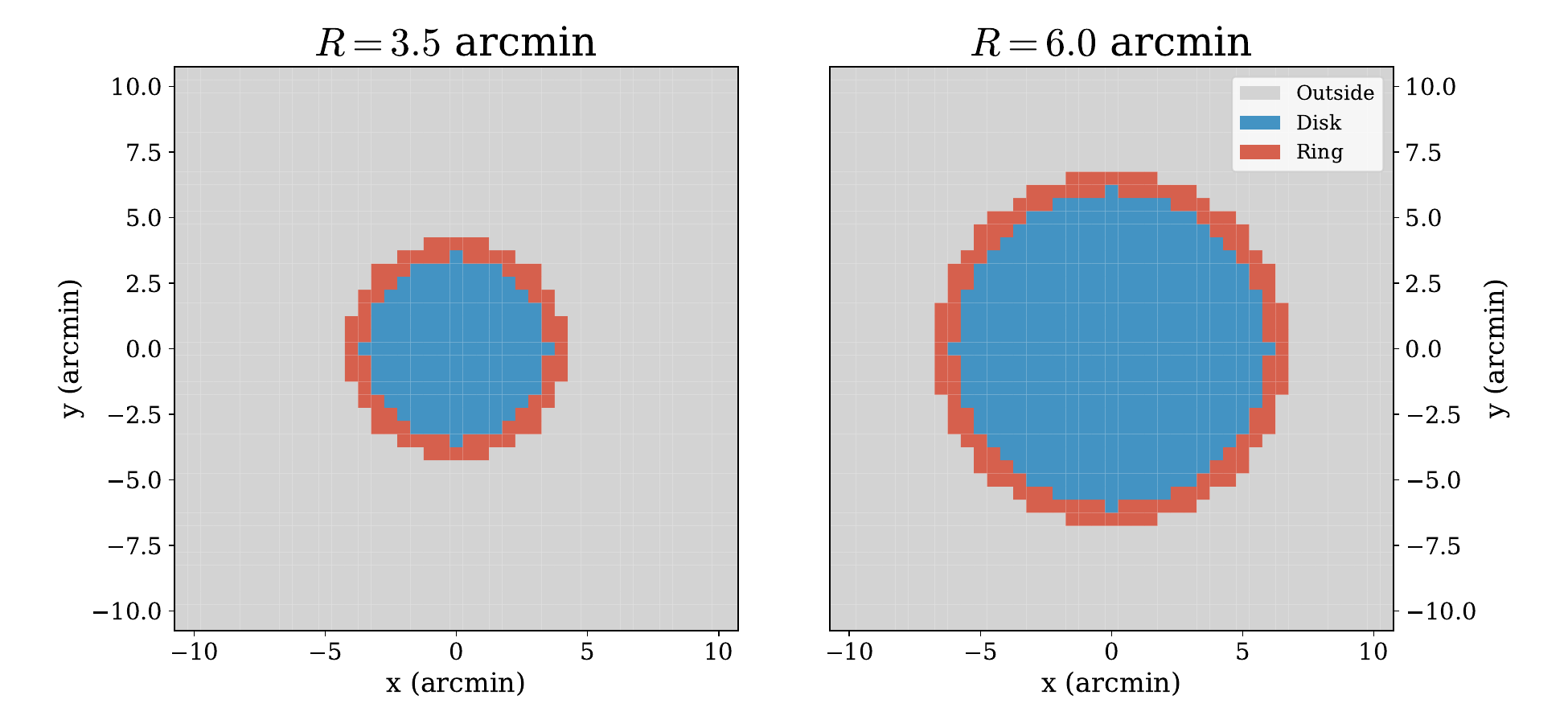}
    \caption{The $\Delta\Sigma$ pixel filter on the CMB maps, evaluated at two example radii. The filter is pixel-averaged over all positive weighted pixels (blue) subtracting all negative weighted pixels (red), achieving a net zero weight to suppress large-scale CMB modes. The filter is evaluated at 9 linearly spaced radii spanning $1$--$6\,\mathrm{arcmin}$, matching the GGL radial bins.}
    \label{fig:dsigma_filter}
\end{figure}

Rather than the Compensated Aperture Photometry (CAP) filter used in prior kSZ measurements with these galaxy samples~\cite{SchaanFerraro2021, Hadzhiyska2024a, RiedGuachalla2025, Qu2026_kSZ, Hadzhiyska2026_kSZ}, we apply the $\Delta\Sigma$ aperture filter to each temperature stamp.  At angular radius $R$ the filter returns the mean temperature within a disk minus the mean temperature in a thin enclosing annulus:
\begin{equation}
\label{eq:dsigma_ksz}
    \Delta\Sigma_{\rm kSZ}(R) \equiv \bar{\Sigma}(<R) - \bar{\Sigma}(R,\,R{+}\delta R),
\end{equation}
where
\begin{equation}
\label{eq:sigma_disk}
    \bar{\Sigma}(<R) = \frac{1}{A(<R)}\int_{|\boldsymbol{\theta}|<R}
        T(\boldsymbol{\theta})\,d^2\boldsymbol{\theta}
\end{equation}
is the mean CMB temperature within radius $R$, and
\begin{equation}
\label{eq:sigma_ring}
    \bar{\Sigma}(R,\,R{+}\delta R) = \frac{1}{A_{\rm ring}}
        \int_{R < |\boldsymbol{\theta}| < R+\delta R}
        T(\boldsymbol{\theta})\,d^2\boldsymbol{\theta}
\end{equation}
is the mean temperature in an annulus of width $\delta R = 0.75\,\mathrm{arcmin}$. This value is chosen because it is larger than the diagonal pixel size on the ACT DR6 CMB maps. The areas $A(<R)$ and $A_{\rm ring}$ are the solid angles of the disk and annulus, respectively, computed using pixel counting. We evaluate $\Delta\Sigma_{\rm kSZ}$ at 9 linearly-spaced radii spanning $1$--$6\,\mathrm{arcmin}$, matching the GGL radial bins described in Section~\ref{sec:methods}. A sample $\Delta\Sigma$ filter is shown in Figure~\ref{fig:dsigma_filter}.

The use of the $\Delta\Sigma$ filter for the kSZ measurements is a key methodological innovation of this work, enabling the bin-by-bin feedback ratio $f_{\rm gas}^{\rm obs}(R)$ without any modelling of the profile shapes, as described in Section~\ref{sec:combine_measurements}. In order to ensure statistical consistency, we evaluate the $\Delta\Sigma$ filter in Appendix~\ref{app:systematic_tests}.

The CAP filter, used in prior work~\cite{SchaanFerraro2021, Amodeo2021, RiedGuachalla2025,Qu2026_kSZ, Hadzhiyska2026_kSZ}, instead integrates the temperature over a disk of radius $\theta_d$ and subtracts the integral over an equal-area annulus:
\begin{equation}
\label{eq:cap_filter}
    T_{\rm CAP}(\theta_d) =
        \int_{|\boldsymbol{\theta}|<\theta_d} T\,d^2\boldsymbol{\theta}
        - \int_{\theta_d < |\boldsymbol{\theta}| < \theta_d\sqrt{2}} T\,d^2\boldsymbol{\theta},
\end{equation}
with output in units of $[\mu\mathrm{K}\cdot\mathrm{arcmin^2}]$.  Both filters satisfy $\int \mathcal{F}\,d^2\boldsymbol{\theta} = 0$, suppressing contributions from large-scale CMB modes, but the $\Delta\Sigma$ filter additionally divides each term by its solid angle, yielding a mean temperature in $[\mu\mathrm{K}]$.  This is the direct CMB-temperature analogue of the lensing observable in Eq.~\ref{eq:dsigma_lensing}, and the shared functional form is what enables the bin-by-bin feedback ratio $f_{\rm gas}^{\rm obs}(R)$ (Eq.~\ref{eq:fgas_ratio}) without any modelling of the profile shape. However, in practice we still need to apply the CMB beam corrections on small scales, discussed further below.

\subsubsection{Covariance Matrix Estimation}
\label{sec:ksz_cov}

Statistical uncertainties on the kSZ $\Delta\Sigma$ profiles are estimated via bootstrap resampling at the galaxy level.
For each galaxy sample, we first apply the catalog mask: objects are required to fall within the ACT footprint, to have their full disk-and-ring aperture free of point-source-masked pixels, and to pass an outlier rejection cut in which any object whose filtered temperature magnitude at any aperture exceeds a sample-size-dependent threshold equivalent to a 5$\sigma$ per-object false-positive probability is discarded.
We then draw $N_{\rm boot} = 10{,}000$ bootstrap resamples with replacement from the $N_{\rm gal}$ objects passing this mask, computing the full velocity-weighted estimator (Eq.~\ref{eq:ksz_estimator}) at all nine aperture radii for each resample.
Each resample is seeded deterministically by its index for reproducibility. 

The $9 \times 9$ bootstrap covariance matrix $\mathbf{C}_{\rm kSZ}$ is then estimated as the sample covariance of the $N_{\rm boot}$ bootstrap profile vectors:
\begin{equation}
\begin{split}
    C_{{\rm kSZ},\,ij} = \frac{1}{N_{\rm boot}-1}\sum_{k=1}^{N_{\rm boot}}
        &\bigl[\hat{T}^{(k)}(R_i) - \bar{T}(R_i)\bigr] \\
        \times &\bigl[\hat{T}^{(k)}(R_j) - \bar{T}(R_j)\bigr],
\end{split}
    \label{eq:ksz_cov}
\end{equation}
where $\hat{T}^{(k)}(R_i)$ is the kSZ profile at aperture $R_i$ for the $k$-th bootstrap resample and $\bar{T}(R_i)$ is the mean across all resamples.
The full matrix is propagated into the uncertainty on $f_{\rm gas}^{\rm obs}(R)$ as described in Section~\ref{sec:combine_measurements}; the diagonal elements give the $1\sigma$ error bars shown on the kSZ profiles in Section~\ref{sec:results}.

\subsection{Combining the Measurements}
\label{sec:combine_measurements}

One of the central methodological contributions of this work is to apply the same $\Delta\Sigma$ filter introduced above to both the velocity-weighted CMB temperature maps and the HSC shear field. This produces kSZ and lensing profiles in identical radial bins and with a shared physical interpretation as projected excess surface density, enabling a direct, bin-by-bin comparison without invoking any particular baryon feedback model.

\subsubsection{Unit Conversion}

The lensing profile $\Delta\Sigma_{\rm lens}(R)$ is measured in units of surface mass density ($M_\odot\,\mathrm{pc}^{-2}$), while the kSZ profile $\Delta\Sigma_{\rm kSZ}(R)$ is measured in CMB temperature units ($\mu\mathrm{K}$). 
To form a one-to-one comparison, we must convert the units of one profile to match the other. We choose to convert the lensing profile into temperature units: given the lensing $\Delta\Sigma$ what kSZ signal would we expect if the gas traced the total matter at the cosmic mean baryon fraction. This choice is motivated by the significantly higher signal-to-noise of the lensing profile, making it a more stable base for the unit conversion.

The kSZ temperature fluctuation for an individual halo $i$ (Eq. \ref{eq:T_ksz_individual}) involves the halo optical depth $\tau_i$. The projected optical depth at angular radius $R$ from the halo centre is related to the gas surface density $\Delta\Sigma_{\rm gas}(R)$ by
\begin{equation}
    \tau(R) = \frac{\sigma_T \, \Delta\Sigma_{\rm gas}(R)}{\mu_e \, m_p},
    \label{eq:tau_sigma}
\end{equation}
where $\sigma_T$ is the Thomson scattering cross-section, $m_p$ is the proton mass, and $\mu_e = 1.14$ is the mean molecular weight per free electron for a fully ionized plasma of primordial composition (hydrogen mass fraction $X \approx 0.76$).


To construct the kSZ prediction under the null hypothesis of no feedback, we assume that the gas traces the total matter distribution at the cosmic mean baryon fraction. In this case, the gas surface density can be expressed as
\begin{equation}
    \Delta\Sigma_{\rm gas}(R) = f_b \, \Delta\Sigma_{\rm lens}(R),
    \label{eq:gas_null}
\end{equation}
where $f_b \equiv \Omega_b / \Omega_m = 0.158$ is the cosmic baryon fraction, adopting the Planck 2018 cosmological parameters stated in Section~\ref{sec:methods}. Substituting Eqs.~(\ref{eq:tau_sigma}) and~(\ref{eq:gas_null}) into the single-halo kSZ expression (Eq.~\ref{eq:T_ksz_individual}), the predicted kSZ profile under this null hypothesis is
\begin{equation}
    \Delta\Sigma_{\rm kSZ}(R) = \alpha_{\rm conv} \, \Delta\Sigma_{\rm phys}(R),
    \label{eq:T_DS_relation}
\end{equation}
where $\Delta\Sigma_{\rm phys}(R)$ is the physical (not comoving) lensing excess surface density, and the conversion coefficient is
\begin{equation}
    \alpha_{\rm conv} = T_{\rm CMB} \, \frac{v_{\rm los}}{c} \cdot \frac{f_b \, \sigma_T}{\mu_e \, m_p}.
    \label{eq:alpha_conv}
\end{equation}
Here $v_{\rm los} = 300\,\mathrm{km\,s^{-1}}$ is the fiducial true root-mean-square velocity scale for the lens sample (following Hadzhiyska et al.~\cite{Hadzhiyska2025_missing_baryons}).
Because the GGL pipeline returns comoving surface densities $\Delta\Sigma_{\rm com}$, we convert to physical units before applying Eq.~(\ref{eq:T_DS_relation}):
\begin{equation}
    \Delta\Sigma_{\rm phys}(R) = (1+z_l)^{2} \, \Delta\Sigma_{\rm com}(R),
    \label{eq:dsigma_phys}
\end{equation}
where $z_l$ is the mean spectroscopic redshift of the lens bin.

\subsubsection{The Gas Fraction}

Because the kSZ effect is sourced by free electrons, an ideal (beam-free) kSZ $\Delta\Sigma$ signal is directly proportional to the gas excess surface density $\Delta\Sigma_{\rm gas}(R)$, with proportionality factor $\alpha_{\rm conv}/f_b = T_{\rm CMB}\,(v_{\rm los}/c)\,\sigma_T/(\mu_e\,m_p)$ following from Eqs.~(\ref{eq:tau_sigma}) and~(\ref{eq:T_ksz_individual}).
Since $\Delta\Sigma_{\rm lens}(R)$ independently measures the total projected matter distribution, the ratio of gas to total matter defines the gas fraction relative to the cosmic baryon fraction:
\begin{equation}
    f_{\rm gas}(R) \equiv \frac{\Delta\Sigma_{\rm gas}(R)}{f_b \, \Delta\Sigma_{\rm lens}(R)}.
    \label{eq:ratio_physical}
\end{equation}
When $f_{\rm gas}(R) = 1$, the gas traces the total matter distribution at the cosmic mean baryon fraction, consistent with negligible feedback. Values $f_{\rm gas}(R) < 1$ indicate baryon depletion at radius $R$: feedback processes have expelled gas beyond the scales being probed. The radial dependence of $f_{\rm gas}(R)$ therefore encodes the scale-dependent redistribution of gas induced by stellar and AGN feedback, without requiring comparison to any specific simulation.

We note that $f_{\rm gas}$, as defined here, is a \emph{differential projected} quantity: it measures the $\Delta\Sigma$-filtered ratio of gas to total matter at each radial bin \rev{in the 2D projected space}, not the traditional cumulative gas fraction $f_{\rm gas}(<R)$ enclosed within a projected \rev{(2D)} or spherical \rev{3D} radius. 
\rev{We note this difference as all three (2D $\Delta \Sigma$, 2D cumulative and 3D cumulative) gas fractions are not identical. The companion paper to this work \cite{LiuInprep2026} as well as other recent work \cite{Kadir2026_SZ} has shown that the 2D cumulative profiles may overestimate gas fractions in comparison to the 3D case.}
However, we do make the note that based on the companion work in Liu et al.~\cite{LiuInprep2026}, the definition of gas fraction in this work is closely related to the cumulative 3D gas fraction $f_{\rm gas}(<R)$ in simulations. We refer the reader to that work for a detailed discussion of the relationship between the measured projected $f_{\rm gas}(R)$ and the cumulative 3D gas fraction $f_{\rm gas}(<R)$, with the note that the ratio observable presented in this work is robust to effects of projection and filtering, especially due to the compensated filter.
The same consideration applies to the observed analogue $f_{\rm gas}^{\rm obs}$ introduced below. This convention follows Hadzhiyska et al.~\cite{Hadzhiyska2026_kSZ, Hadzhiyska2025_missing_baryons}, though we make use of the $\Delta\Sigma$ filter in this work as opposed to the CAP filter.

In practice, however, the kSZ signal measured from the ACT CMB maps is convolved with the instrumental beam, so the ratio of the observed kSZ and lensing profiles gives not $f_{\rm gas}(R)$ but a beam-suppressed observable $f_{\rm gas}^{\rm obs}(R)$, defined in the following subsection.

\subsubsection{The Observed Gas Fraction and Beam Effects}
\label{sec:beam_effects}

The quantity directly accessible from our measurements is the ratio of the observed kSZ $\Delta\Sigma$ profile to the lensing $\Delta\Sigma$ profile. Unlike $\Delta\Sigma_{\rm lens}$, the kSZ profile is convolved with the ACT instrumental beam, so this ratio does not directly yield the true gas fraction. We call this observed quantity the \emph{observed gas fraction}:
\begin{equation}
\label{eq:fgas_ratio}
    f_{\rm gas}^{\rm obs}(R) \equiv \frac{\Delta \Sigma_{\rm kSZ}(R)}{\alpha_{\rm conv} \, \Delta\Sigma_{\rm lens}(R)}.
\end{equation}
The difference between $f_{\rm gas}^{\rm obs}$ and the true gas fraction $f_{\rm gas}$ (Eq.~\ref{eq:ratio_physical}) is due to the finite resolution of the ACT CMB maps. The ACT DR6 maps used in this work have a pixel resolution of $0.5\,\mathrm{arcmin}$ and a Gaussian beam with FWHM $= 1.6\,\mathrm{arcmin}$, which smooths the CMB temperature field and suppresses $\Delta\Sigma_{\rm kSZ}(R)$ relative to its ideal value $\Delta\Sigma_{\rm kSZ}^{\rm ideal}(R)$. The lensing measurement, derived from HSC optical imaging, is unaffected by the CMB beam. As a result, $f_{\rm gas}^{\rm obs}(R) < f_{\rm gas}(R)$ in general. The suppression is strongest at angular scales $R \lesssim 2\,\mathrm{arcmin}$, comparable to the beam FWHM, and falls below $\sim$10\% at larger radii where the beam transfer function approaches unity.

Crucially, the beam suppression factor $f_{\rm gas}^{\rm obs}(R)/f_{\rm gas}(R)$ is largely independent of the underlying feedback model. Its magnitude is set primarily by the beam profile and the angular scale of the measurement, not by the astrophysics of gas redistribution around haloes. This feedback-independence is demonstrated using cosmological hydrodynamical simulations in Section~\ref{sec:simulations} (Figure~\ref{fig:beam_compensation_factor}), where six simulations from the Illustris, IllustrisTNG, SIMBA, and FLAMINGO suites, spanning a wide range of AGN feedback strengths and implementations, predict nearly identical beam suppression curves.

Since the beam suppression factor is feedback-independent, it is both amenable to forward-modelling in simulations and correctable via a simulation-calibrated factor. In Section~\ref{sec:simulated_observations}, we take the former approach: the ACT beam is convolved into the simulated kSZ profiles so that data and simulations reflect $f_{\rm gas}^{\rm obs}$ on equal footing. In Section~\ref{sec:beam_compensated_fgas}, we take the latter, applying a simulation-calibrated compensation factor to recover the beam-compensated gas fraction $f_{\rm gas}(R)$ from our measurements.

\subsubsection{Error Propagation}
\label{sec:error_propagation}

\rev{The CMB noise in the ACT maps and the shape noise in the HSC catalogue are independent as both come from different observables, so a cross-covariance between $\Delta\Sigma_{\rm kSZ}$ and $\Delta\Sigma_{\rm lens}$ is statistically negligible.}
The covariance of $f_{\rm gas}^{\rm obs}(R)$ can therefore be propagated from the two independent covariance matrices, $\mathbf{C}^{\rm kSZ}$ (Section~\ref{sec:ksz_cov}) and $\mathbf{C}^{\rm lens}$ (Section~\ref{sec:lensing_profiles}), via first-order Gaussian error propagation. Writing the $i$-th radial bin estimator as the ratio $f_i \equiv y_{1,i} / y_{2,i}$, where $y_{1,i} = \Delta\Sigma_{\rm kSZ}(R_i)$ and $y_{2,i} = \alpha_{\rm conv}\,\Delta\Sigma_{\rm lens}(R_i)$, the propagated covariance is
\begin{equation}
\label{eq:fgas_cov}
    C^{f_{\rm gas}}_{ij} = \frac{C^{\rm kSZ}_{ij}}{y_{2,i}\, y_{2,j}}
    + \frac{y_{1,i}\, y_{1,j}}{y_{2,i}^2\, y_{2,j}^2}\, C^{\rm lens}_{ij}.
\end{equation}
In matrix form, $\mathbf{C}^{f_{\rm gas}} = \mathbf{J}\, \mathbf{C}_{\rm block}\, \mathbf{J}^{\rm T}$, where $\mathbf{C}_{\rm block} = \mathrm{diag}(\mathbf{C}^{\rm kSZ}, \mathbf{C}^{\rm lens})$ is the block-diagonal joint covariance and $\mathbf{J}$ is the $N \times 2N$ Jacobian with elements $\partial f_i / \partial y_{1,j} = \delta_{ij}/y_{2,i}$ and $\partial f_i / \partial y_{2,j} = -\delta_{ij}\, y_{1,i}/y_{2,i}^{2}$. Before forming the ratio, the lensing profile $\Delta\Sigma_{\rm lens}$ is converted to temperature units via Eqs.~(\ref{eq:T_DS_relation})--(\ref{eq:dsigma_phys}), so that $y_1$ and $y_2$, and their covariances $\mathbf{C}^{\rm kSZ}$ and $\mathbf{C}^{\rm lens}$, are expressed in consistent units ($\mu\mathrm{K}$) and Eq.~(\ref{eq:fgas_cov}) yields a dimensionless covariance for $f_{\rm gas}^{\rm obs}$.





\section{Results}
\label{sec:results}

In this section, we present the measurements of the kSZ $\Delta\Sigma$ profiles and the galaxy-galaxy lensing $\Delta\Sigma$ profiles for the DESI BGS and LRG samples, and combine them to extract information about baryon feedback following Section~\ref{sec:combine_measurements}.

\subsection{Measurements of the kSZ Signal}

\begin{figure}
    \centering
    \includegraphics[width=\columnwidth]{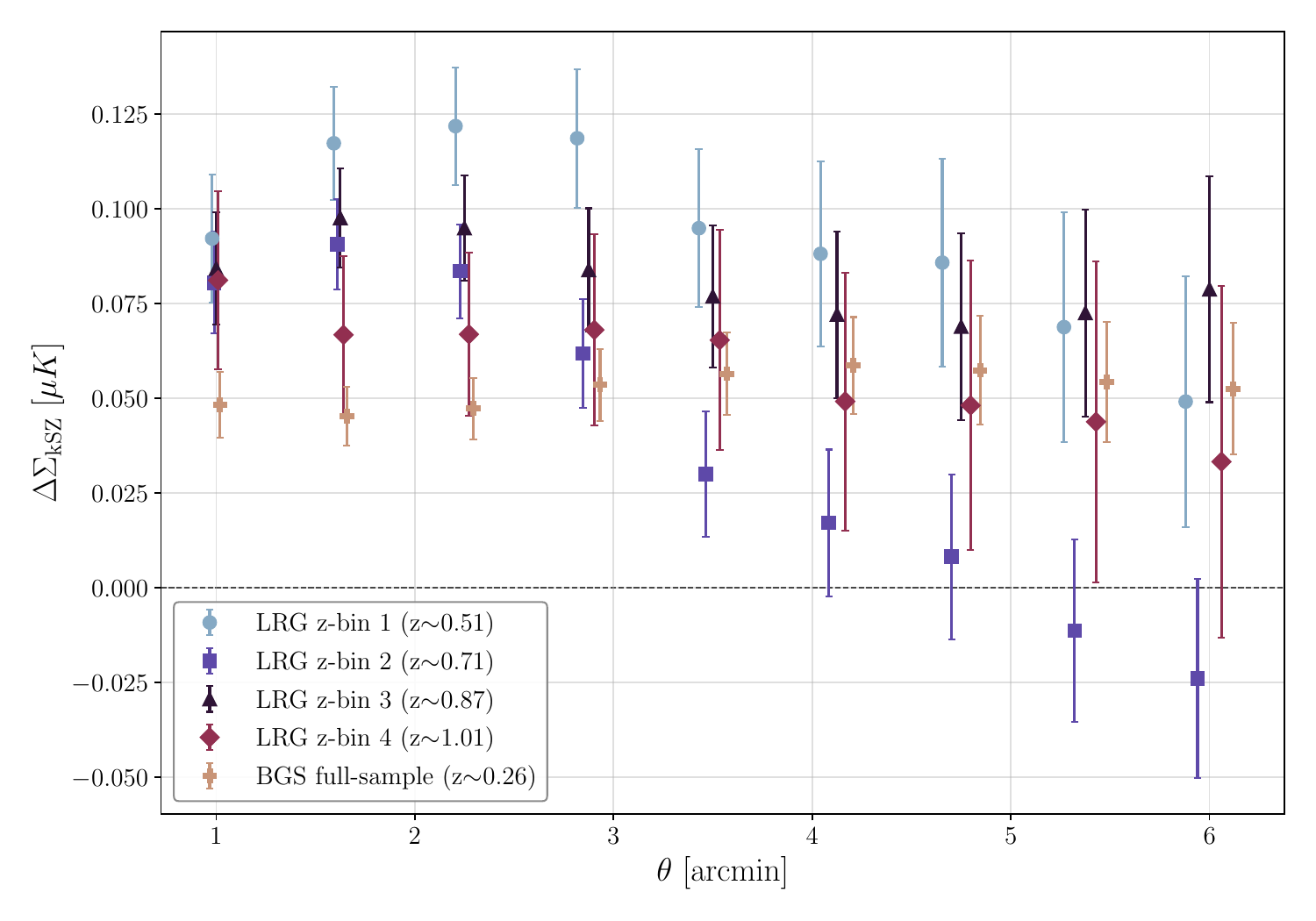}
    \caption{Measurements of the kSZ $\Delta\Sigma$ profiles for the DESI BGS and LRG samples, obtained by applying the $\Delta\Sigma$ aperture filter to velocity-weighted ACT DR6 ILC temperature stamps. The error bars are $1\sigma$ statistical uncertainties from bootstrap resampling over 10000 resamples. 
    Though the different galaxy samples are stacked over the same angular scale, the corresponding physical scales differ due to the varying redshifts of the samples.}
    \label{fig:ksz_profiles}
\end{figure}

\begin{figure}
    \centering
    \includegraphics[width=0.85\columnwidth]{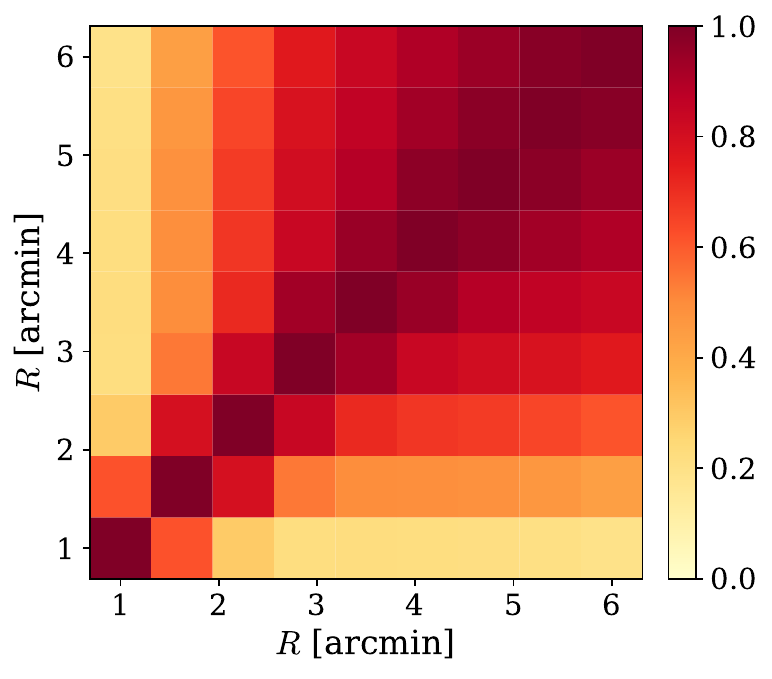}
    \caption{Correlation Matrix obtained from bootstrap covariance for the $\Delta\Sigma$-filtered kSZ profiles for the DESI BGS sample. The covariance is estimated from 10000 bootstrap resamples of the lens sample, as described in Section~\ref{sec:ksz_cov}. The off-diagonal elements indicate correlations between different radial bins, which arise from the shared galaxy sample and the nature of the $\Delta\Sigma$ filter. Similar covariance matrices are obtained for the LRG samples.}
    \label{fig:ksz_cov}
\end{figure}

The measurements of the kSZ $\Delta\Sigma$ profiles for the DESI BGS and LRG samples are shown in Figure~\ref{fig:ksz_profiles}, while the correlation matrix for the BGS sample is shown in Figure~\ref{fig:ksz_cov} (the other samples have similar correlation matrices so are not shown).

\begin{table}
    \centering
    \setlength{\tabcolsep}{10pt}  
    \begin{tabular}{l|ccc}
    \hline\hline
     Sample & $\chi^2_{\rm null}$ & dof &$\mathrm{SNR}_{\rm null}$\\
    \hline
     BGS       &   59.2 & 9 &  7.69 \\
     LRG bin 1 &  107.5 & 9 & 10.37 \\
     LRG bin 2 &  114.0 & 9 & 10.68 \\
     LRG bin 3 &   71.5 & 9 &  8.46 \\
     LRG bin 4 &   27.7 & 9 &  5.26 \\
    \hline\hline
    \end{tabular}
    \caption{$\chi^2$ and signal to noise of the kSZ measurements. The $\mathrm{SNR}_{\rm null} = \sqrt{\chi^2_{\rm null}}$ is computed with respect to the null hypothesis of no kSZ signal. The degrees of freedom (dof) are equal to the number of radial bins (9) in the $\Delta\Sigma$ profiles.
    }
    \label{tab:SNR}
\end{table}



We note the unphysical negative signal at large aperture for LRG bin~2, which will be further discussed in the combined signal below, as well as in Appendix \ref{app:consistency_tests}.

\subsection{Measurements of the Galaxy-Galaxy Lensing Signal}
\label{sec:lensing_profiles}

\begin{figure}[h]
    \centering
    \includegraphics[width=\columnwidth]{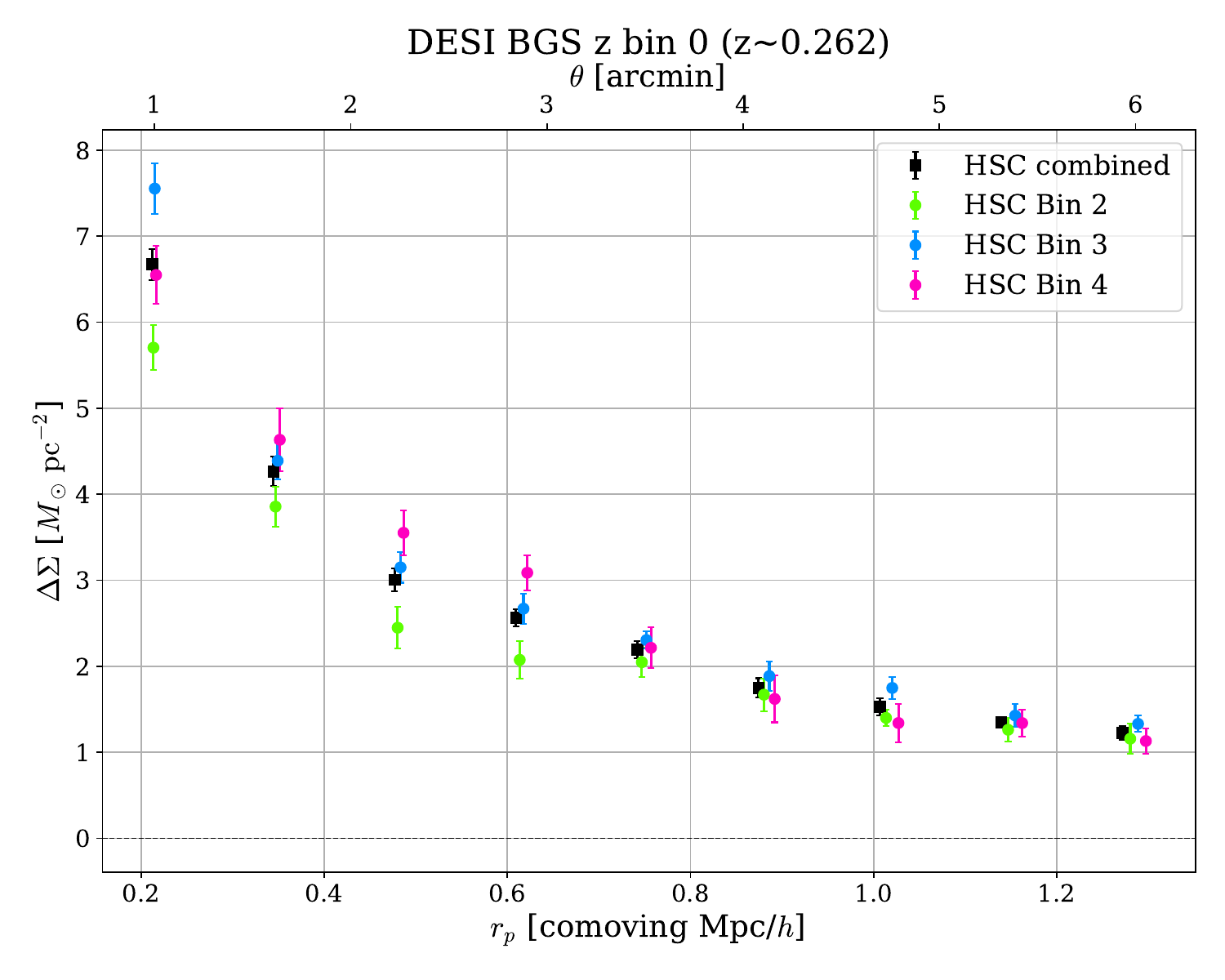}
    \caption{Galaxy-galaxy lensing $\Delta\Sigma$ profiles for the DESI BGS sample. Coloured points show individual HSC source bins (bin~2 in green, bin~3 in blue, bin~4 in pink); black squares show the combined measurement. 
    Error bars are $1\sigma$ from jackknife resampling.}
    \label{fig:BGS_lensing_profiles}
\end{figure}

\begin{figure}[h]
    \centering
    \includegraphics[width=\columnwidth]{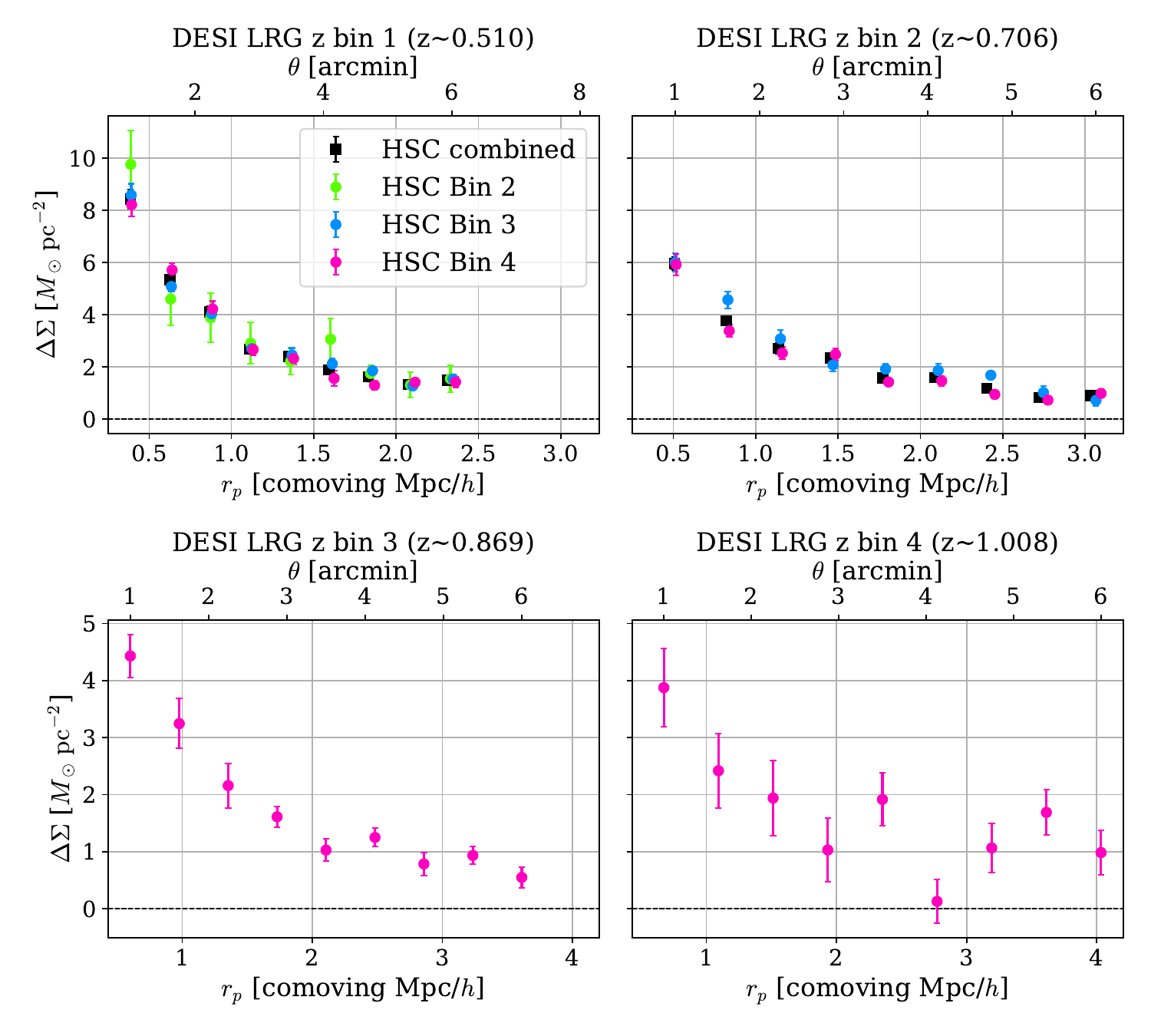}
    \caption{Galaxy-galaxy lensing $\Delta\Sigma$ profiles for the DESI LRG sample in four redshift bins. Bin~1 uses HSC source bins 2, 3, and 4; bin~2 uses HSC source bins~3, and~4 (coloured points) with a combined measurement in black; bins~3 and~4 use only HSC bin~4 due to the higher lens redshifts. Statistical uncertainties increase with lens redshift as the number of usable lens-source pairs decreases. Error bars are $1\sigma$ from jackknife resampling.}
    \label{fig:LRG_lensing_profiles}
\end{figure}

\begin{figure}[h]
    \centering
    \includegraphics[width=0.85\columnwidth]{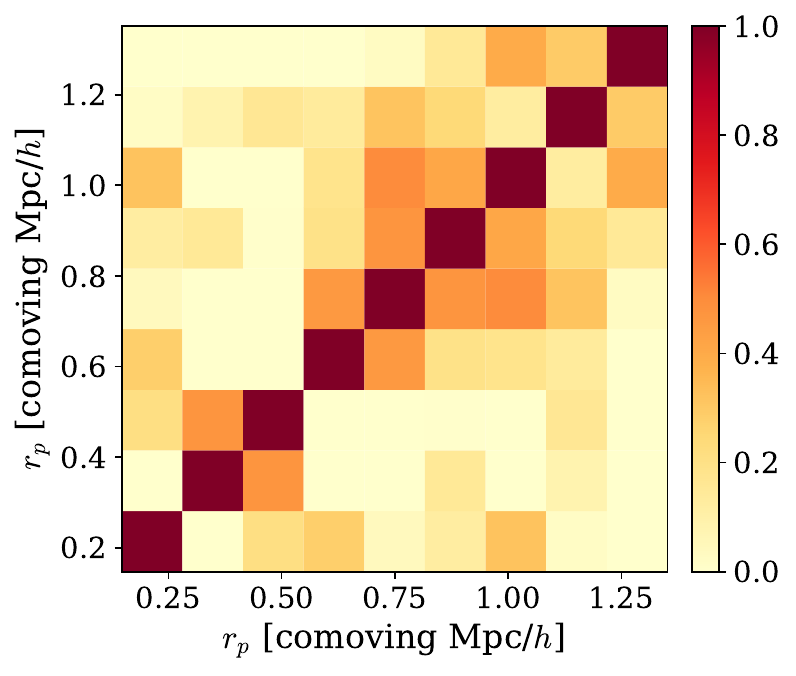}
    \caption{Lensing correlation matrix for the DESI BGS sample, estimated via jackknife resampling. Unlike the kSZ signal, the lensing signal shows less correlation between different radial bins. Similar covariance matrices are obtained for the LRG samples.}
    \label{fig:BGS_lensing_cov}
\end{figure}

The measurements of the galaxy-galaxy lensing $\Delta\Sigma$ profiles for the DESI BGS and LRG samples are shown in Figs.~\ref{fig:BGS_lensing_profiles} and \ref{fig:LRG_lensing_profiles}, respectively. The covariance matrices for the BGS measurements are shown in Fig.~\ref{fig:BGS_lensing_cov}, whereas the LRG covariance is omitted for its visual similarity. In addition to the individual source-bin measurements, the combined lensing measurement for each lens sample is shown in black. This combined measurement is obtained by summing the weighted lens-source pair statistics accumulated during precomputation across all eligible HSC source bins before computing $\Delta\Sigma$ and its jackknife covariance. This procedure is equivalent to processing all source galaxies from all bins in a single pass, with each pair naturally weighted by its source weight and lensing geometry.

The individual HSC source bins show a modest systematic ordering at small projected separations, most clearly visible for the BGS sample: HSC bin~2 yields a systematically lower $\Delta\Sigma$ than bins~3 and~4.
The lensing profiles are broadly consistent with those measured by Heydenreich et al.~\cite{Heydenreich2025} using a similar DESI$\times$HSC analysis pipeline. We note that LRG bin~4 has significantly higher noise than the other bins, primarily due to its high redshift relative to the HSC source sample, which reduces the number of usable lens-source pairs. 

\subsection{Combined Measurement of Baryon Feedback}
\label{sec:combined_results}

\begin{figure}[h]
    \centering
    \includegraphics[width=\columnwidth]{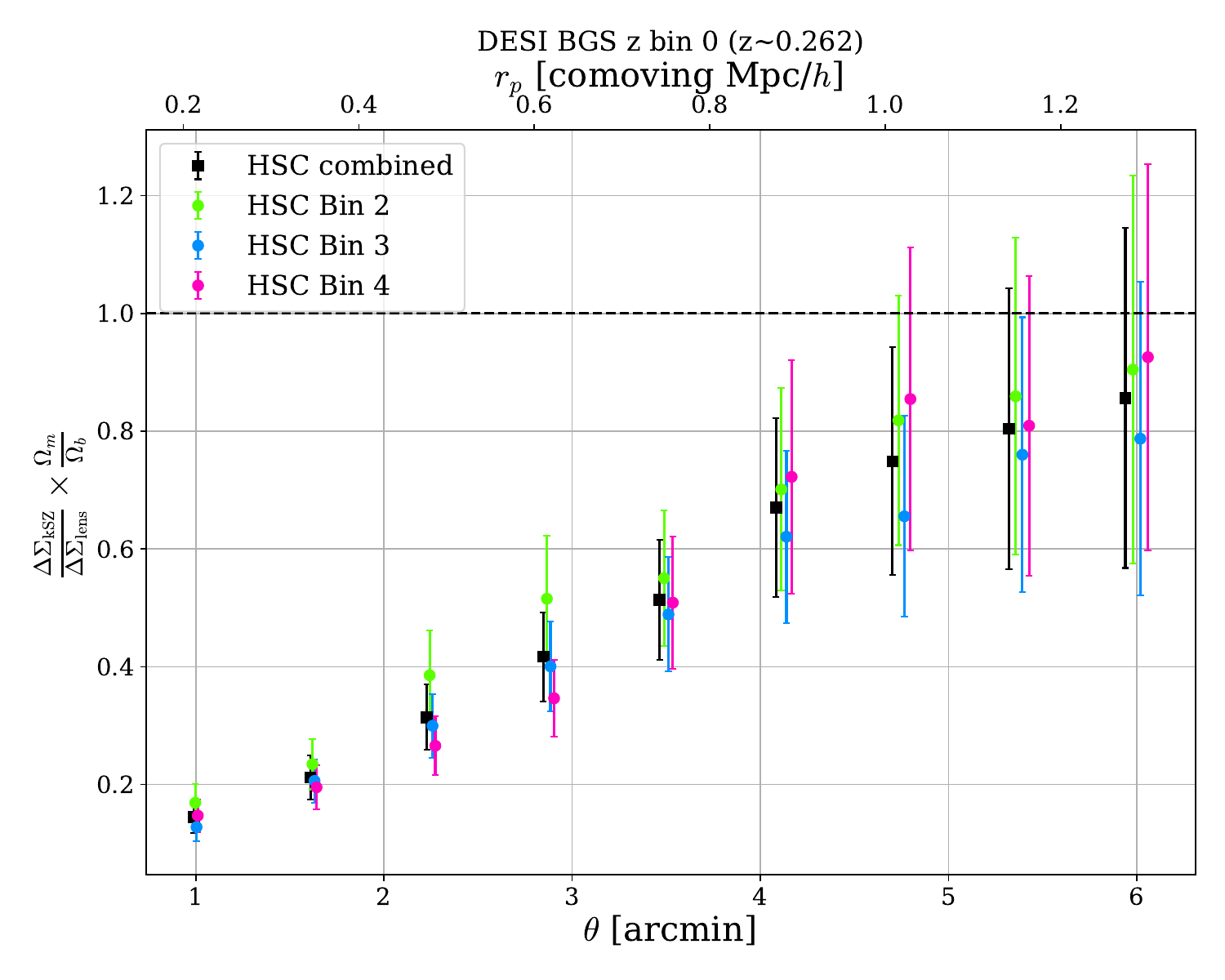}
    \caption{Observed gas fraction $f_{\rm gas}^{\rm obs}(R)$ (Eq.~\ref{eq:fgas_ratio}) for the DESI BGS sample. The dashed line at unity marks the null hypothesis of no feedback, in which gas traces the total matter at the cosmic baryon fraction. The suppression at $R \lesssim 2\,\mathrm{arcmin}$ is enhanced by the ACT beam and should not be interpreted as the true gas fraction at those radii. At larger radii, $f_{\rm gas}^{\rm obs}$ rises toward unity, indicating recovery of the cosmic baryon fraction. Black squares show the combined HSC measurement; coloured points show individual source bins.}
    \label{fig:combined_BGS}
\end{figure}

\begin{figure}[h]
    \centering
    \includegraphics[width=\columnwidth]{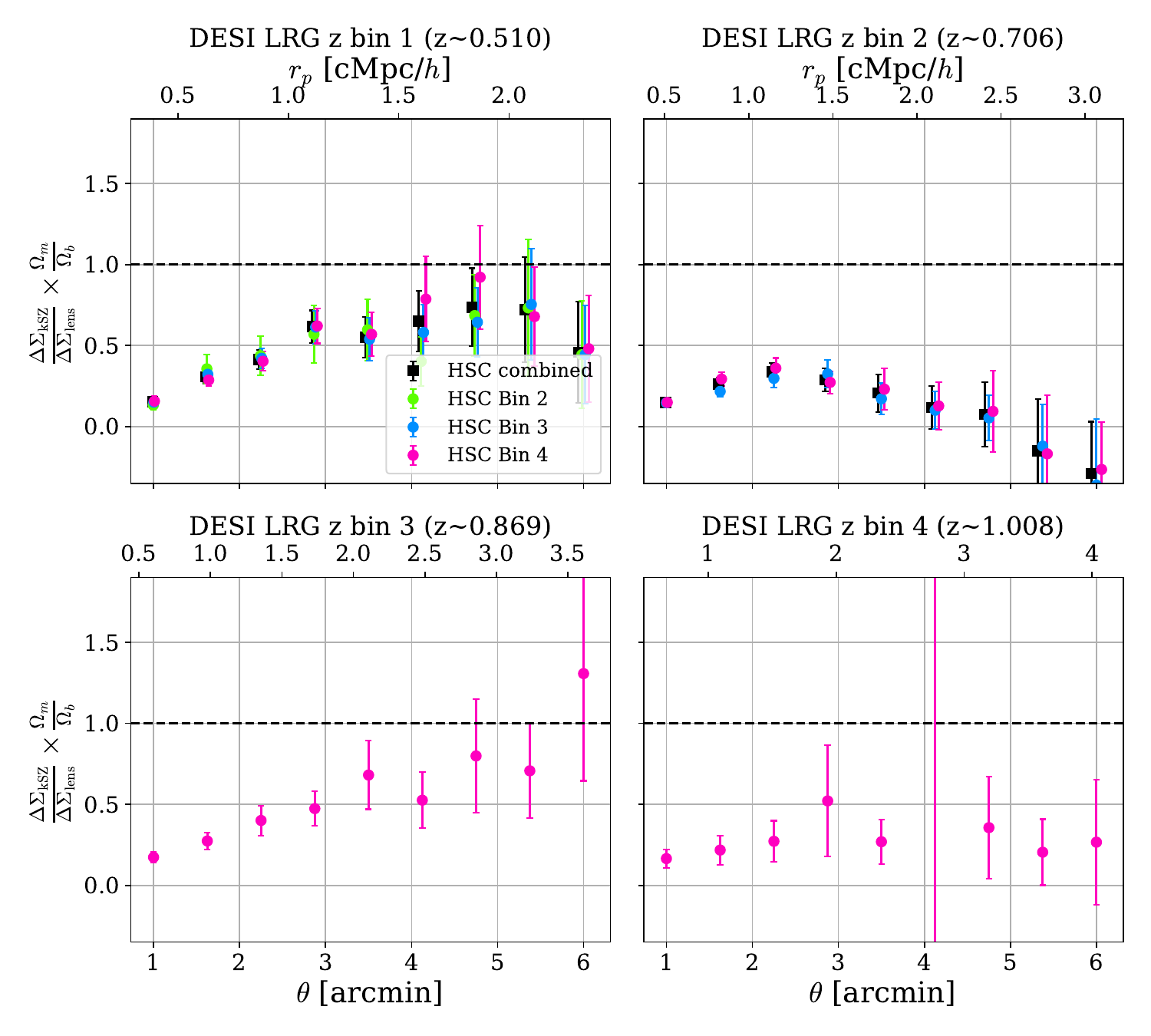}
    \caption{Observed gas fraction $f_{\rm gas}^{\rm obs}(R)$ for the DESI LRG sample in four redshift bins. The dashed line at unity marks the no-feedback null hypothesis. As in Figure~\ref{fig:combined_BGS}, ACT beam suppression contributes to the low values of $f_{\rm gas}^{\rm obs}$ at small radii. The combined HSC measurement (black) is shown only for bins~1 and~2, where multiple source bins contribute; for bins~3 and~4 it is redundant with the single available source bin (HSC bin~4, pink). The observed spike at $R\sim 4\,\mathrm{arcmin}$ in LRG bin~4 is likely a noise artefact caused by a lower lensing measurement; it is not concerning given the large error bars at these scales.}
    \label{fig:combined_LRG}
\end{figure}

Making use of the combination method described in Section~\ref{sec:combine_measurements}, we combine the kSZ $\Delta\Sigma$ profiles and the galaxy-galaxy lensing $\Delta\Sigma$ profiles to extract information about baryon feedback for the DESI BGS and LRG samples. The resulting measurements of the observed gas fraction $f_{\rm gas}^{\rm obs}(R)$ (Eq.~\ref{eq:fgas_ratio}) are shown in Figs.~\ref{fig:combined_BGS} and \ref{fig:combined_LRG} for the BGS and LRG samples, respectively. We emphasise that the profiles shown here are $f_{\rm gas}^{\rm obs}$, the directly measured kSZ to galaxy-galaxy lensing ratio, and represent the first measurement of this quantity for these samples. These profiles are subject to ACT beam suppression and should be distinguished from the beam-compensated gas fraction $f_{\rm gas}(R)$ presented in Section~\ref{sec:beam_compensated_fgas}. The error bars are propagated assuming the kSZ and lensing measurements are independent, following the error propagation method described in Section~\ref{sec:combine_measurements}.

As discussed in Section~\ref{sec:beam_effects}, the ACT beam (FWHM $= 1.6\,\mathrm{arcmin}$) smooths the CMB temperature field and suppresses the kSZ signal at small angular scales. The low $f_{\rm gas}^{\rm obs}$ values ($\approx 0.15$--$0.3$) at $R \lesssim 2\,\mathrm{arcmin}$ are therefore partly instrumental and should not be taken at face value as reflecting the true gas fraction at those radii. Although these small-scale measurements still encode information about baryon feedback (as demonstrated in the simulation comparison of Section~\ref{sec:simulations}), their interpretation requires forward-modelling the beam convolution. At larger radii where beam effects are sub-dominant, $f_{\rm gas}^{\rm obs}$ can be more directly interpreted: values significantly below unity indicate that feedback has redistributed gas to larger scales, while values near unity indicate that the gas traces the total matter at the cosmic baryon fraction.


For the BGS sample and the lowest-redshift LRG bin ($z \sim 0.51$), the measured $f_{\rm gas}^{\rm obs}$ rises from very low values at the smallest apertures to values consistent with unity at large radii, indicating that the gas distribution recovers the cosmic baryon fraction on these scales.
LRG bin~2 is an outlier in this trend, with $f_{\rm gas}^{\rm obs}$ remaining below unity at all radii and even showing negative values at the two largest apertures, which is unphysical. As discussed in Appendix~\ref{app:consistency_tests}, this anomalous behaviour is localised to the SGC field for this redshift bin; the NGC-only result for LRG bin~2 is consistent with the other LRG bins. A similar feature was reported previously by Qu et al.~\cite{Qu2026_kSZ}. Quantitatively, we observe a $1.7\sigma$ discrepancy between the NGC and SGC measurements in this bin, with the NGC result lying closer to the measurements of the other LRG bins. This indicates an unexplained systematic effect or a statistical fluctuation in the SGC for this specific redshift bin.

The other LRG bins show a qualitatively similar trend, with $f_{\rm gas}^{\rm obs}$ increasing with radius. LRG bin~4 ($z \sim 1.01$), which relies on a single HSC source bin, has the largest statistical uncertainties and is the most difficult to interpret.

An important internal consistency check is provided by the agreement between different HSC source bins. For a given lens sample, the $f_{\rm gas}^{\rm obs}$ values derived from HSC bins~2, 3, and~4 are broadly consistent with one another and with the combined measurement. Since these source bins have different redshift distributions and different sensitivities to photometric redshift systematics, their mutual agreement argues against HSC photo-$z$ calibration errors dominating the measured signal.

Comparing the BGS and LRG samples, the BGS galaxies (lower halo mass, $z \sim 0.26$) recover $f_{\rm gas}^{\rm obs} \sim 1$ at relatively smaller comoving radii than the LRG sample (higher halo mass, $z \sim 0.5$--$1.0$), which shows more persistent gas depletion extending to larger scales. However, we note that the effects of the CMB beam make it hard to compare the BGS and LRG samples at small angular scales, meaning the physical scales at which the gas fraction recovers the cosmic baryon fraction are not directly comparable between the two samples. For more detailed comparison of the mass and redshift dependence of gas depletion, it is necessary to forward-model the beam convolution in simulations, as presented in Section~\ref{sec:simulations}.

Our results for $f_{\rm gas}^{\rm obs}$ are broadly consistent with the measurements of Hadzhiyska et al.~\cite{Hadzhiyska2025_missing_baryons} and Hadzhiyska et al.~\cite{Hadzhiyska2026_kSZ} for similar galaxy samples, though a detailed comparison is complicated by differences in the aperture filters used (the $\Delta\Sigma$ filter in this work versus the CAP filter in those works) and the treatment of beam effects. A more direct comparison with these works will be possible after forward-modelling the beam convolution into simulations, as presented in Section~\ref{sec:simulated_observations}; a qualitative comparison of the beam-compensated $f_{\rm gas}(R)$ with Hadzhiyska et al.~\cite{Hadzhiyska2025_missing_baryons} is given in Appendix~\ref{app:comparisons_fgas}.


\subsubsection{Systematic Effects}



Our measurements of $f_{\rm gas}^{\rm obs}(R)$ are subject to several potential systematic effects that could bias the interpretation of the results. We discuss these effects here and assess their potential impact on our conclusions. For a detailed discussion of our treatment of these systematics, see the appendices in order.

Both the kSZ and the lensing measurements are subject to independent systematic effects that could bias each of the potential profiles. At the same time, both are also subject to combined systematics in the form of halo miscentering. Miscentering of the halo position relative to the galaxy position used for stacking can suppress the measured profiles at small radii, mimicking the effect of gas depletion. 
In order to consider the effects of this, in Appendix \ref{app:miscentering} we consider methods of pruning the sample to remove close pairs of galaxies from our catalogue. We find that the kSZ and lensing profiles are robust to the choice of pruning method, and that the resulting $f_{\rm gas}^{\rm obs}$ profiles are consistent with one another within the statistical uncertainties. However, pruning reduces the statistical power of the measurements, so we do not apply it to our fiducial results. We also note that the beam suppression at small radii is much stronger than any potential miscentering suppression, so the interpretation of $f_{\rm gas}^{\rm obs}$ at $R \lesssim 2\,\mathrm{arcmin}$ is dominated by beam effects rather than miscentering, which is correctable via simulation forward-modelling and simulation-calibrated compensation factors, as shown in Sec.~\ref{sec:beam_compensated_fgas}.
At larger radii where beam effects are sub-dominant, the impact of miscentering is also reduced, and the consistency between pruned and unpruned results suggests that miscentering does not significantly bias our conclusions about gas depletion at these scales.

The kSZ measurements could also be biased by residual contamination from the thermal Sunyaev-Zel'dovich (tSZ) effect, which arises from the inverse Compton scattering of CMB photons by hot electrons in galaxy clusters, as well as dust emissions from the Cosmic Infrared Background (CIB). In Appendix \ref{app:systematic_tests}, we assess the potential impact of tSZ contamination on our kSZ $\Delta\Sigma$ profiles by performing shuffled velocity null tests. Since both the tSZ and CIB signals are uncorrelated with the line-of-sight velocity field, shuffling the velocities should eliminate the kSZ signal while preserving any residual tSZ or CIB contamination. We find that the shuffled velocity null tests yield $\Delta\Sigma$ profiles consistent with zero, indicating that residual tSZ and CIB contamination in the ACT DR6 ILC map does not significantly bias our kSZ measurements or the resulting $f_{\rm gas}^{\rm obs}$ profiles.


The fiducial line-of-sight velocity $v_{\rm los} = 300\,\mathrm{km\,s^{-1}}$ and the velocity reconstruction factor $r_V$ enter the conversion coefficient $\alpha_{\rm conv}$ (Eq.~\ref{eq:alpha_conv}) and the kSZ estimator normalisation (Eq.~\ref{eq:ksz_estimator}) as overall multiplicative factors. A change in either quantity therefore shifts $f_{\rm gas}^{\rm obs}$ uniformly at all radii without altering the radial shape of the profile. The values adopted here ($r_V = 0.65$ for LRGs, $r_V = 0.64$ for BGS) are calibrated on $N$-body simulations~\cite{RiedGuachalla2024, Hadzhiyska2024b}. \rev{Though both are potential amplitude corrections to our signal, neither affects the shape of the $f_{\rm gas}^{\rm obs}$ profile.} 

On the lensing side, systematic uncertainties in the photometric redshift calibration of the HSC source galaxies could bias the $\Delta\Sigma_{\rm lens}$ profiles and thus $f_{\rm gas}^{\rm obs}$. However, the consistency between different HSC source bins (Section~\ref{sec:combined_results}) argues against photo-$z$ errors dominating the measured signal. Another potential bias in the lensing measurement is our distance conversion methodology. By taking the mean redshift of the lens sample to compute the angular diameter distance for converting $\Delta\Sigma$ to physical units, we assume that all lenses lie at the same redshift. In reality, the lens samples have a finite redshift distribution, and this approximation could bias the lensing profiles. However, since the lens redshift distributions are relatively narrow and the lensing efficiency does not vary rapidly with redshift at these scales, we expect this bias to be sub-dominant compared to our statistical uncertainties.

Lastly, as discussed in Section~\ref{sec:lensing_methods}, we do not apply a boost factor correction; the effect is percent-level \citep{Lange2024, Heydenreich2025} and sub-dominant relative to our statistical uncertainties. 

\subsection{Gas Fraction Dependence on Stellar Mass}
\label{sec:mass_bins}


\begin{figure}[h]
    \centering
    \includegraphics[width=\columnwidth]{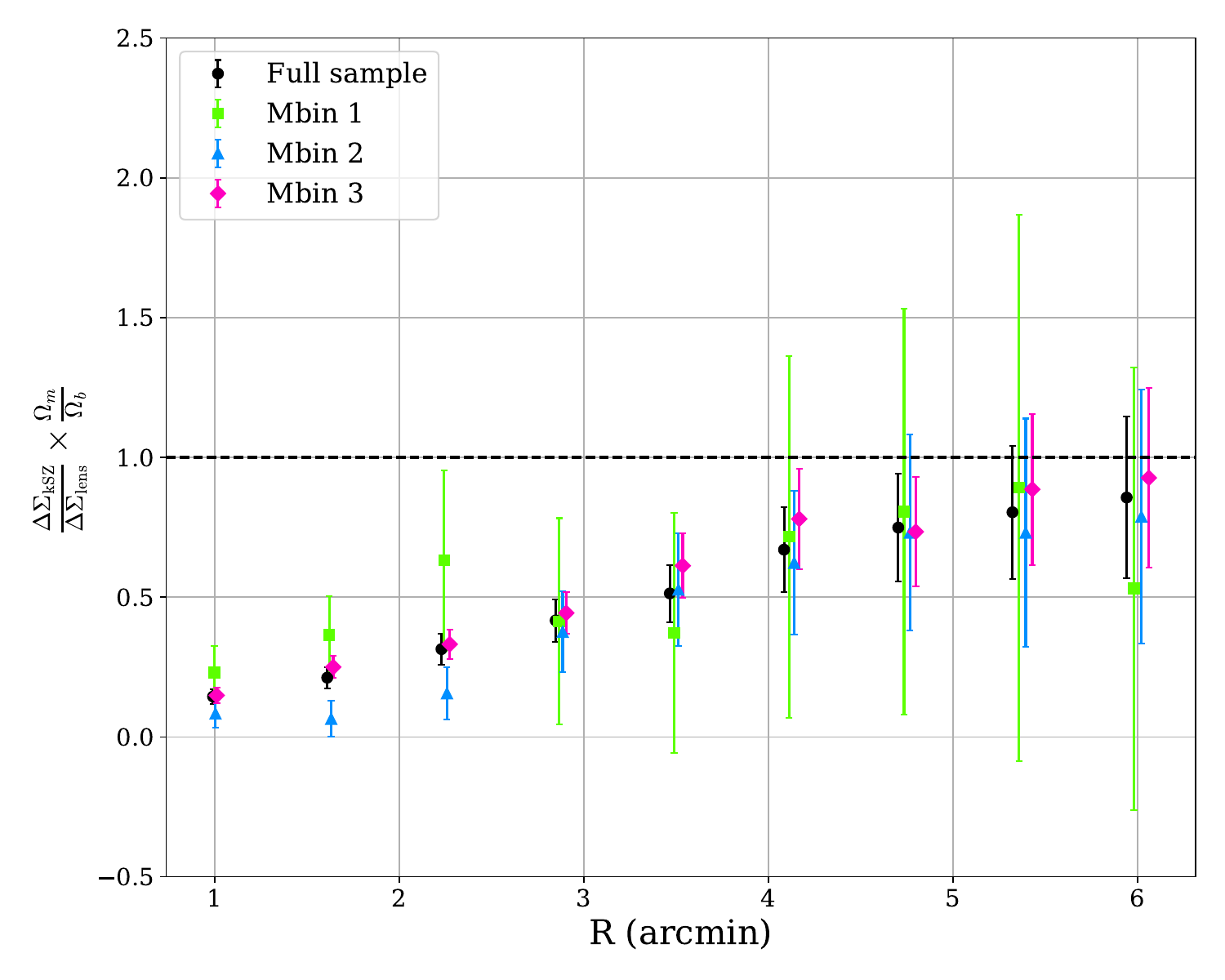}
    \caption{Observed gas fraction $f_{\rm gas}^{\rm obs}(R)$ for the DESI BGS sample in three stellar mass bins, \rev{with stellar mass increasing from Mbin 1 to 3}. The lack of a clear mass dependence in the $f_{\rm gas}^{\rm obs}$ profiles suggests that the gas fraction around haloes may not have a strong dependence on halo mass within the range probed by our BGS sample, though more precise measurements would be needed to confirm this trend.}
    \label{fig:combined_BGS_mass_bins}
\end{figure}

\begin{figure}
    \centering
    \includegraphics[width=\columnwidth]{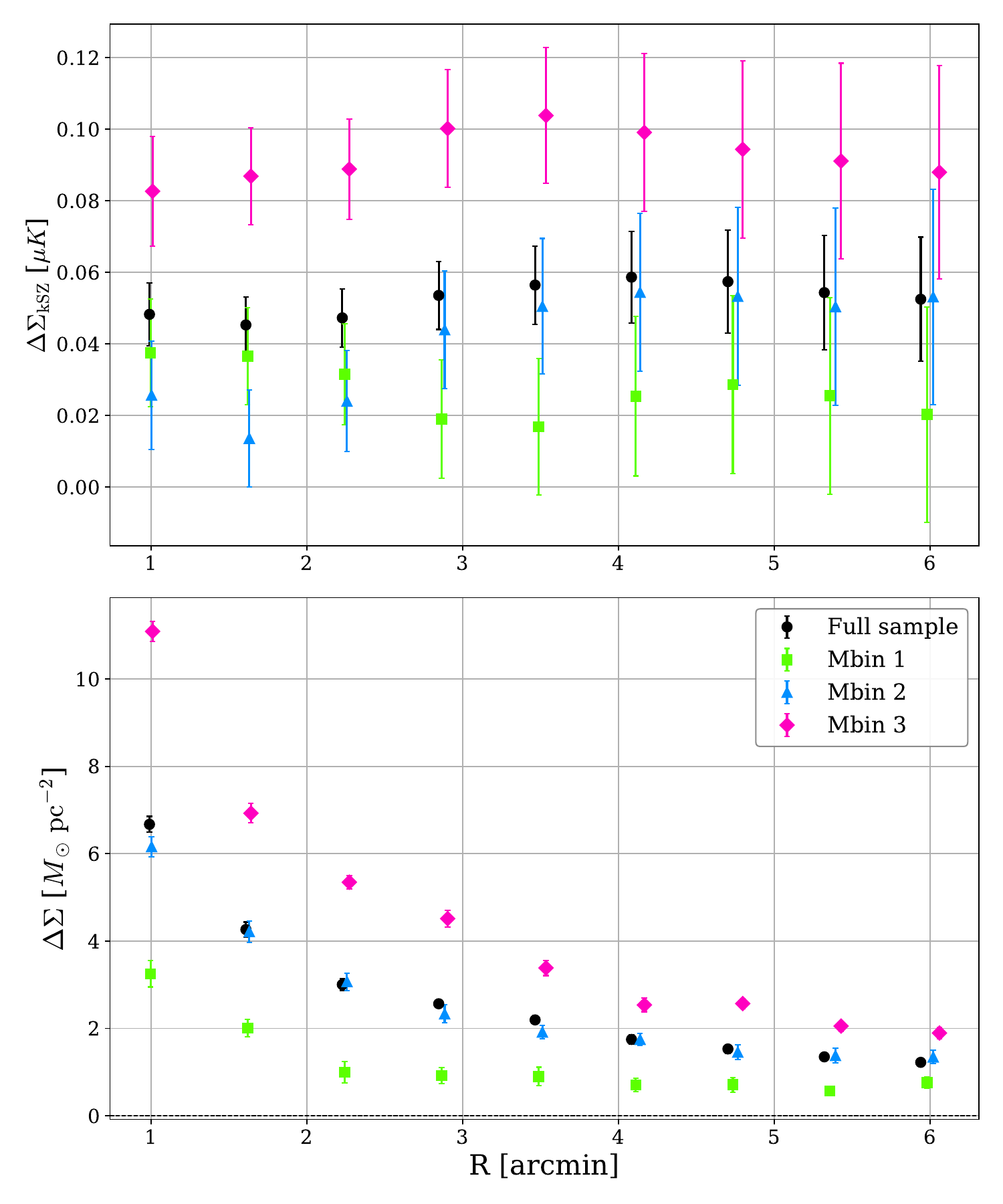}
    \caption{kSZ (top) and lensing (bottom) measurements for the DESI BGS sample in three stellar mass bins, \rev{with stellar mass increasing from Mbin 1 to 3}. Though the individual kSZ and lensing signals show a clear mass dependence with higher mass bins, the ratio of these two signals, shown in Figure~\ref{fig:combined_BGS_mass_bins}, does not show a strong mass dependence within the statistical uncertainties of our measurements.}
    \label{fig:sz_lensing_BGS_mass_bins}
\end{figure}

\begin{figure}[h]
    \centering
    \includegraphics[width=\columnwidth]{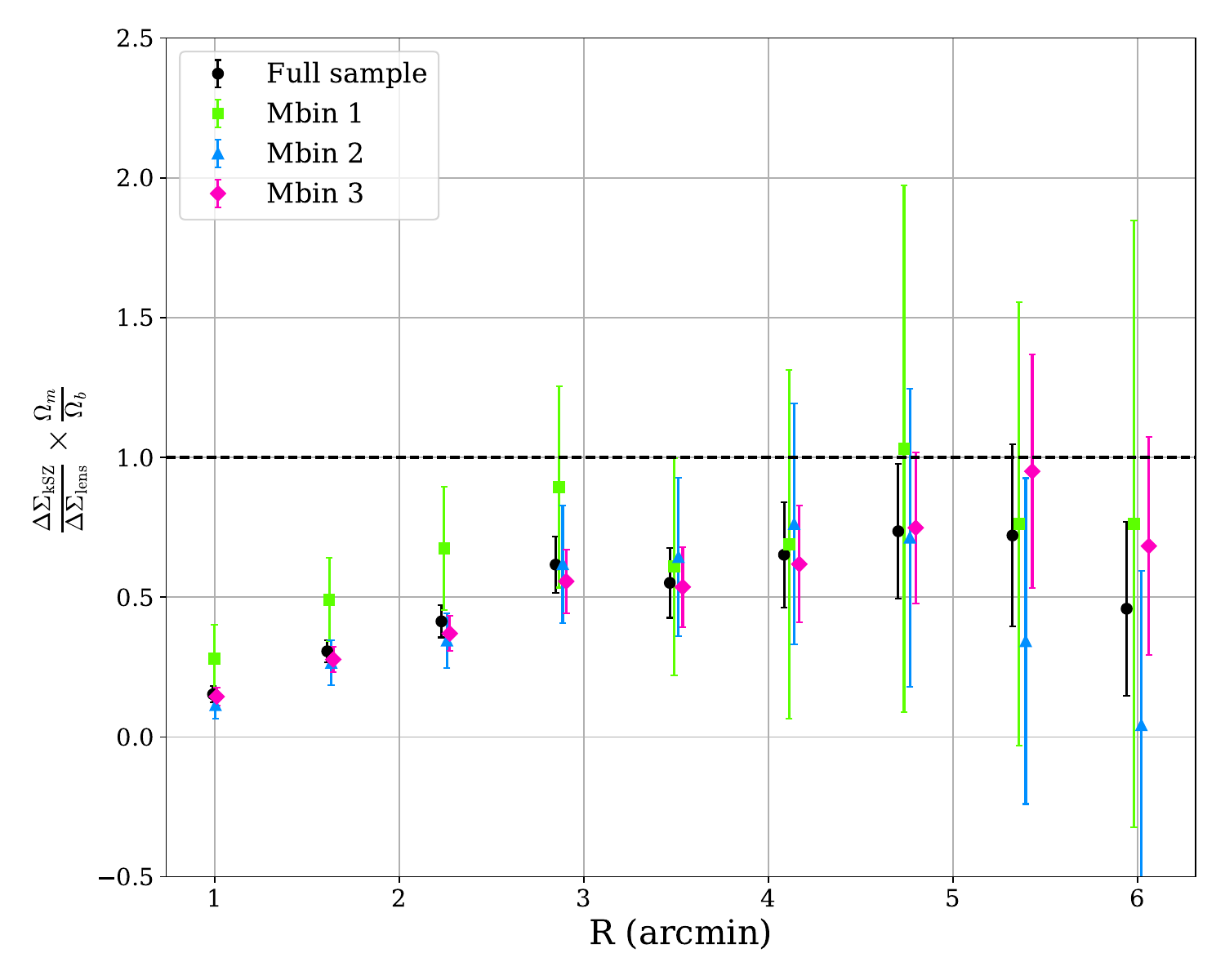}
    \caption{Observed gas fraction $f_{\rm gas}^{\rm obs}(R)$ for the DESI LRG sample in three stellar mass bins, \rev{with stellar mass increasing from Mbin 1 to 3}. As with the BGS case, we see a lack of a clear mass dependence in the $f_{\rm gas}^{\rm obs}$ profiles.}
    \label{fig:combined_LRGz1_mass_bins}
\end{figure}

\begin{figure}
    \centering
    \includegraphics[width=\columnwidth]{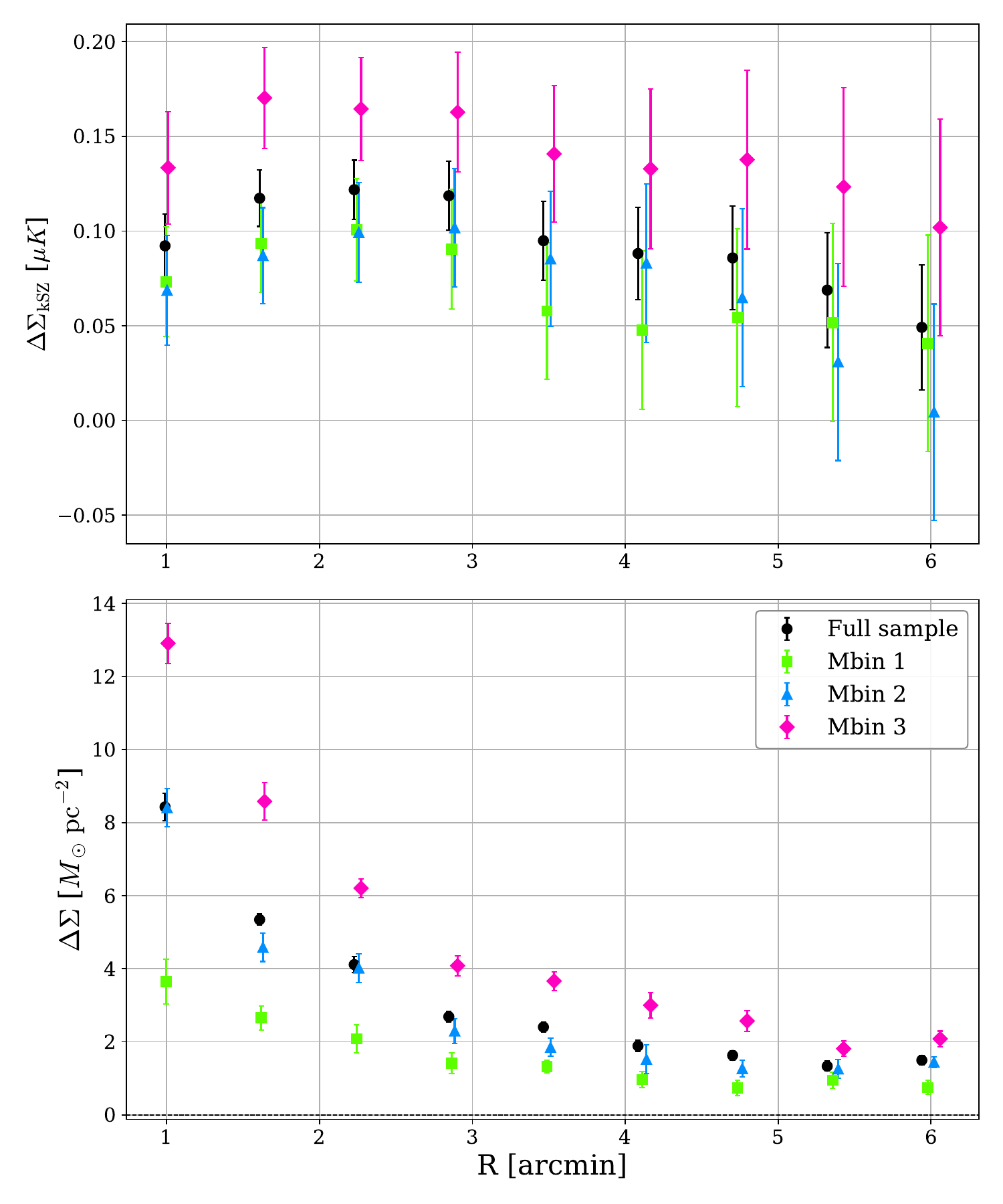}
    \caption{kSZ (top) and lensing (bottom) measurements for the DESI LRG sample in three stellar mass bins, \rev{with stellar mass increasing from Mbin 1 to 3}. As with the BGS case, the individual measurements show a mass dependence, especially in the highest mass bin, though this mass dependence is less evident in Figure~\ref{fig:combined_LRGz1_mass_bins}.}
    \label{fig:sz_lensing_LRGz1_mass_bins}
\end{figure}

To explore the dependence of $f_{\rm gas}^{\rm obs}$ on stellar mass, we further bin our galaxy samples by stellar mass and measure $f_{\rm gas}^{\rm obs}(R)$ for each bin. This analysis provides insight into whether the gas depletion profile varies with halo mass, and how feedback redistributes gas across the halo mass range probed by our samples.

We perform this analysis for the BGS sample as well as the LRG sample (specifically LRG bin 1, the lowest-redshift LRG bin, which we also use for the simulation comparison).
The stellar mass estimates for the DESI BGS sample are obtained from the CIGALE code \cite{Siudek2024_CIGALE,Siudek2025_CIGALE}, which provided stellar mass estimates based on fits to galaxy spectral energy distributions from the optical to the infrared using stellar population synthesis, delayed star formation histories with optional bursts, nebular emission, dust attenuation and re-emission, and AGN templates.
The stellar mass estimates for the DESI LRG sample are obtained from the DESI Legacy Imaging Survey (LS) photometric redshift catalogues first used in \cite{Zhou2023_DESILRG}, which applied a random forest algorithm trained on DESI Legacy Imaging Survey photometry with stellar masses from \cite{Bundy2015} as the training set.

We split both the LRG and BGS samples into three stellar mass bins of equal size, with mean stellar mass and sample size shown in Table \ref{tab:mass_bins}. \rev{We choose equal-sized bins as a compromise between having sufficient signal-to-noise in each bin and adequately sampling the stellar mass range.} We measure $f_{\rm gas}^{\rm obs}(R)$ for each bin using the same methods as described in the main text. The resulting $f_{\rm gas}^{\rm obs}$ profiles for each mass bin are shown in Figure~\ref{fig:combined_BGS_mass_bins} for the BGS sample and Figure~\ref{fig:combined_LRGz1_mass_bins} for the LRG bin 1 sample.


\begin{table}[htbp!]
    \centering
    \setlength{\tabcolsep}{6.3pt}  
    \begin{tabular}{l | cc | cc}
        \hline \hline
        & \multicolumn{2}{c|}{BGS}
        & \multicolumn{2}{c}{LRG Bin 1} \\
        Sample
        & $\log{\overline{M}_{*}}$ & $N_{\rm objects}$
        & $\log{\overline{M}_{*}}$ & $N_{\rm objects}$ \\
        \hline
        Mbin 1  &  10.29 & 740,119 & 11.09 & 199,036  \\
        Mbin 2  &  10.69 & 740,119 & 11.35 & 199,034  \\
        Mbin 3  &  11.14 & 740,121 & 11.58 & 199,039  \\
        \hline \hline
    \end{tabular}
    \caption{Mean stellar mass $\log_{10}(\overline{M}_{*}/M_\odot)$ and number of objects $N_{\rm objects}$ in each stellar mass bin for the DESI BGS and LRG samples. The three mass bins are of equal size, with the mean stellar mass increasing from Mbin 1 to Mbin 3.}
    \label{tab:mass_bins}
\end{table}

As is expected, the lowest mass bin has the lowest SNR, and the highest mass bin has the highest SNR. However, it is interesting to note that we don't see a clear mass dependence in the $f_{\rm gas}^{\rm obs}$ profiles, as the three mass bins show broadly similar trends with radius. The individual lensing and kSZ signals, plotted in Figures \ref{fig:sz_lensing_BGS_mass_bins} and \ref{fig:sz_lensing_LRGz1_mass_bins}, do show a clear mass dependence with higher mass bins. However, the ratio of these two signals, which gives $f_{\rm gas}^{\rm obs}$, does not show a strong mass dependence within the statistical uncertainties of our measurements. This result suggests that the gas fraction around haloes may not have a strong dependence on halo mass within the range probed by our BGS sample, though more precise measurements would be needed to confirm this trend.

It is also important to note the implications of a mass-independent $f_{\rm gas}^{\rm obs}$ profile for our use of simulations to interpret our feedback measurements. Simulation comparisons to lensing or kSZ measurements are often affected by the systematics of simulation subhalo selection, and the selection of subhaloes to match the observed galaxy sample can be non-trivial. However, if the $f_{\rm gas}^{\rm obs}$ profile is largely independent of mass, then the interpretation of our feedback measurements in terms of gas fractions becomes more robust to the details of subhalo selection in simulations, as the gas fraction profiles would be similar across a range of halo masses. 

\section{Comparison to Simulated Profiles}
\label{sec:simulations}

\subsection{Simulations}

We compare our observational $f_{\rm gas}^{\rm obs}$ profiles with predictions from six cosmological hydrodynamical simulations drawn from four suites, spanning a representative range of AGN and stellar feedback prescriptions: Illustris-1 from the Illustris suite, TNG300-1 from IllustrisTNG, SIMBA-100 from SIMBA, and three runs of the FLAMINGO suite in its L1\_m9 configuration, with the fiducial model and the fgas$-8\sigma$ and Jet\_fgas$-4\sigma$ feedback variations.

The Illustris \cite{Vogelsberger2014_Illustris, Genel2014_Illustris, Nelson2015_Illustris} and IllustrisTNG \cite{Nelson2019_IllustrisTNG, Pillepich2018_IllustrisTNG_stellar_mass, Springel2018_IllustrisTNG_clustering, Naiman2018_IllustrisTNG_elements, Nelson2018_IllustrisTNG_colors, Marinacci2018_IllustrisTNG_radio_haloes} suites share a similar code framework but differ in their feedback implementations. Illustris employs a strong AGN feedback model that drives large-scale outflows and has been shown to over-deplete gas around massive haloes relative to X-ray observations \cite{Nelson2015_Illustris}; IllustrisTNG revised this model, yielding gas fractions more consistent with observations. The two suites also assume different cosmological parameters: Illustris adopts WMAP-9 values \cite{WMAP2013_cosmological_parameters}, while IllustrisTNG adopts Planck 2015 values \cite{Planck2016_cosmological_parameters}. We use the Illustris-1 run ($106.5\,\mathrm{cMpc}$ box, baryon mass resolution $\sim$$1.3 \times 10^6\,M_{\odot}$) and the TNG300-1 run ($302.6\,\mathrm{cMpc}$ box, baryon mass resolution $\sim$$1.1 \times 10^7\,M_{\odot}$).

The SIMBA simulation \cite{Dave2019_SIMBA} is built on the GIZMO code \cite{Hopkins2015_GIZMO} and adopts Planck 2015 cosmological parameters. Its primary distinguishing feature relative to Illustris and IllustrisTNG is a kinetic bipolar jet model for AGN feedback, rather than a spherically symmetric implementation. This jet prescription produces markedly different gas distributions around massive haloes, 
making SIMBA a useful point of contrast. We use the SIMBA-100 run ($100\,h^{-1}\,\mathrm{cMpc}$ box, baryon mass resolution $\sim$$1.8 \times 10^7\,M_{\odot}$).

The FLAMINGO suite \cite{Schaye2023_FLAMINGO, Helly2026_FLAMINGO} comprises large-volume cosmological hydrodynamical simulations run with the SWIFT code \cite{Schaller2024_SWIFT}, with subgrid stellar and AGN feedback parameters calibrated, via machine-learning emulators, to reproduce the observed low-redshift galaxy stellar mass function and cluster gas fractions \cite{Kugel2023_FLAMINGO}. All runs used in this work adopt the suite's fiducial cosmology, the DES Y3 ``3$\times$2pt + All Ext.''\ parameters \cite{Abbott2022_DES}; the feedback variations alter only the calibration targets, not the cosmology. We use three runs in the L1\_m9 configuration ($1\,\mathrm{cGpc}$ box, baryon mass resolution $\sim$$1.1 \times 10^9\,M_{\odot}$): the fiducial run, in which AGN feedback is implemented as thermal energy injection with parameters calibrated to the observational data (labelled ``FLAMINGO L1-m9'' in our figures, as the feedback variations share the same box and resolution); fgas$-8\sigma$, in which the AGN feedback is recalibrated such that the predicted cluster gas fractions lie $8\sigma$ below the calibration data, representing substantially stronger feedback; and Jet\_fgas$-4\sigma$, in which AGN feedback is instead implemented as kinetic jets \cite{Husko2022_FLAMINGO}, calibrated to cluster gas fractions $4\sigma$ below the data. These variations allow us to test controlled changes in both the strength and the implementation of AGN feedback within a single simulation framework. Although the FLAMINGO runs have a coarser mass resolution than the other simulations used here, their much larger volume provides correspondingly larger galaxy samples.

\rev{The cosmologies between the simulations used differ modestly: $\sigma_8$ spans $0.807$ to $0.820$, and $\Omega_m$ spans $0.2726$ (Illustris, WMAP-9) to $0.3089$ (IllustrisTNG, Planck 2015), against $\Omega_m = 0.3097$ for the Planck 2018 cosmology assumed for the data. The implied cosmic baryon fractions $\Omega_b/\Omega_m$ range from $0.157$ (IllustrisTNG) to $0.167$ (Illustris). However, as the simulated $f_{\rm gas}^{\rm obs}$ profiles are normalized by the cosmic baryon fraction of each simulation, these differences do not affect the comparison of the simulated and observed $f_{\rm gas}^{\rm obs}$ profiles.}

For the BGS comparison at $z = 0.26$, we use snapshots 80 (TNG300-1), 116 (Illustris-1), and 136 (SIMBA-100); for the LRG comparison at $z = 0.5$, we use snapshots 67 (TNG300-1), 103 (Illustris-1), 125 (SIMBA-100), and 67 (FLAMINGO). These snapshots are chosen to match the mean spectroscopic redshifts of the respective lens samples. The FLAMINGO outputs do not include a snapshot at $z = 0.26$, so for the BGS comparison we use the closest available output, snapshot 71 at $z = 0.30$, and adopt $z = 0.30$ consistently when converting the FLAMINGO comoving separations to angular scales.

\subsection{Constructing Simulated Observations}
\label{sec:simulated_observations}

\begin{figure*}[htbp!]
  \centering
  \begin{subfigure}{0.49\textwidth}
    \centering
    \includegraphics[width=\textwidth]{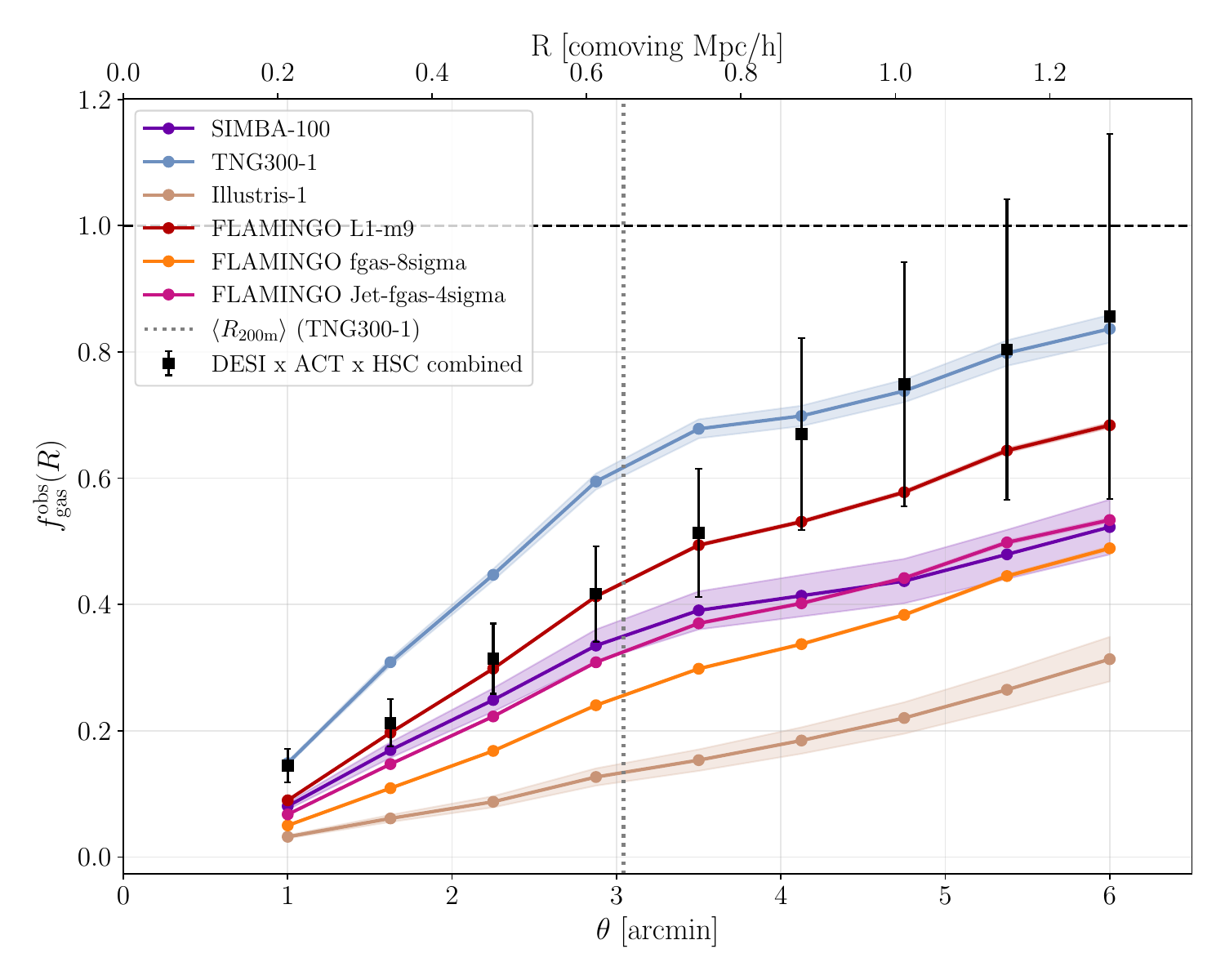}
    \caption{DESI BGS sample at $z = 0.26$.}
    \label{fig:simulation_compare_BGS}
  \end{subfigure}
  \hfill
  \begin{subfigure}{0.49\textwidth}
    \centering
    \includegraphics[width=\textwidth]{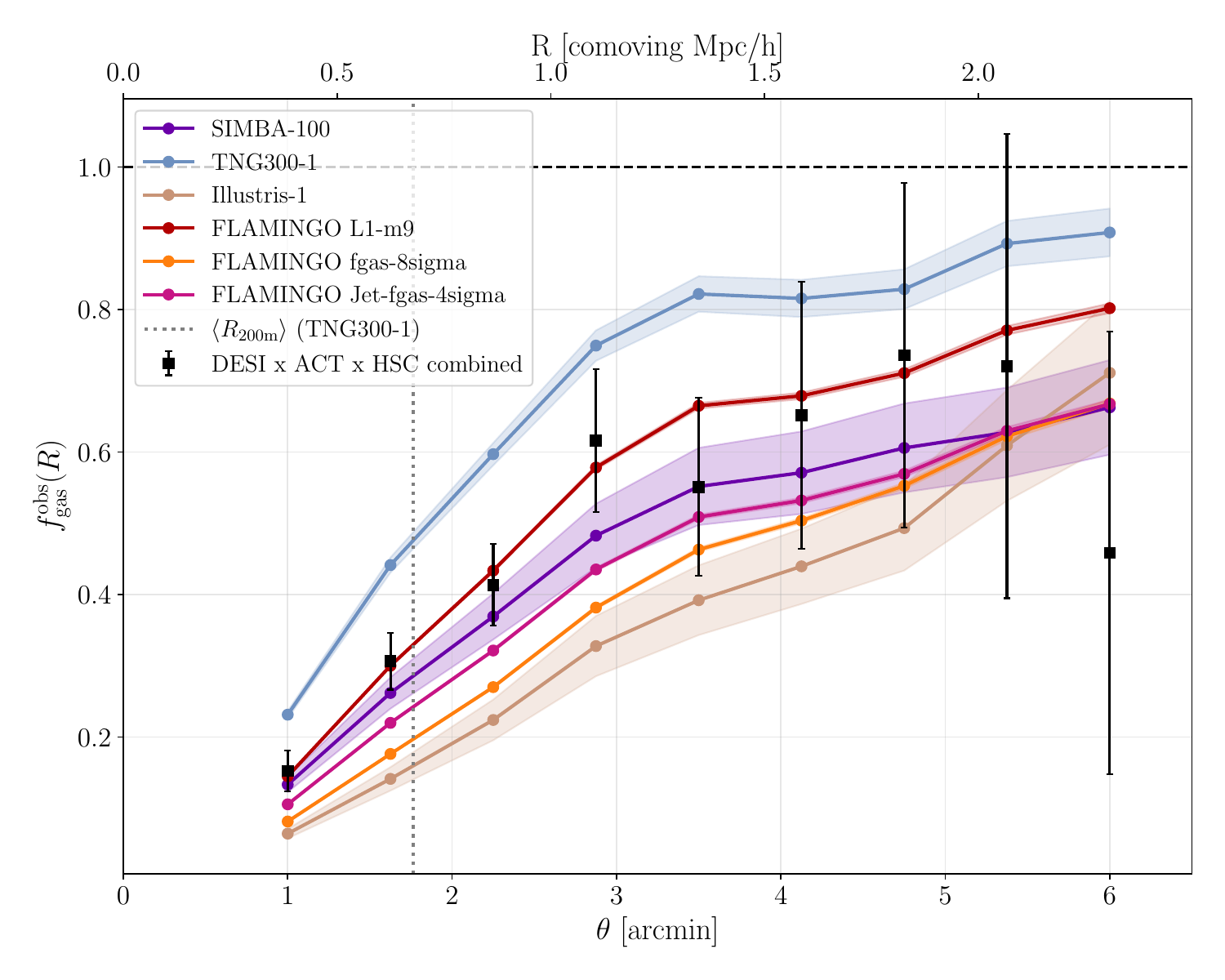}
    \caption{DESI LRG bin 1 at $z = 0.5$.}
    \label{fig:simulation_compare_LRG}
  \end{subfigure}
  \caption{Comparison of the observed (CMB beam convolved) gas fraction $f_{\rm gas}^{\rm obs}(R)$ (Eq.~\ref{eq:fgas_ratio}) for the DESI BGS sample (black squares, left) and LRG bin~1 (right) with predictions from six cosmological hydrodynamical simulations. To match the observational conditions, simulated kSZ $\Delta\Sigma$ profiles have been convolved with the ACT beam (FWHM $= 1.6\,\mathrm{arcmin}$), so both the data and simulations shown here reflect beam-suppressed $f_{\rm gas}^{\rm obs}$ rather than the intrinsic gas fraction. The dashed horizontal line at unity marks the null hypothesis in which the gas traces the total matter at the cosmic baryon fraction $\Omega_b/\Omega_m$. The dotted vertical line marks the virial radii ($R_{200m}$) of the haloes selected in TNG300 as a point of reference; the other simulations have similar halo mass distributions and thus similar virial radii. 
  Shaded bands show the $1\sigma$ uncertainty on the simulated mean. The bottom axis shows angular separation in arcminutes; the top axis shows the corresponding comoving separation in $\mathrm{Mpc}/h$.}
  \label{fig:simulation_compare}
\end{figure*}

In order to compare our observational measurements with simulations, we first need to construct simulated observations that mimic the conditions of our actual measurements. We start by constructing a 2D projected map of the optical depth or total matter distribution from a simulation snapshot, projecting the full simulation box along one axis. We select similar galaxies to those in the observational sample using subhalo abundance matching.
We then apply the same $\Delta\Sigma$ filter to these maps to the selected galaxies to obtain simulated kSZ and lensing profiles. Finally, we compute the ratio of these profiles to obtain a simulated $f_{\rm gas}^{\rm obs}$ profile that can be directly compared to our observational measurements. We describe this process briefly here. For a detailed prescription of constructing our simulated observations, see \cite{LiuInprep2026}.

We construct the total matter maps by projecting the 3D distribution of all simulation particles (dark matter, gas, stars, and black holes) onto a 2D grid. The ionized gas maps (which the kSZ signal traces) are constructed by projecting only the gas particles, weighted by the ionization fraction of the simulation gas particles.
\rev{The projection depth of each simulation depends on the simulation box size, and thus differs between simulations. However, this does not impact the observations: structures uncorrelated with the stacked haloes contribute only a uniform density, which is removed by the compensated $\Delta\Sigma$ filter.}
To select galaxies in the simulation that are comparable to our observational samples for the stacking procedure, we use the subhalo abundance matching method. 

We rank-order the subhaloes by their subhalo stellar mass and select those above a threshold corresponding to an abundance of $5\times 10^{-4} \; h^3 \mathrm{Mpc}^{-3}$ for the DESI LRG comparison, and $1 \times 10^{-3} \; h^3 \mathrm{Mpc}^{-3}$ for the BGS comparison. To simulate the observational conditions where large clusters are masked, we further make masking cuts to remove haloes with host halo masses above $5 \times 10^{14}\,h^{-1} M_\odot$ from our selected sample before the subhalo abundance matching.

\rev{Ranking by stellar mass is a proxy for the observational selection, which for the LRGs is based on colours and magnitudes rather than stellar mass directly~\cite{Zhou2023_DESILRG}, and the simulated stellar masses themselves differ between codes. We do not attempt a closer match because the ratio observable is insensitive to it: the stellar-mass splits in Sec.~\ref{sec:mass_bins} show no significant dependence of $f_{\rm gas}^{\rm obs}$ on mass in the data, and Appendix~\ref{sec:mass_bins_sims} shows the same for variations of the number-density threshold in the simulations.}




One important detail in our simulation comparison is the accurate treatment of the ACT beam effects on the kSZ $\Delta\Sigma$ profiles. As mentioned in Section~\ref{sec:combine_measurements}, the finite resolution of the ACT CMB maps can suppress the kSZ signal at small radii. In order to account for this effect in our simulated observations, we need to treat our simulated kSZ profiles differently than our simulated lensing profiles. First, we project the simulation distances to angular units on the sky. To do this, we convert the comoving transverse lengths to angular separations via the small-angle relation
\begin{equation}
\label{eq:angular_size}
    \theta = \frac{\ell_{\mathrm{com}}}{D_M(z)},
\end{equation}
where $\ell_{\mathrm{com}}$ is the comoving transverse extent of the quantity of interest and $D_M(z)$ is the comoving transverse distance (also called the proper motion distance) to redshift $z$ \citep{Hogg1999}. For a spatially flat cosmology, $D_M$ reduces to the comoving line-of-sight distance $D_C(z) = (c/H_0)\int_0^z \mathrm{d}z'/E(z')$, where $E(z) \equiv H(z)/H_0$. The factors of $(1+z)$ arising from the comoving-to-physical length conversion ($\ell_{\mathrm{phys}} = \ell_{\mathrm{com}}/(1+z)$) and from the angular diameter distance ($D_A = D_M/(1+z)$) cancel identically in Eq.~\ref{eq:angular_size}, such that the angular size of a comoving structure is simply the ratio of its comoving extent to the comoving transverse distance.

To mimic the beam effects on the kSZ profiles, we bin the simulation particles to a pixel resolution of $0.5 \;\mathrm{arcmin}$, matching the pixel resolution of the ACT DR6 CMB maps. We then apply a Gaussian beam in Fourier space and convolve the binned kSZ maps with the ACT beam profile, which is characterized by a FWHM of $1.6 \;\mathrm{arcmin}$. On the other hand, to construct the higher resolution simulated lensing profiles, we bin the simulation particles to a finer pixel resolution of $0.2 \;\mathrm{arcmin}$, and do not apply any beam convolution to these maps. By treating the simulated kSZ and lensing profiles differently in this way, we can ensure that our simulated observations accurately reflect the conditions of our actual measurements, allowing for a meaningful comparison between the two.

\subsection{Simulation Comparison Results}



We choose to make simulation comparisons with the combined BGS sample at $z = 0.26$ and the lowest-redshift LRG bin at $z = 0.5$, as these samples have the highest signal-to-noise measurements of $f_{\rm gas}^{\rm obs}$ and thus provide the most stringent tests of feedback models. The results are shown in Figure~\ref{fig:simulation_compare}.

Figures~\ref{fig:simulation_compare_BGS} and~\ref{fig:simulation_compare_LRG} compare the predicted $f_{\rm gas}^{\rm obs}$ profiles from the six simulations described above with our observational measurements for the DESI full BGS and LRG bin 1 samples. In these comparisons, the simulated kSZ profiles have been convolved with the ACT beam so that both the data and simulations reflect beam-suppressed $f_{\rm gas}^{\rm obs}$ on equal footing. For both samples, the measured $f_{\rm gas}^{\rm obs}$ values are generally below unity and rise with radius, in qualitative agreement with the simulations. Of the six, the fiducial FLAMINGO run tracks the data most closely across the radial range in both samples.

At large apertures, our measurements of $f_{\rm gas}^{\rm obs}$ recover toward the cosmic baryon fraction. In this regime TNG300-1 best matches the BGS data ($z=0.26$), predicting $f_{\rm gas}^{\rm obs}$ values close to unity, while for the LRG sample ($z=0.5$) the fiducial FLAMINGO run tracks the large-aperture measurements most closely, with TNG300-1 predicting somewhat higher values. For the BGS sample, Illustris-1 predicts $f_{\rm gas}^{\rm obs}$ values below those observed at all radii: its feedback model depletes gas more aggressively than observed throughout the full radial range probed. For the LRG sample, the Illustris-1 profile rises more steeply with radius and approaches the data only at the largest apertures probed ($R \sim 6\,\mathrm{arcmin}$), where measurement uncertainties are largest. The strong-feedback FLAMINGO variation fgas$-8\sigma$ behaves qualitatively similarly, predicting $f_{\rm gas}^{\rm obs}$ below the measurements at most radii, though less severely than Illustris-1. SIMBA-100 and the FLAMINGO Jet\_fgas$-4\sigma$ run predict remarkably similar profiles despite their different codes and calibration strategies; both show reasonable agreement with the data at $z=0.5$, with Jet\_fgas$-4\sigma$ falling somewhat below the data at the smallest radii, but are less consistent at the lower redshift of the BGS sample.
At smaller apertures, the data shows stronger suppression of $f_{\rm gas}^{\rm obs}$ than predicted by TNG300-1, while Illustris-1 predicts the strongest suppression of the six simulations. The radial ordering at small scales is therefore Illustris-1 $<$ data $<$ TNG300-1, with the fiducial FLAMINGO run and SIMBA-100 tracing the data most closely in this regime. With the data bracketed between the most and least aggressive feedback prescriptions at small radii, the fiducial FLAMINGO run follows the measurements most closely overall, falling below the data only in the innermost BGS bin, while at large apertures in the BGS sample TNG300-1 lies closer to the measured values.
The finding that the data lie below TNG300-1 at small radii is also consistent with Hadzhiyska et al.~\cite{Hadzhiyska2025_missing_baryons}. Within the FLAMINGO suite, our measurements lie closest to the fiducial calibration and above the stronger-feedback fgas$-8\sigma$ variation at most radii; we note that McCarthy et al.~\cite{McCarthy2025_FLAMINGO}, comparing FLAMINGO to joint kSZ and lensing measurements, found a preference for feedback stronger than the fiducial calibration.

We note that many additional factors in simulations can affect the predicted $f_{\rm gas}^{\rm obs}$ profiles, including stellar fractions, neutral gas content, halo selection criteria, and background or foreground contamination. For a detailed discussion of these effects on mock simulation observations, see \cite{LiuInprep2026}. Nevertheless, the comparisons presented here suggest that current feedback models capture the qualitative trend of gas depletion around haloes, with the fiducial FLAMINGO calibration closest to our measurements.



\subsection{Simulation Beam-Compensated Gas Fractions}
\label{sec:beam_compensated_fgas}

\begin{figure}[h]
    \centering
    \includegraphics[width=0.9\columnwidth]{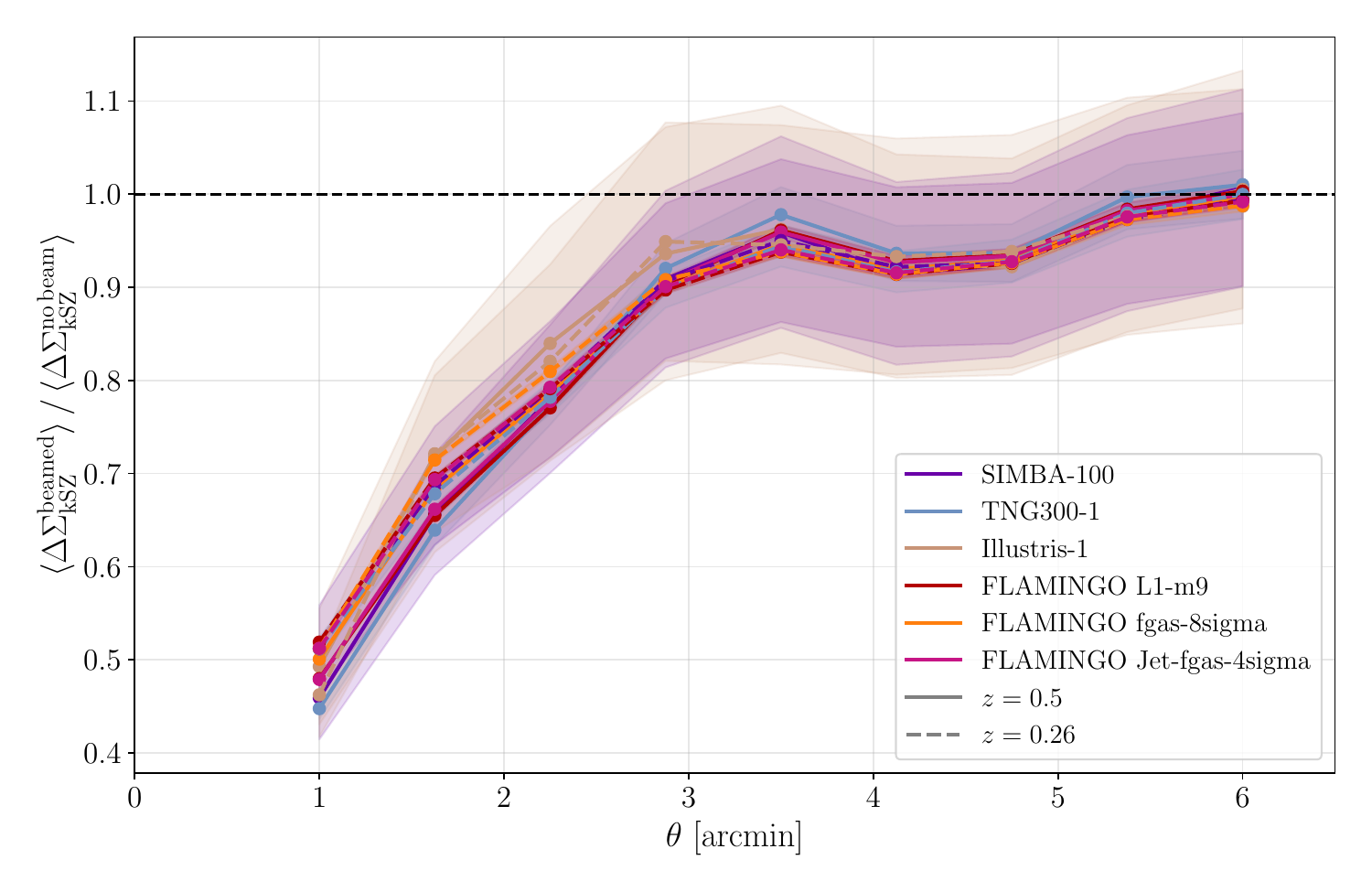}
    \caption{Simulation mock of beam effects on the kSZ $\Delta\Sigma$ profiles, shown as the ratio of the convolved kSZ $\Delta\Sigma$ profile to the unconvolved kSZ $\Delta\Sigma$ profile for each simulation. Despite the simulations' different feedback prescriptions (shown in Figure~\ref{fig:simulation_compare}), the beam suppression is largely independent of feedback model. The beam compensation factor $C_{\rm beam}(R)$ is defined as the inverse of the mean of these ratios across simulations (so that $C_{\rm beam} \geq 1$), and is applied to the observed gas fraction $f_{\rm gas}^{\rm obs}$ to recover the beam-compensated gas fraction $f_{\rm gas}$ (Eq.~\ref{eq:fgas_compensated}).
    The error bars represent the deviations of profiles within each simulation.
    }
    \label{fig:beam_compensation_factor}
\end{figure}

\begin{figure*}[htbp!]
    \centering
    \begin{subfigure}{0.49\textwidth}
        \centering
        \includegraphics[width=\textwidth]{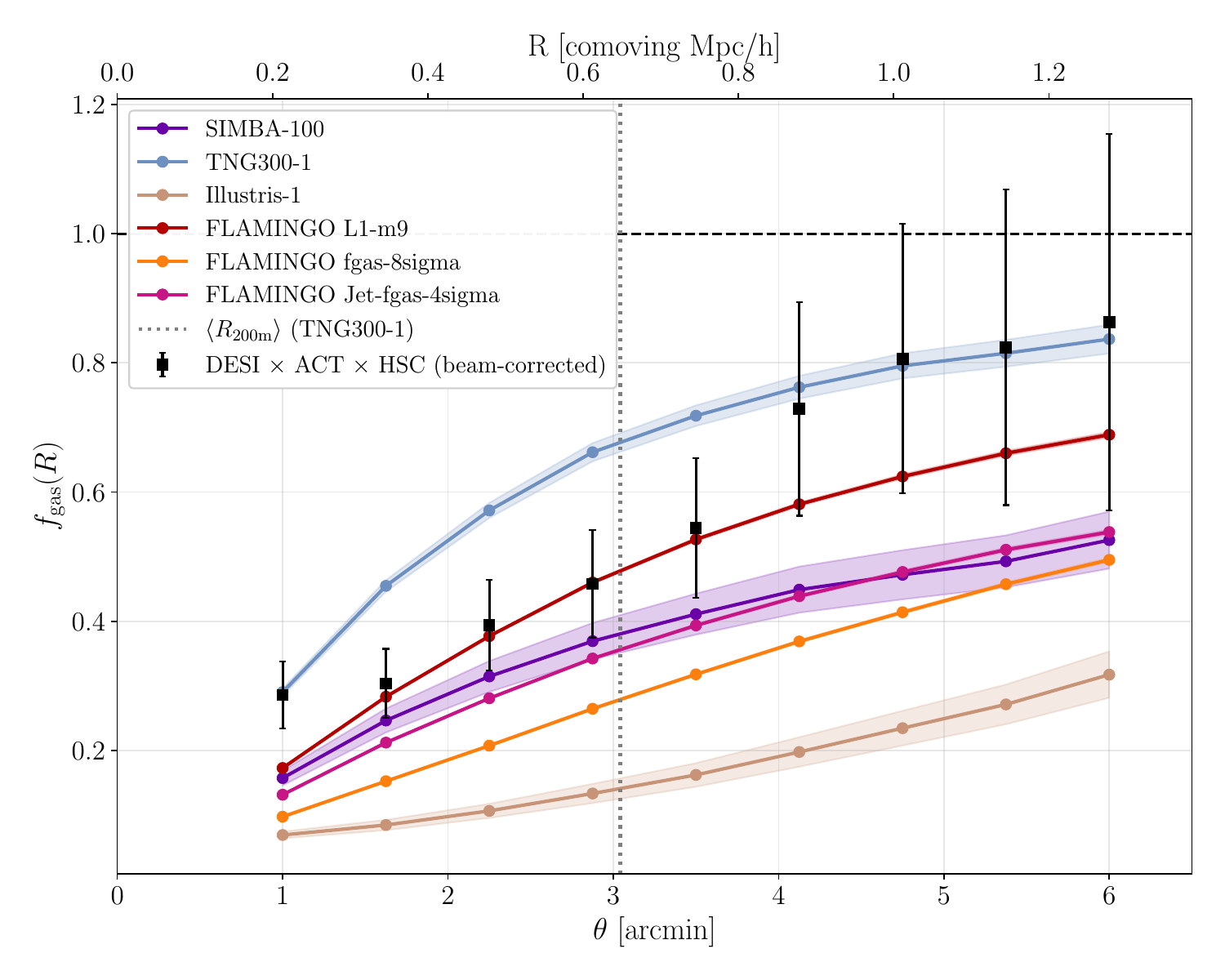}
        \caption{DESI BGS sample at $z = 0.26$. Detection significance against the no-feedback null ($f_{\rm gas} = 1$ at all radii): $\mathrm{SNR}_{\rm BGS} = 17.2$.}
        \label{fig:beam_compensated_BGS}
    \end{subfigure}
    \hfill
    \begin{subfigure}{0.49\textwidth}
        \centering
        \includegraphics[width=\textwidth]{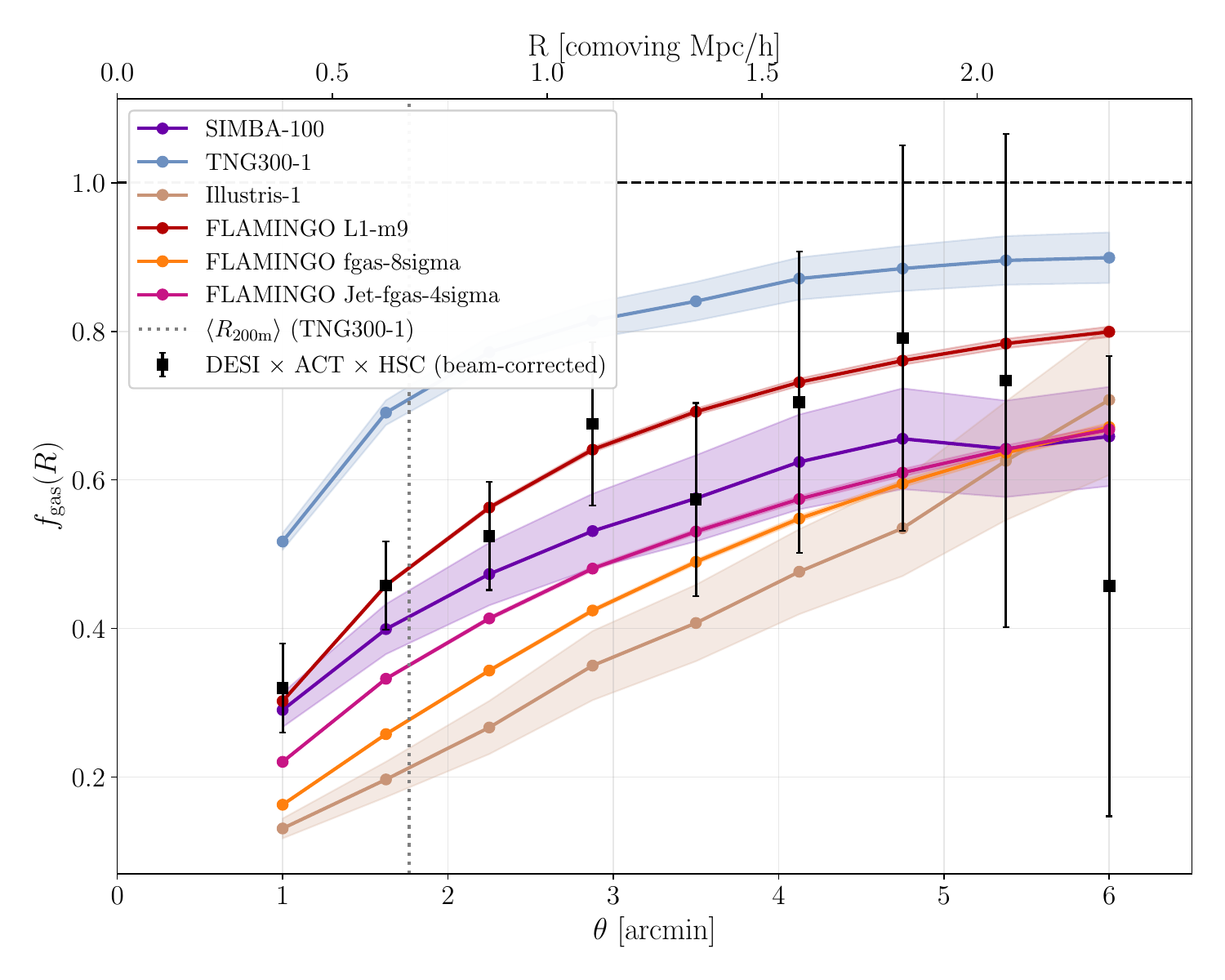}
        \caption{DESI LRG bin 1 at $z = 0.5$. Detection significance against the no-feedback null ($f_{\rm gas} = 1$ at all radii): $\mathrm{SNR}_{\rm LRG} = 13.8$.}
        \label{fig:beam_compensated_LRG}
    \end{subfigure}
    \caption{Beam-compensated gas fraction $f_{\rm gas}(R) = C_{\rm beam}(R) \times f_{\rm gas}^{\rm obs}(R)$ for the DESI BGS (left, $z=0.26$) and LRG bin~1 (right, $z=0.5$) lens samples. Observational measurements (black squares) have been corrected for ACT beam suppression using the compensation factor $C_{\rm beam}(R)$ derived from simulations (Figure~\ref{fig:beam_compensation_factor}); the simulation curves show the corresponding intrinsic (unconvolved) gas fraction profiles. 
    The dashed horizontal line at unity marks the null hypothesis in which ionized gas traces the total matter distribution at the cosmic baryon fraction $\Omega_b/\Omega_m$. The dotted vertical line represents the virial radius ($R_{200m}$) of the haloes selected in TNG300 as a point of reference; the other simulations have similar halo mass distributions and thus similar virial radii.
    Shaded bands show the $1\sigma$ uncertainty on the simulated mean. The bottom axis shows projected angular separation; the top axis shows the corresponding comoving transverse separation in $\mathrm{Mpc}/h$ for the mean redshift of each sample. 
    }
    \label{fig:beam_compensated}
\end{figure*}

\begin{figure*}[htbp!]
    \centering
    \begin{subfigure}{0.49\textwidth}
        \centering
        \includegraphics[width=\textwidth]{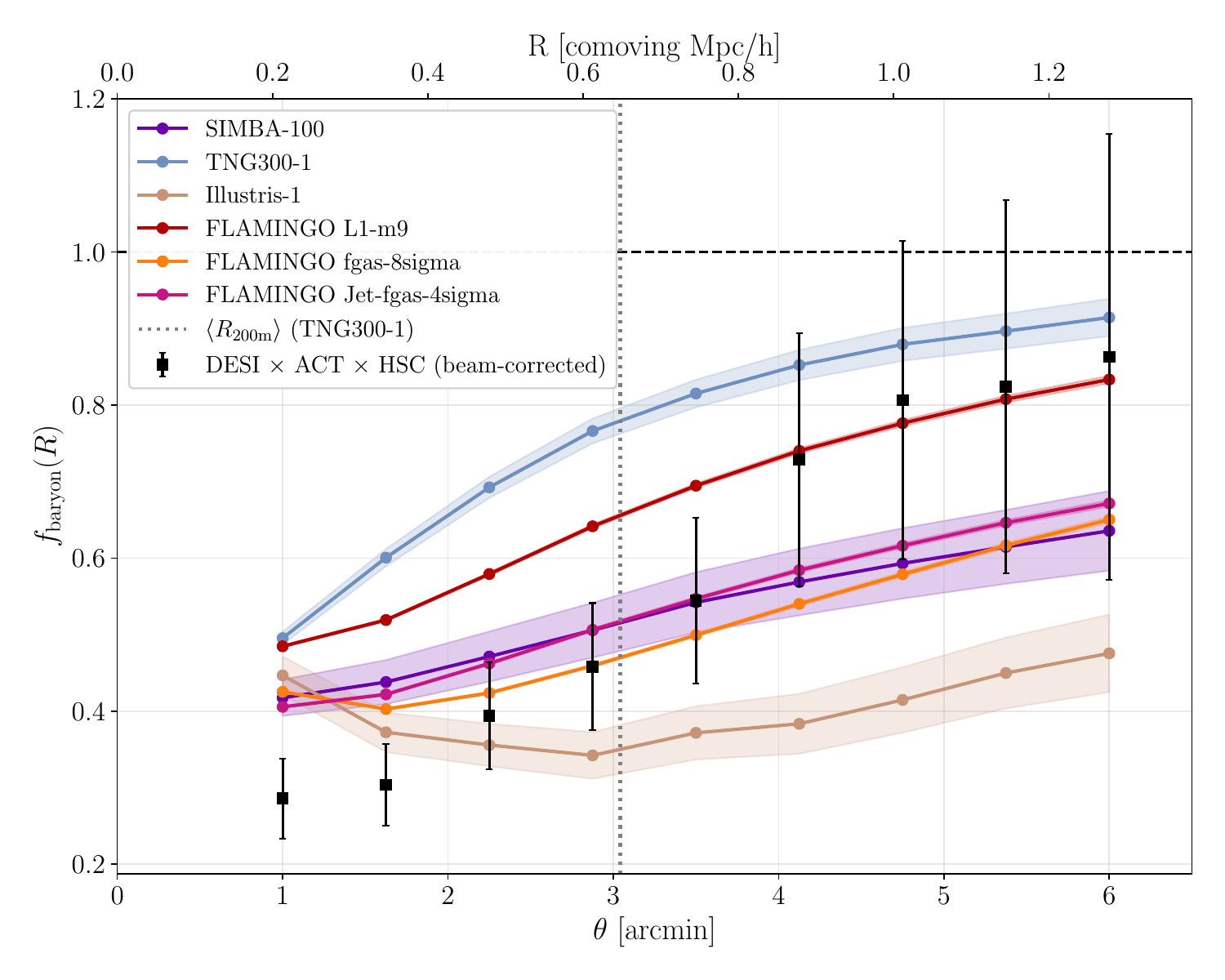}
        \caption{DESI BGS sample at $z = 0.26$.}
        \label{fig:beam_compensated_BGS_baryon}
    \end{subfigure}
    \hfill
    \begin{subfigure}{0.49\textwidth}
        \centering
        \includegraphics[width=\textwidth]{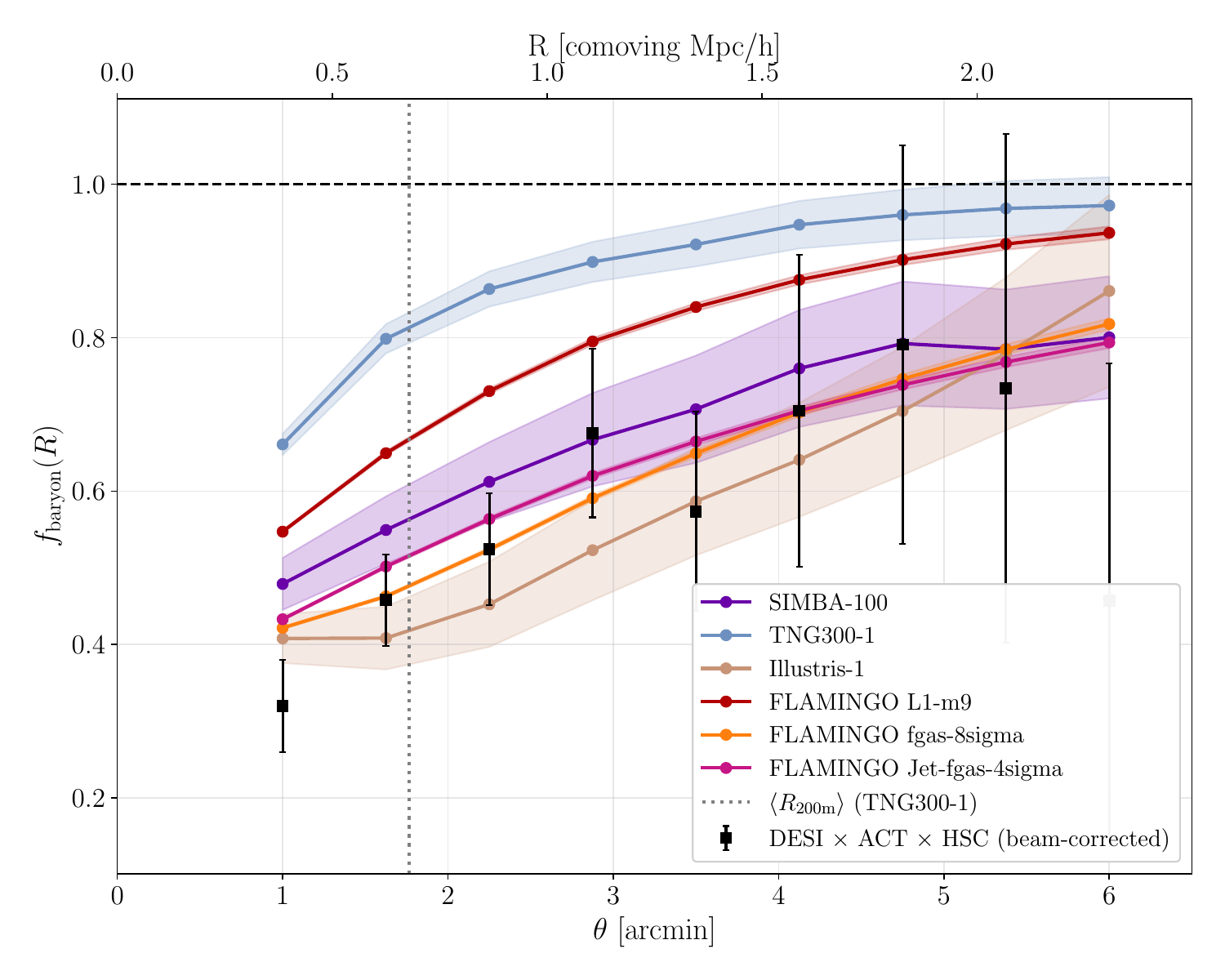}
        \caption{DESI LRG bin 1 at $z = 0.5$.}
        \label{fig:beam_compensated_LRG_baryon}
    \end{subfigure}
    \caption{Beam-compensated gas fraction measurements compared with simulated baryon fraction profiles $f_{\rm baryon}(R)$ for the DESI BGS (left, $z=0.26$) and LRG bin~1 (right, $z=0.5$) lens samples.
    The measurements are identical to those in Figure~\ref{fig:beam_compensated}, since the kSZ signal is sourced only by free electrons; the simulation curves, however, now show the total baryon fraction, including neutral gas and stars as well as ionized gas.
    The curves are therefore not a like-for-like comparison to the data, but the difference between each simulation's curves here and in Figure~\ref{fig:beam_compensated} is itself informative: it is precisely the non-ionized baryon budget of that simulation, and shows that the ionized gas we probe accounts for only part of the total baryon content around these haloes.
    All other plot elements are as in Figure~\ref{fig:beam_compensated}.
    }

    \label{fig:beam_compensated_baryon}
\end{figure*}


Having presented the observed gas fraction $f_{\rm gas}^{\rm obs}(R)$ and compared it with beam-forward-modelled simulations, we now derive the beam-compensated gas fraction $f_{\rm gas}(R)$ using simulations to correct for the ACT beam suppression. The beam-compensated profiles provide a more direct interpretation of the underlying gas fractions around haloes.

Figure~\ref{fig:simulation_compare} illustrated the beam-suppressed $f_{\rm gas}^{\rm obs}$ profiles, in which the simulated kSZ profiles have been convolved with the ACT beam to match the conditions of our measurements.
Though informative, these profiles are not directly comparable to the true underlying gas fractions around haloes, as the beam convolution suppresses the kSZ signal at small radii.

The beam suppression of our measured $f_{\rm gas}^{\rm obs}$ profiles can be quantified in simulations, as shown in Figure~\ref{fig:beam_compensation_factor}. As established in Section~\ref{sec:beam_effects}, this suppression is feedback-independent and can be used to derive a beam compensation factor $C_{\rm beam}(R)$ at each radius, which is then applied to our measured $f_{\rm gas}^{\rm obs}$ profiles to obtain beam-compensated profiles that more directly reflect the underlying gas fractions.

To obtain the beam-compensation factor $C_{\rm beam}(R)$, we compute the ratio of the unconvolved kSZ $\Delta\Sigma$ profile to the convolved kSZ $\Delta\Sigma$ profile for each simulation, and take the mean of these ratios across the six simulations to obtain a robust estimate of the beam compensation factor as a function of radius. We then apply this factor to our measured $f_{\rm gas}^{\rm obs}$ profiles to obtain the beam-compensated gas fraction:
\begin{equation}
    f_{\rm gas}(R) = C_{\rm beam}(R) \times f_{\rm gas}^{\rm obs}(R).
    \label{eq:fgas_compensated}
\end{equation} 

The resulting beam-compensated gas fraction profiles for the DESI BGS and LRG samples are shown in Figure~\ref{fig:beam_compensated}. These profiles provide a more direct interpretation of the underlying observed gas fractions around haloes, though the nature of the simulation beam compensation procedure means that the profiles shown are no longer strictly model-independent. However, we see that the beam-compensated profiles still show a significant baryon deficiency at the centres of these haloes.

Furthermore, the beam-compensated profiles allow for a more direct comparison at physical scales between the DESI BGS and LRG samples, as each sample is corrected for its own redshift-dependent beam suppression and plotted against comoving transverse separation (top axis of Figure~\ref{fig:beam_compensated}). Comparing the two samples at the same physical scales, we see similar baryon suppression within the virial radii. Though the BGS sample seems to recover to cosmic baryon fraction at a faster rate outside the virial radii, we note the consistency of this measurement with the LRG measurement given the larger error bars in this regime. 
The consistency of the BGS and LRG measurements further supports the interpretation of relative mass-independence for the gas fraction profiles as discussed in Section \ref{sec:mass_bins}, as the BGS sample probes lower mass haloes than the LRG sample.

We can further compare the beam-compensated profiles with the intrinsic (unconvolved) simulation curves. To quantify the statistical significance of the measured baryon depletion, we compute $\chi^2 = (\mathbf{f}_{\rm gas} - \mathbf{1})^T \mathbf{C}^{-1} (\mathbf{f}_{\rm gas} - \mathbf{1})$, where $\mathbf{f}_{\rm gas}$ is the vector of beam-compensated gas fraction values across the 9 radial bins, $\mathbf{1}$ is the no-feedback null vector ($f_{\rm gas} = 1$ at all radii), and $\mathbf{C}$ is the full propagated covariance of $f_{\rm gas}$ (Eq.~\ref{eq:fgas_cov}), with the Hartlap correction applied \cite{Hartlap2007}. 
Defining $\mathrm{SNR} \equiv \sqrt{\chi^2}$, we detect baryon depletion relative to the cosmic mean at $\mathrm{SNR}_{\rm BGS} = 17.2$ for the BGS sample and $\mathrm{SNR}_{\rm LRG} = 13.8$ for LRG bin~1.

We compare our beam-compensated measurements with the intrinsic gas fraction profiles predicted by the simulations, which are shown as the curves in Figure~\ref{fig:beam_compensated}. We see consistent results when compared to Figure~\ref{fig:simulation_compare}: at small apertures the data remain bracketed by the strongest and weakest feedback prescriptions, with SIMBA-100 and the fiducial FLAMINGO run tracing the data most closely, while at large apertures TNG300-1 provides the better match for the BGS sample and the fiducial FLAMINGO run for the LRG sample.

However, we note here that the ionized gas, which the kSZ signal probes, is only one component of the total baryon budget around haloes. Other components, such as neutral gas and stars, also contribute to the total baryon content and can be probed by other observational techniques. 
This distinction bears directly on the interpretation of Figure~\ref{fig:beam_compensated}: a simulation can predict a low ionized-gas fraction either because feedback has ejected gas from the halo, or because a larger fraction of its baryons are locked in stars or remain neutral. These possibilities are physically distinct but produce an identical kSZ signature, so the spread between simulations in Figure~\ref{fig:beam_compensated} cannot be attributed to differences in feedback strength alone.
To illustrate the potential impact of these other baryon components, Figure~\ref{fig:beam_compensated_baryon} repeats the comparison of Figure~\ref{fig:beam_compensated} with the simulation curves replaced by the intrinsic baryon fraction profiles $f_{\rm baryon}(R)$, which include all gas (ionized and neutral) as well as stars. The data points are identical to those in Figure~\ref{fig:beam_compensated}, as the kSZ signal is sourced only by free electrons; the difference between each simulation's baryon and ionized-gas curves, which grows toward small radii, therefore reflects the stellar and neutral-gas contributions to the baryon budget that our measurement does not probe. We emphasise, however, that the stellar and neutral-gas content itself differs substantially between simulations: Comparing Figures \ref{fig:beam_compensated} and \ref{fig:beam_compensated_baryon}, we see that the FLAMINGO runs appear to contain roughly a factor of two more non-ionized baryons (stars and neutral gas) than IllustrisTNG for the halo samples considered here. Since stars are invisible to the kSZ effect, this uncertainty in the stellar fraction directly complicates the interpretation of feedback strength from baryon-fraction comparisons such as Figure~\ref{fig:beam_compensated_baryon}. 
The difference between Figures~\ref{fig:beam_compensated} and~\ref{fig:beam_compensated_baryon} for a given simulation is therefore precisely its non-ionized baryon budget.
We refer the reader to \cite{LiuInprep2026} for a detailed discussion of the contributions of these different baryon components, including the stellar fraction, to the total baryon budget around haloes, and their implications for feedback constraints.




\section{Discussion and Conclusion}
\label{sec:conclusion}

In this work, we present model-independent measurements of the gas fraction around the DESI BGS and LRG galaxy samples, combining kSZ measurements from DESI and ACT with galaxy-galaxy lensing from DESI and HSC. Applying the same $\Delta\Sigma$ aperture filter to both observables, we take their ratio to obtain the observed gas fraction $f_{\rm gas}^{\rm obs}(R)$ directly, without profile fitting or simulation calibration. 
To account for the ACT CMB beam, we apply a simulation-derived, feedback-independent beam-compensation factor to recover the beam-corrected gas fraction $f_{\rm gas}(R)$ for the highest signal bins (DESI BGS and LRG bin~1). After this correction, we detect baryon depletion relative to the cosmic mean baryon fraction at $\mathrm{SNR} = 17.2$ for the BGS sample ($z \sim 0.26$) and $\mathrm{SNR} = 13.8$ for the lowest-redshift LRG bin ($z \sim 0.5$). In both samples $f_{\rm gas}(R)$ is strongly suppressed at small radii and rises toward unity at larger radii, providing direct, model-independent evidence that feedback redistributes gas from the inner regions of haloes outward.

Comparing these measurements with six simulations from the Illustris, IllustrisTNG, SIMBA, and FLAMINGO suites, we find broad qualitative agreement, with a gas fraction below unity at small radii that recovers at large radii, and the fiducial FLAMINGO run tracks the data most closely in both samples.
TNG300-1 matches the large-aperture recovery best in the BGS sample, while the fiducial FLAMINGO run matches the LRG bin~1 measurements best at large apertures; at small apertures the data lie closest to the fiducial FLAMINGO run and SIMBA-100. Illustris-1, with its aggressive feedback, depletes gas more than observed at all radii for the BGS sample and at small radii for the LRG sample, behaviour shared to a lesser degree by the strong-feedback FLAMINGO variation fgas$-8\sigma$.
These results are consistent with, and complementary to, recent joint analyses of the kSZ effect and CMB lensing by Hadzhiyska et al.~\cite{Hadzhiyska2025_missing_baryons},
as well as work showing the relative mass-independence of the gas fractions at group-sized haloes \cite{Roper2025}, 
with the distinction that our $\Delta\Sigma$-ratio yields the gas fraction directly, 
because the total matter is measured by lensing around the same galaxies with the same filter, so that no halo mass calibration or assumed profile shape intervenes between the two observables.

Our comparison is also in line with some recent X-ray studies of galaxy groups. The X-ray scaling relations of these groups disfavour highly ejective feedback models~\cite{Eckert2026, Seppi2026}, with the intermediate-strength FLAMINGO variant fgas$-2\sigma$ (not included in this work) preferred and the most ejective variant fgas$-8\sigma$ excluded~\cite{Seppi2026}, while hot gas fractions measured out to the virial radius lie closest to the fiducial FLAMINGO calibration~\cite{Khalil2026}, as does the cross-correlation between diffuse X-ray emission and cosmic shear~\cite{McDonald2026_Xray}. These measurements trace the hot, X-ray-emitting gas within the virial radius of X-ray-selected, predominantly low-redshift systems, whereas the kSZ signal is sensitive to all ionized gas regardless of temperature, around optically selected galaxies at higher redshift and out beyond the virial radius. The two probes are therefore complementary, and in both the most ejective FLAMINGO variant falls below the measured gas content.

The simulated profiles used in this comparison are constructed following our companion paper~\cite{LiuInprep2026}, which examines the same simulations in detail. That work shows that the spread in predicted gas fractions across simulations is driven not only by the strength of feedback but also by several effects to which the kSZ signal responds in subtle ways: the fraction of baryons locked in stars, which are invisible to the SZ effect and vary by factors of a few across codes; the treatment of neutral gas, notably SIMBA's on-the-fly molecular partitioning, which removes gas from the ionized budget; and the contribution of diffuse, unbound gas beyond the virial radius, which escapes halo-based baryon-painting models. Because these modelling choices can mimic or mask differences in feedback strength, the companion paper argues that robust constraints require model-independent measurements of the kind presented here. Our $f_{\rm gas}^{\rm obs}(R)$ measurement is precisely such a probe, and the spread between simulations illustrates the challenge of model-dependent inference: part of this scatter reflects variation in baryon-budget bookkeeping rather than feedback physics alone, so even the close agreement with the fiducial FLAMINGO run does not by itself validate its feedback model.

Given the novel nature of the $\Delta\Sigma$-ratio measurement, we perform a suite of systematic and consistency tests to validate our results in the appendices.
This includes null tests in which stacking on random positions and on shuffled velocities both yield $\Delta\Sigma$ profiles consistent with zero (Appendix~\ref{app:systematic_tests}), indicating that residual foregrounds and velocity-independent contaminants such as the tSZ effect and the CIB do not bias the kSZ signal. Splitting the samples by stellar mass (Section~\ref{sec:mass_bins}), we find that $f_{\rm gas}^{\rm obs}(R)$ is largely independent of mass even though the individual kSZ and lensing signals are not, which makes the result robust to the details of subhalo selection in the simulation comparison and echoes the weak dependence on halo-selection method found in the companion paper. Finally, pruning close pairs to remove satellites (Appendix~\ref{app:miscentering}) leaves a central-dominated sample whose $f_{\rm gas}^{\rm obs}$ profile is unchanged within the uncertainties, demonstrating that our results are robust both to halo miscentering and to the central-versus-satellite composition of the sample.

We have discussed several potential sources of systematics for our work.
Our raw model-independent measurement is the beam-suppressed $f_{\rm gas}^{\rm obs}(R)$, with a strong beam effect suppressing the gas fraction at small apertures.
The simulation-calibrated beam compensation method, though appearing feedback independent, is still dependent on simulations and introduces its own modelling into the measurement process, so the beam-corrected $f_{\rm gas}(R)$ is no longer strictly model-independent at the smallest radii where the correction is largest.
The fiducial line-of-sight velocity $v_{\rm los} = 300\,\mathrm{km\,s^{-1}}$ and the velocity-reconstruction factor $r_V$ enter as overall multiplicative normalizations, shifting $f_{\rm gas}^{\rm obs}$ uniformly without changing its radial shape. On the lensing side, we do not apply a lens-magnification correction and we approximate each lens bin by its mean redshift; both effects are expected to be subdominant to our statistical uncertainties. 
One notable feature is the LRG bin~2 kSZ profile, which turns negative at large apertures. As shown in Appendix~\ref{app:consistency_tests}, this originates in the SGC field, where the NGC and SGC measurements differ at the $1.7\sigma$ level; the NGC-only result is consistent with the other samples. A similar feature was reported by Qu et al.~\cite{Qu2026_kSZ}. We attribute the anomaly to a residual systematic or a statistical fluctuation in the SGC kSZ measurement for this bin.

Our configuration-space measurements of $f_{\rm gas}(R)$ have direct implications for the suppression of the matter power spectrum by baryonic feedback in Fourier space, one of the leading systematic uncertainties for weak-lensing cosmology. Because feedback that redistributes gas around haloes also suppresses small-scale power, the measured gas profiles can be mapped onto a predicted power-spectrum suppression, either analytically through halo models fit to our profiles or by training emulators on simulation suites such as CAMELS \cite{Villaescusa-Navarro2021_CAMELS} that span a wide range of feedback parameters. 
These model-independent measurements can also be used to calibrate and constrain the next generation of cosmological hydrodynamical simulations, which are rapidly improving in resolution and physical fidelity but still show significant discrepancies in their feedback prescriptions and their impact on the gas distribution around haloes.

\rev{These prospects will be sharpened by the next generation of lensing and CMB data. The depth that makes HSC the only current shear catalogue able to reach the full DESI LRG redshift range also makes this measurement a preview of what will be possible with the Rubin Observatory. The HSC Y3 catalogue reaches a $5\sigma$ $i$-band point-source depth of $i \approx 26$ with an effective source density of $n_{\rm eff} = 19.9 \,\mathrm{arcmin}^{-2}$ \cite{Li2022_HSCY3_ShearCatalog}, between the effective source densities of 10 and 27 $\mathrm{arcmin}^{-2}$ assumed for LSST Y1 and Y10 in the LSST DESC Science Requirements Document \cite{LSST2018}.
For the BGS and lowest-redshift LRG samples the uncertainty on $f_{\rm gas}^{\rm obs}$ is dominated by the kSZ rather than the lensing, so Rubin would bring only a modest gain in precision for these bins. The highest-redshift LRG bins, by contrast, are lensing-limited (see Fig.~\ref{fig:LRG_lensing_profiles}), and Rubin's depth and source density would extend the model-independent measurement cleanly to $z \sim 1$ and allow finer splits by stellar mass and redshift. 
Combined with the Simons Observatory \cite{SO2019}, with further gains from the larger spectroscopic samples of future DESI releases, this would turn $f_{\rm gas}(R)$ into an even more precise, model-independent prior on baryonic feedback \cite{Wayland2025}.}

Finally, the kSZ signal probes only the ionized gas, a substantial but incomplete component of the total baryon budget. The comparison with simulated baryon fractions in Figure~\ref{fig:beam_compensated_baryon} shows that the neutral-gas and stellar components contribute an additional, code-dependent fraction. Combining our ionized-gas $f_{\rm gas}(R)$ with independent constraints on these components, including neutral gas from 21\,cm measurements, and stars from stellar-mass estimates, alongside cross-checks on the ionized component from FRB dispersion measures and X-rays, offers a route to closing the full baryon census around galaxies.

In summary, we have established $\Delta\Sigma$ filtering of the kSZ and weak-lensing signals as a direct, model-independent probe of the gas distribution around galaxies, and used it to measure the radial gas fraction and detect baryon depletion at high significance in the DESI BGS and LRG samples. These measurements offer a robust route to constraining baryonic feedback and its impact on cosmological observables, complementing the simulation-based perspective of our companion paper~\cite{LiuInprep2026}.

\section*{Data Availability}
Data points for the figures are available in digital format at \url{https://zenodo.org/records/22952603}.

The code used to generate the figures and perform the analysis is available upon reasonable request to the corresponding author.

\begin{acknowledgments}
We thank Calvin Leung and Linda Jin for helpful discussions and feedback on this work.
We acknowledge the use of Claude for code implementation assistance and for proofreading the manuscript. All scientific content, analysis choices, and results and conclusions are the authors’ own, and the authors take full responsibility for the contents of this paper.

RHL is supported by the Postgraduate-Doctoral Scholarship from the Natural Sciences and Engineering Research Council of Canada (NSERC), funding reference number PGSD-567923-2022.

This material is based upon work supported by the U.S. Department of Energy (DOE), Office of Science, Office of High-Energy Physics, under Contract No. DE–AC02–05CH11231, and by the National Energy Research Scientific Computing Center, a DOE Office of Science User Facility under the same contract. Additional support for DESI was provided by the U.S. National Science Foundation (NSF), Division of Astronomical Sciences under Contract No. AST-0950945 to the NSF’s National Optical-Infrared Astronomy Research Laboratory; the Science and Technology Facilities Council of the United Kingdom; the Gordon and Betty Moore Foundation; the Heising-Simons Foundation; the French Alternative Energies and Atomic Energy Commission (CEA); the Secretariat of Science, Humanities, Technology and Innovation (SECIHTI) of Mexico; the Ministry of Science, Innovation and Universities of Spain (MICIU/AEI/10.13039/501100011033), and by the DESI Member Institutions: \url{https://www.desi.lbl.gov/collaborating-institutions}.
Any opinions, findings, and conclusions or recommendations expressed in this material are those of the author(s) and do not necessarily reflect the views of the U. S. National Science Foundation, the U. S. Department of Energy, or any of the listed funding agencies.

The authors are honored to be permitted to conduct scientific research on Iolkam Du'ag (Kitt Peak), a mountain with particular significance to the Tohono O'odham Nation.

We acknowledge the use of public ACT data products made available through the National Energy Research Scientific Computing Center (NERSC), a U.S. Department of Energy Office of Science User Facility operated under Contract No. DE-AC02-05CH11231.

The Hyper Suprime-Cam (HSC) collaboration includes the astronomical communities of Japan and Taiwan, and Princeton University. The HSC instrumentation and software were developed by the National Astronomical Observatory of Japan (NAOJ), the Kavli Institute for the Physics and Mathematics of the Universe (Kavli IPMU), the University of Tokyo, the High Energy Accelerator Research Organization (KEK), the Academia Sinica Institute for Astronomy and Astrophysics in Taiwan (ASIAA), and Princeton University. Funding was contributed by the FIRST program from Japanese Cabinet Office, the Ministry of Education, Culture, Sports, Science and Technology (MEXT), the Japan Society for the Promotion of Science (JSPS), Japan Science and Technology Agency (JST), the Toray Science Foundation, NAOJ, Kavli IPMU, KEK, ASIAA, and Princeton University. 
This paper makes use of software developed for the Large Synoptic Survey Telescope. We thank the LSST Project for making their code available as free software at  http://dm.lsst.org
This paper is based on data collected at the Subaru Telescope and retrieved from the HSC data archive system, which is operated by the Subaru Telescope and Astronomy Data Center (ADC) at NAOJ. Data analysis was in part carried out with the cooperation of Center for Computational Astrophysics (CfCA), NAOJ. We are honored and grateful for the opportunity of observing the Universe from Maunakea, which has the cultural, historical and natural significance in Hawaii. 

We acknowledge the Virgo Consortium for making their simulation data available. The flamingo simulations were performed using the Durham Memory Intensive system managed by the Institute for Computational Cosmology on behalf of the STFC DiRAC facility (\url{www.dirac.ac.uk}).

\end{acknowledgments}

\appendix

\section{Effects of Miscentering}
\label{app:miscentering}

\begin{figure}[h]
    \centering
    \includegraphics[width=\columnwidth]{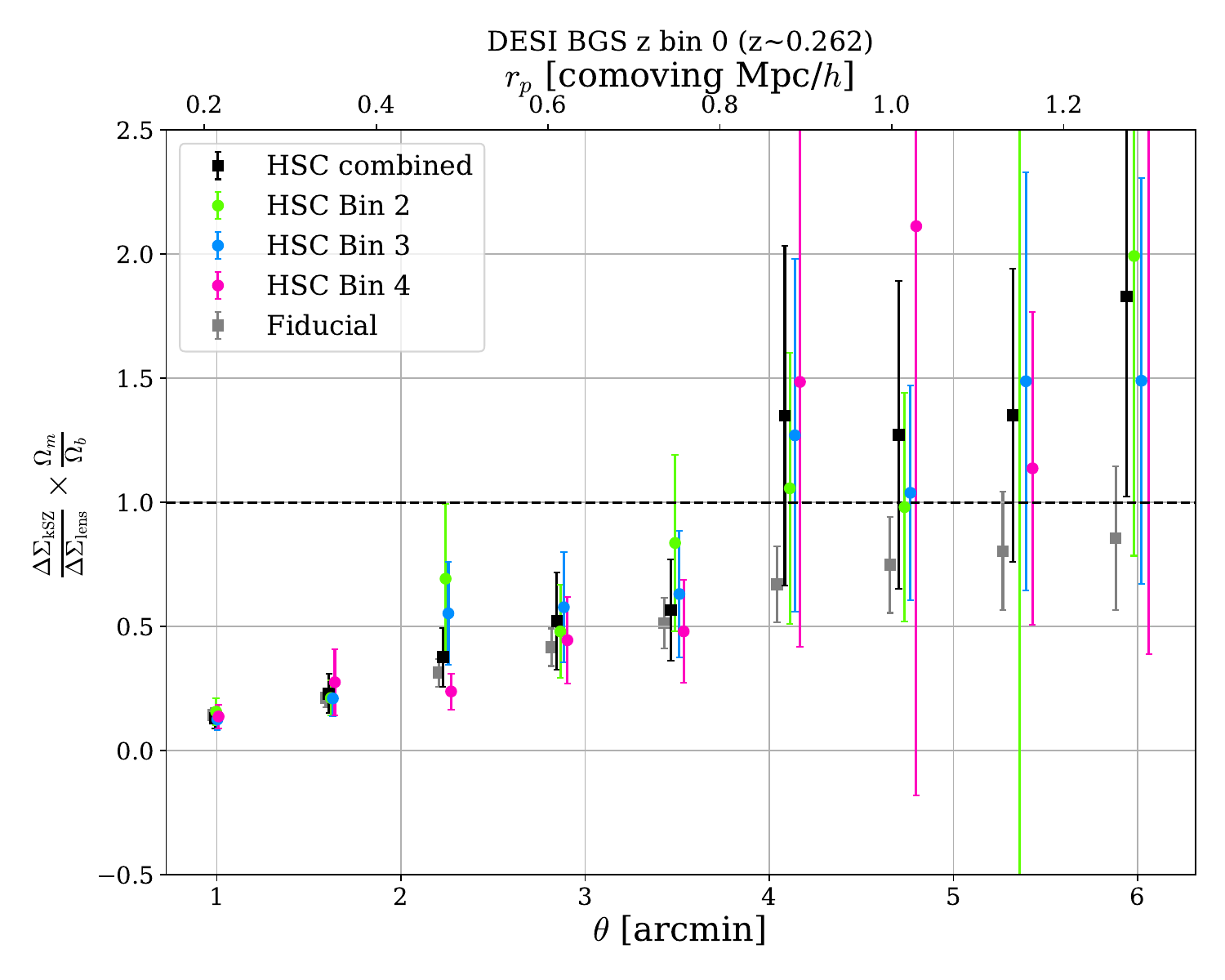}
    \caption{Observed gas fraction $f_{\rm gas}^{\rm obs}(R)$ for the DESI BGS sample with a method of pruning close pairs of galaxies to mitigate halo miscentering effects (separation threshold $0.5\,\mathrm{Mpc}$).
    The grey points show the fiducial unpruned measurement (HSC combined, $\pm1\sigma$) from Fig.~\ref{fig:combined_BGS} for reference.}
    \label{fig:combined_BGS_nopairs}
\end{figure}   

\begin{figure}[h]
    \centering
    \includegraphics[width=\columnwidth]{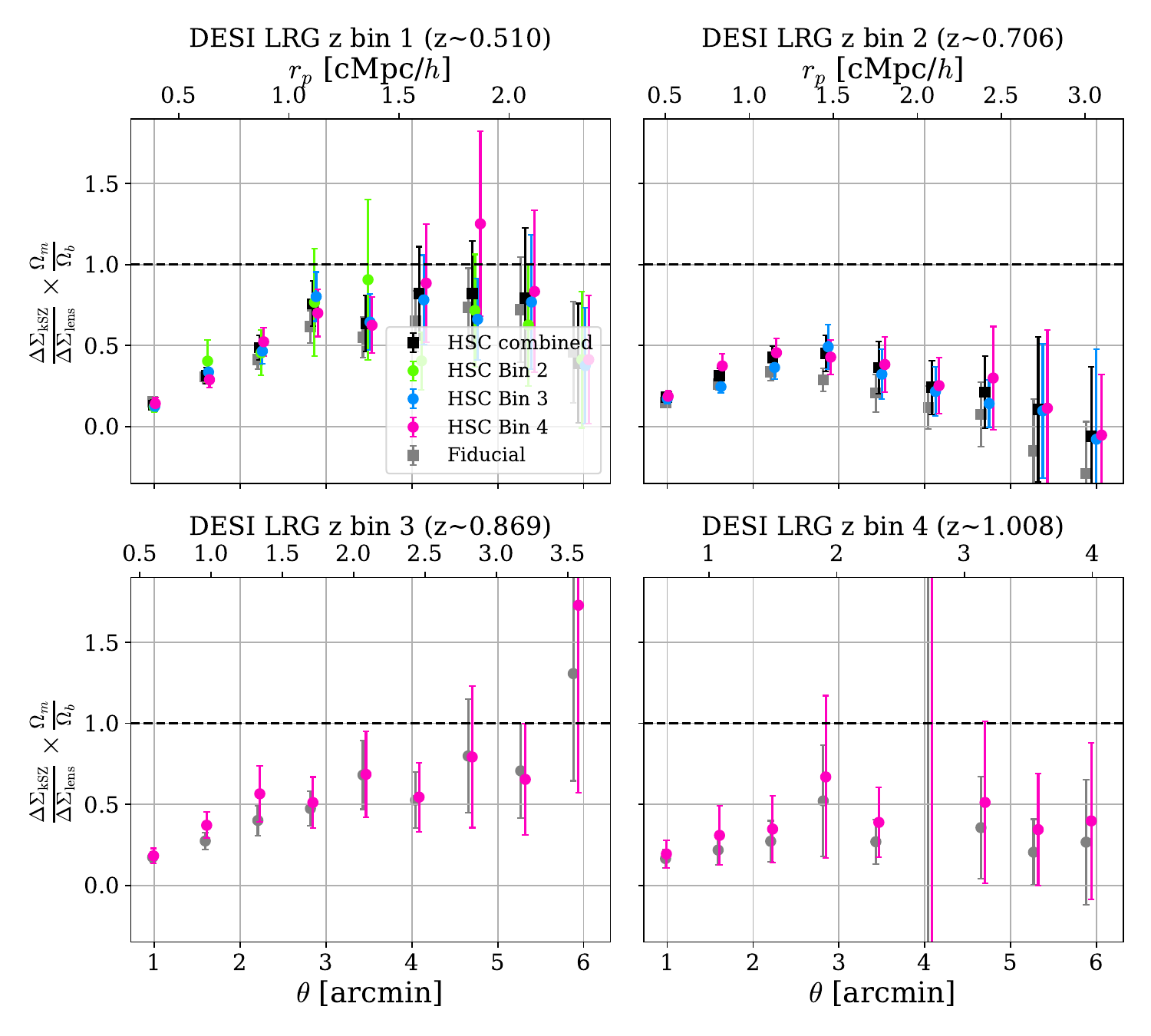}
    \caption{Observed gas fraction $f_{\rm gas}^{\rm obs}(R)$ for the DESI LRG sample in four redshift bins, with a method of pruning close pairs of galaxies to mitigate halo miscentering effects (separation threshold $0.5\,\mathrm{Mpc}$).
    \rev{The grey points shows the fiducial unpruned measurement (HSC combined, $\pm1\sigma$) from Fig.~\ref{fig:combined_LRG} for reference.}}
    \label{fig:combined_LRG_nopairs}
\end{figure}    

One important systematic effect that can bias both our kSZ and lensing measurements is halo miscentering, which occurs when the galaxy position used for stacking does not coincide with the true center of the halo. This miscentering can suppress the measured profiles at small radii, mimicking the effect of gas depletion. To mitigate this effect, we can apply pruning methods to remove close pairs of galaxies from our catalogue, which are more likely to be affected by miscentering. In this appendix, we explore removing close pairs of galaxies to only perform the analysis on isolated objects, and assess their impact on our measurements of $f_{\rm gas}^{\rm obs}(R)$ for both the DESI BGS and LRG samples.

Figures \ref{fig:combined_BGS_nopairs} and \ref{fig:combined_LRG_nopairs} show the $f_{\rm gas}^{\rm obs}(R)$ profiles for the BGS and LRG samples, respectively, both pruned to remove close pairs of galaxies at $0.5\,\mathrm{Mpc}$. As this physical distance represents a larger angular scale for the closer redshift bins, this pruning method is more aggressive for the BGS sample than for the LRG sample, and more aggressive for the lower redshift LRG bins than for the higher redshift LRG bins. We see that this pruning method does not significantly change our measurements of $f_{\rm gas}^{\rm obs}(R)$, especially at smaller $R$ where one expects a larger measurement effect from miscentered haloes.
This is in some ways expected, as the ACT beam size of $1.6\,\mathrm{arcmin}$ corresponds to a comoving scale of $\sim$$0.5\,\mathrm{Mpc}$ at $z \sim 0.3$, already smoothing and mitigating the most dramatic effects of miscentering in our beam-suppressed measurements shown.
However, this pruning method significantly increases the noise of measurement especially at larger apertures. As the miscentering effect is not expected to be significant at larger apertures, this increase in noise is not justified by any change in the measured profiles, and we therefore do not apply this pruning method in our main analysis.

\section{Halo Selection Tests in Simulations}
\label{sec:mass_bins_sims}

\begin{figure}
    \centering
    \includegraphics[width=\columnwidth]{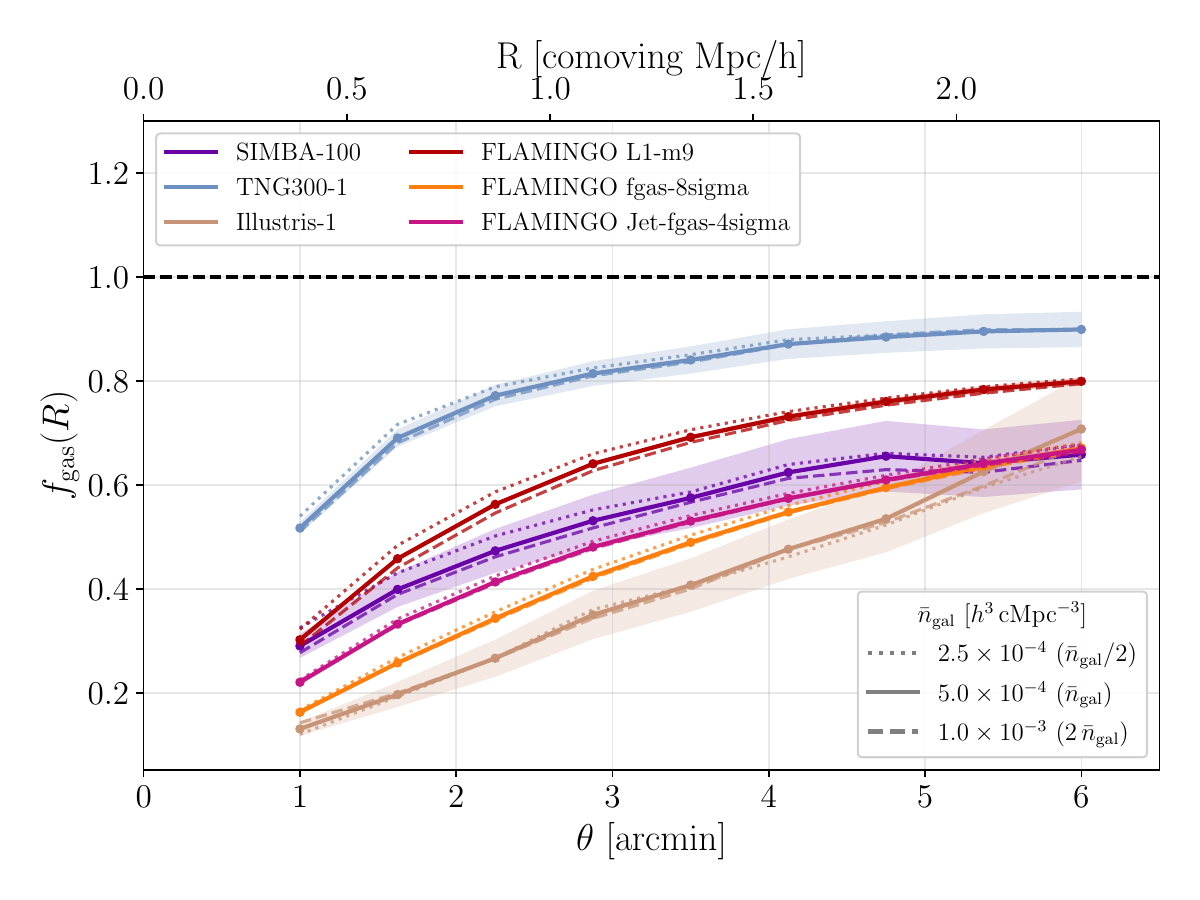}
    \caption{The intrinsic (unconvolved) $f_{\rm gas}(R)$ profiles at $z=0.5$ for the six simulations, for galaxy selections with number densities $\bar{n}_{\rm gal}/2$, $\bar{n}_{\rm gal}$ and $2\bar{n}_{\rm gal}$ (with $\bar{n}_{\rm gal} = 5\times10^{-4}\,h^3\,\mathrm{cMpc}^{-3}$, the fiducial LRG value), corresponding to different stellar-mass thresholds. The profiles are largely consistent across the different thresholds, and the little variation that is present is subdominant compared to the statistical uncertainties as shown in Figure \ref{fig:beam_compensated}. This demonstrates that our $\Delta\Sigma$-ratio measurement is robust to halo selection variations in the simulations.}
    \label{fig:mass_abundance_sims}
\end{figure}

One core methodological advantage of our $\Delta\Sigma$-ratio measurement is that it is largely independent of halo mass, even though the individual kSZ and lensing signals are not. This property is empirically shown in data in Section~\ref{sec:mass_bins}, where we split the DESI BGS and LRG samples into stellar-mass bins and find that the resulting $f_{\rm gas}^{\rm obs}(R)$ profiles are consistent across mass bins. However, this mass independence is not guaranteed a priori, and it is important to verify that it holds in simulations as well.

We test this mass independence in simulations by varying the number-density threshold of the simulated galaxy selection and measuring the resulting $f_{\rm gas}(R)$ profiles. Figure~\ref{fig:mass_abundance_sims} shows that the profiles are largely consistent across the different thresholds, indicating that our $\Delta\Sigma$-ratio measurement is robust to variations in halo mass.

Though our observable $f_{\rm gas}(R)$ is not strictly independent of halo mass, the $\Delta\Sigma$-ratio measurement effectively cancels out much of the mass dependence. Furthermore, the existing mass dependence is subdominant compared to the statistical uncertainties in our measurements, as shown in Figure \ref{fig:beam_compensated}. As a result, the simulation comparison in Section~\ref{sec:simulations} is not significantly biased by uncertainties in the halo mass selection of our samples, and our conclusions regarding feedback strength remain robust.





\section{Comparison with Previous Gas Fraction Measurements}
\label{app:comparisons_fgas}

\begin{figure}
    \centering
    \includegraphics[width=0.9\columnwidth]{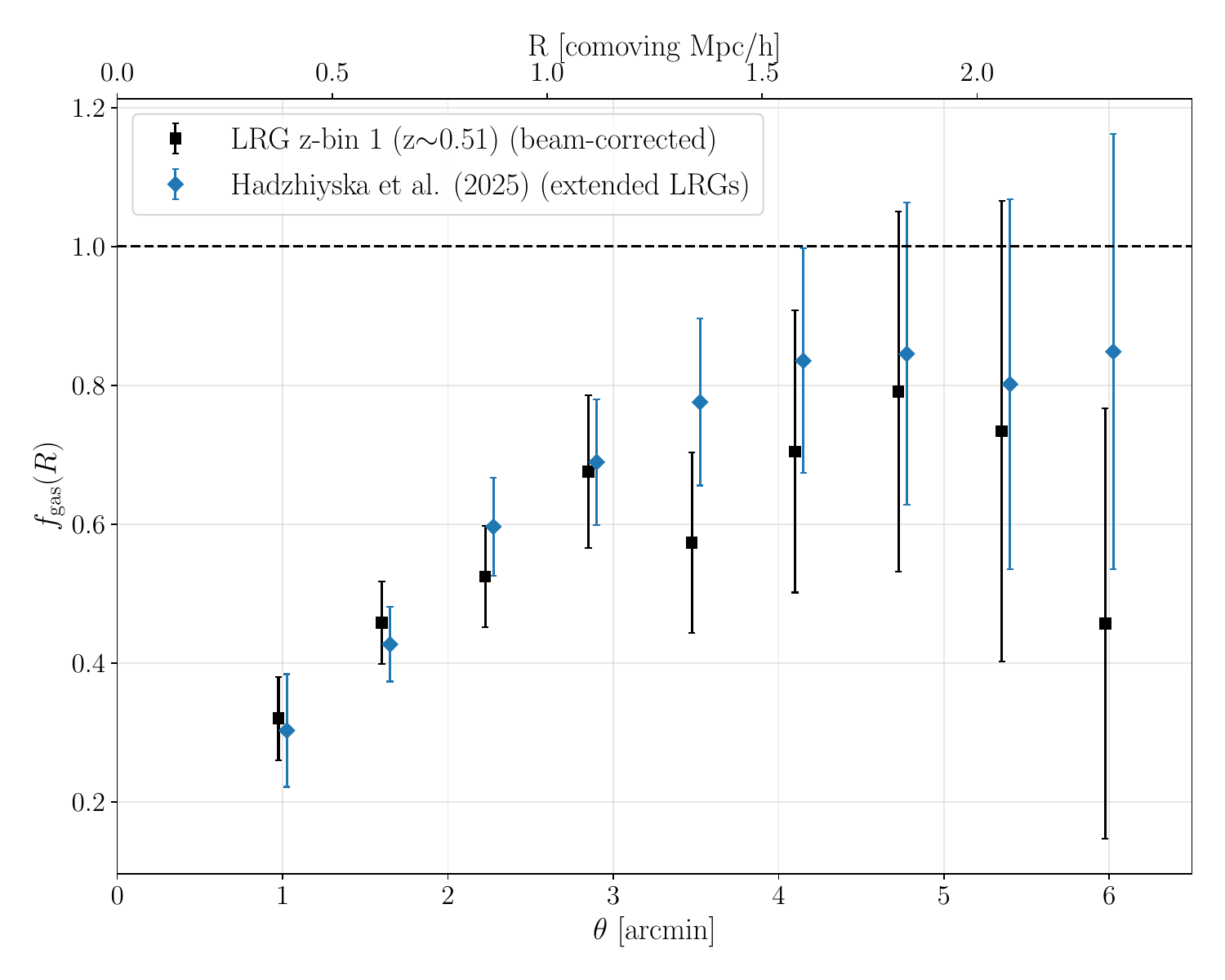}
    \caption{Beam-compensated gas fraction $f_{\rm gas}(R)$ for DESI LRG bin~1 from this work (black, as in Figure~\ref{fig:beam_compensated_LRG}) compared with the gas fraction of the photometric Extended LRG sample from Figure~6 of Hadzhiyska et al.~\cite{Hadzhiyska2025_missing_baryons} (blue), obtained from the ratio of CAP-filtered kSZ profiles to a no-feedback prediction calibrated on ACT DR6 CMB lensing. Both are shown at the same nine angular apertures, offset slightly in $\theta$ for readability. The dashed line at unity marks the null hypothesis in which gas traces the total matter at the cosmic baryon fraction. The upper axis gives the comoving separation at $z = 0.51$ and applies only to the black points; the Extended sample has a mean redshift near $z \sim 0.7$. Error bars are $1\sigma$.}
    \label{fig:compare_hadzhiyska}
\end{figure}

Figure~\ref{fig:compare_hadzhiyska} compares our beam-compensated $f_{\rm gas}(R)$ for LRG bin~1 with the measurement of Hadzhiyska et al.~\cite{Hadzhiyska2025_missing_baryons}, the closest analogue to this work. That analysis divides the CAP-filtered kSZ profiles of photometric DESI LRGs~\cite{Hadzhiyska2024a} by a no-feedback prediction derived from ACT DR6 CMB lensing. The prediction is convolved with the ACT beam before filtering, so the beam largely cancels in their ratio and the result is a beam-corrected quantity like our $f_{\rm gas}(R)$, rather than the beam-suppressed $f_{\rm gas}^{\rm obs}(R)$.
Like ours, the gas fraction in their analysis is obtained from a compensated-filter ratio (Section~\ref{sec:combine_measurements}). We show their Extended LRG sample, which has the higher signal-to-noise of their two LRG samples. The two profiles are consistent within the uncertainties at every aperture, both suppressed to $f_{\rm gas} \approx 0.3$ at $R = 1\,\mathrm{arcmin}$ and rising to $f_{\rm gas} \approx 0.8$ by $R \gtrsim 4\,\mathrm{arcmin}$.

The most important difference is how the total matter profile is obtained. In this work, it is measured directly by galaxy-galaxy lensing around the same galaxy sample with the same filter, so $f_{\rm gas}(R)$ is a ratio of two direct measurements. 
The ACT DR6 convergence maps used by Hadzhiyska et al. are limited to $\ell < 3000$, a cut imposed to limit foreground leakage from the tSZ effect and the CIB, which roughly coincides with the scale at which the one-halo term takes over, so the stacked convergence profiles constrain the halo mass but do not reach the scales probed by the smallest kSZ apertures. 
The matter profile in the denominator is instead obtained by fitting a halo occupation distribution model to the $\ell < 3000$ measurement in the AbacusSummit $N$-body simulations and extending the best-fitting mock sample to smaller scales. 
Their gas fraction is thus the ratio of a measurement to a simulation-derived profile calibrated on data, and is thus not a model independent measurement of gas fraction. 

The two analyses also differ slightly in sample selection, redshift, and filter.
The photometric Extended LRG sample~\cite{Zhou2023_DESILRG} is denser and less massive than the DESI LRG target selection, combines all four photometric redshift bins with a mean redshift near $z \sim 0.7$, and relies on photometric rather than spectroscopic velocity reconstruction, with a correspondingly lower velocity correlation coefficient. Because the comparison is made at fixed angular scale, a given aperture corresponds to a larger comoving separation for their points than the upper axis of Figure~\ref{fig:compare_hadzhiyska} indicates.
The CAP and $\Delta\Sigma$ filters also weight the projected profiles differently, though we expect this to matter less for the ratio than for the individual profiles, since each analysis applies the same filter to gas and matter; the companion paper~\cite{LiuInprep2026} examines this further.
The agreement should therefore be read as a qualitative consistency check rather than a quantitative test. It is nevertheless a reassuring agreement, and serves as an independent check to the validity of our measured gas profiles.

\section{Null and Systematic Tests}
\label{app:systematic_tests}

The novel component of this work is our use of the $\Delta\Sigma$ aperture filter to measure the kSZ signal. In this appendix, we perform a variety of null tests to ensure that our measurements are robust and not contaminated by systematic effects. These tests include stacking on random positions and shuffling velocities to check for tSZ leakage.

\begin{table}[htbp!]
    \centering
    \setlength{\tabcolsep}{6.3pt}  
    \begin{tabular}{l | ccc | ccc}
        \hline \hline
        & \multicolumn{3}{c|}{Random positions}
        & \multicolumn{3}{c}{Shuffled velocities} \\
        Sample
        & $\chi^2_{\rm null}$ & dof & PTE
        & $\chi^2_{\rm null}$ & dof & PTE \\
        \hline
        BGS        & 16.32 & 9 & 0.060 & 9.62 & 9 & 0.382 \\
        LRG bin 1  &  8.76 & 9 & 0.460 & 8.48 & 9 & 0.486 \\
        LRG bin 2  &  8.22 & 9 & 0.512 & 4.40 & 9 & 0.883 \\
        LRG bin 3  &  9.03 & 9 & 0.434 & 3.86 & 9 & 0.921 \\
        LRG bin 4  &  6.19 & 9 & 0.721 & 8.97 & 9 & 0.440 \\
        \hline \hline
    \end{tabular}
    \caption{Null test results for the kSZ $\Delta\Sigma$ profiles. The table shows the $\chi^2_{\rm null}$, degrees of freedom (dof), and PTE values (when compared to the expected zero) for each galaxy sample when stacking on random positions and when shuffling velocities. The PTE values indicate that the profiles are consistent with zero across all angular scales.}
    \label{tab:null_tests}
\end{table}

\subsection{Random Position Stacking}

\begin{figure}
    \centering
    \includegraphics[width=\columnwidth]{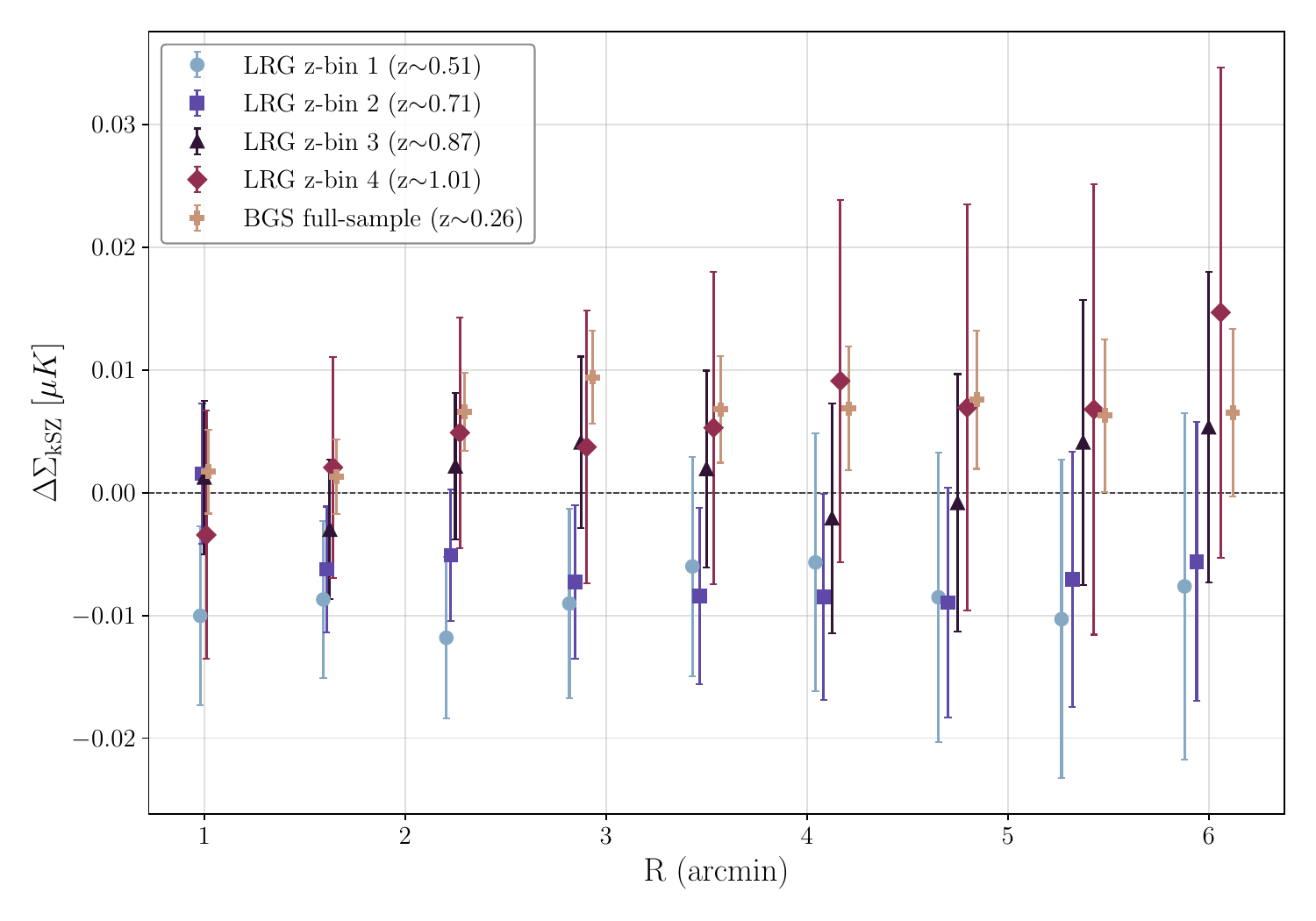}
    \caption{Null test of the kSZ $\Delta\Sigma$ profiles obtained by stacking on random instead of galaxy positions. The profiles are consistent with zero across all angular scales, indicating that our kSZ measurements are not contaminated by systematics. Error bars are $1\sigma$ statistical uncertainties from bootstrap resampling over 10000 samples.}
    \label{fig:ksz_random_positions}
\end{figure}

We perform a null test of our kSZ $\Delta\Sigma$ profiles by stacking on random positions instead of galaxy positions, using the DESI random catalogues. This test checks for any systematic effects associated with the stacking procedure, such as residual foreground contamination or instrumental systematics in the ACT CMB maps. If our kSZ measurements were contaminated by such effects, we would expect to see a significant signal in the random position stacks. However, as shown in Figure~\ref{fig:ksz_random_positions}, the kSZ $\Delta\Sigma$ profiles obtained from stacking on random positions are consistent with zero across all angular scales, with acceptable Probability to Exceed \rev{($\rm{PTE} \geq 0.05$) }values listed in Table \ref{tab:null_tests}. 
\rev{The BGS sample has the lowest PTE (0.060), with several radial bins scattering slightly above zero in Fig.~\ref{fig:ksz_random_positions}. However we do note that this is still above our threshold of 0.05.}
This result indicates that our kSZ measurements are not contaminated by systematics, providing confidence in the robustness of our results.

\subsection{Shuffling Velocities}

To test for potential tSZ contamination as well as other velocity-independent systematics in our kSZ measurements, we perform a null test by shuffling the line-of-sight velocities assigned to each galaxy in our sample. By shuffling the velocities while keeping the galaxy positions fixed, we can check whether any residual tSZ signal or other systematics that do not depend on velocity are present in our kSZ $\Delta\Sigma$ profiles. If the shuffled velocity profiles show a significant signal, this would indicate that tSZ leakage or other systematics are contaminating our kSZ measurements. Conversely, if the shuffled profiles are consistent with zero, this would provide evidence that our kSZ measurements are robust against such contamination.

\begin{figure}
    \centering
    \includegraphics[width=\columnwidth]{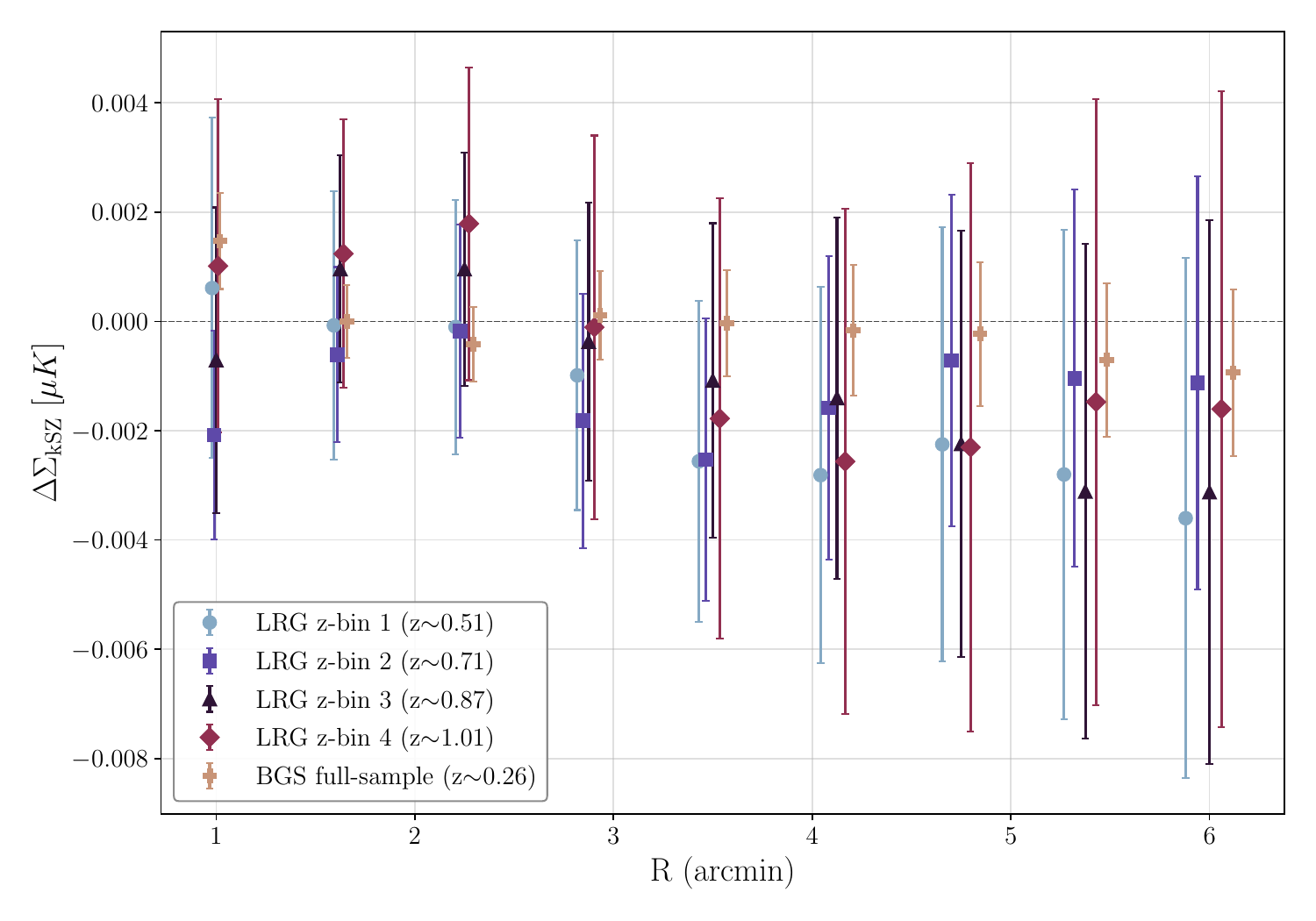}
    \caption{Null test of the kSZ $\Delta\Sigma$ profiles obtained by shuffling the line-of-sight velocities assigned to each galaxy in the sample. The profiles are consistent with zero, indicating that our kSZ measurements are not contaminated by residual tSZ leakage or other velocity-independent systematics. Error bars are $1\sigma$ statistical uncertainties estimated as the standard error of the mean across 100 shuffled realizations.}
    \label{fig:ksz_shuffled_velocities}
\end{figure}

A single shuffled velocity catalogue would introduce substantial realization noise to the resulting null test, so we perform 100 shuffles for each catalogue sample (the four LRG redshift bins as well as the BGS catalogue) and average the resulting kSZ $\Delta\Sigma$ profiles to reduce this noise. The error bars on the shuffled profiles are given by the standard error of the mean over the 100 shuffled realizations, which captures the statistical uncertainty from the shuffling procedure. The resulting averaged shuffled profiles are shown in Figure~\ref{fig:ksz_shuffled_velocities}. The profiles are consistent with zero, with acceptable PTE values listed in Table \ref{tab:null_tests}, indicating that our kSZ measurements are not contaminated by residual tSZ leakage or other velocity-independent systematics.

\section{Consistency Tests Across Survey Footprints}
\label{app:consistency_tests}

\rev{Because the kSZ and lensing profiles are measured over different sky areas, we test in this appendix both that the DESI lens population is homogeneous across the footprint, so that the ratio of Eq. \ref{eq:fgas_ratio} is well defined, and that the kSZ measurement itself is consistent between the two DESI survey regions.}

\subsection{Homogeneity of the lens sample across the sky}
\label{app:lensing_consistency}

\begin{figure}
    \centering
    \includegraphics[width=\columnwidth]{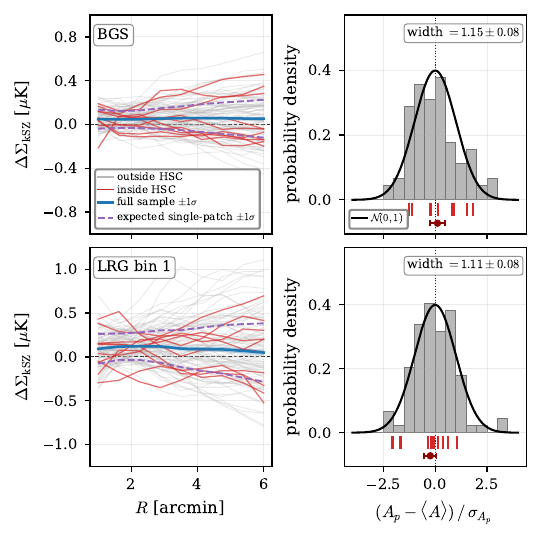}
    \caption{\rev{kSZ $\Delta\Sigma$ profiles resolved into $100$ contiguous sky patches for the BGS sample (top row) and LRG bin~1 (bottom row). Left panels: per-patch profiles, with the ten patches inside the HSC footprint in red, the remaining $90$ in grey, the full-sample measurement in blue, and the expected $\pm1\sigma$ scatter of a single patch shown by the dashed lines. Right panels: the distribution of standardised patch amplitudes $(A_p - \langle A \rangle)/\sigma_{A_p}$, compared to a unit normal; red ticks mark the HSC patches and the dark red point their mean. The HSC patches lie well within the bulk of the distribution.}}
    \label{fig:lensing_homogeneity}
\end{figure}

\rev{Equation~\ref{eq:fgas_ratio} divides a kSZ profile measured over DESI\,$\cap$\,ACT by a lensing profile measured over DESI\,$\cap$\,HSC, so it is well defined only if the DESI lens population is homogeneous across the two regions. We test this by splitting each lens sample into the galaxies that contribute to the lensing measurement: those with at least one HSC source pair, ${\sim}9\%$ of each stacked sample, and the remainder of the ACT footprint, and comparing the two both at catalogue level and in the kSZ signal, the only signal-level comparison possible since lensing exists only within the HSC footprint.


At the level of the signal, we measure the kSZ $\Delta\Sigma$ profile separately inside and outside the HSC footprint and test the difference against zero, using the same statistic as Table \ref{tab:consistency_tests}. As shown in Table \ref{tab:homogeneity}, all five samples are consistent. Furthermore, at catalogue level the two populations are indistinguishable in the properties that set the signal in terms of stellar mass and redshifts.

Figure~\ref{fig:lensing_homogeneity} resolves the same measurement into $100$ contiguous sky patches for the BGS sample and LRG bin~1, the two samples compared with simulations in Sec.~\ref{sec:simulations}. Patches are defined by $k$-means clustering of the lens positions, constrained so that each lies wholly inside or outside the HSC footprint. Individual patches are noise dominated by construction, and the HSC patches sit well within the bulk of the patch amplitude distribution.
}
\begin{table}
    \begin{ruledtabular}
    \begin{tabular}{lcccc c}
    Sample & $\chi^2_{\rm null}$ & dof & PTE & $\sigma_{\rm PTE}$ & $\Delta A$ \\
    \hline
    BGS          &  5.19 & 9 & 0.818 & 0.23 & $-0.05 \pm 0.45$ \\
    LRG bin 1    &  3.63 & 9 & 0.934 & 0.08 & $-0.15 \pm 0.34$ \\
    LRG bin 2    & 10.16 & 9 & 0.338 & 0.96 & $+0.00 \pm 0.33$ \\
    LRG bin 3    &  5.56 & 9 & 0.783 & 0.28 & $+0.78 \pm 0.41$ \\
    LRG bin 4    &  7.53 & 9 & 0.582 & 0.55 & $-0.20 \pm 0.82$ \\
    \end{tabular}
    \end{ruledtabular}
\caption{\rev{Consistency test results comparing kSZ $\Delta\Sigma$ profiles measured inside and outside the HSC lensing footprint. For each sample, $\chi^2_{\rm null}$ is computed from the profile difference $\delta = \Delta\Sigma_{\rm in} - \Delta\Sigma_{\rm out}$ with covariance $C_{\rm diff} = C_{\rm in} + C_{\rm out}$, under the null hypothesis that the two regions are consistent. $\Delta A$ is the difference in best-fit overall amplitude relative to the full-sample profile shape, and quantifies the size of difference the test could have detected. All samples are within 2$\sigma$ of the expected scatter.}}
\label{tab:homogeneity} 
\end{table}

\subsection{Consistency between the North and South Galactic Caps in DESI}
\label{app:ngc_sgc_consistency}

The kSZ $\Delta\Sigma$ profiles and combined $f_{\rm gas}^{\rm obs}$ measurements presented in Figures~\ref{fig:ksz_profiles}, \ref{fig:combined_LRG}, and \ref{fig:combined_LRG_nopairs} reveal an anomaly in LRG bin~2: the measured $\Delta\Sigma_{\rm kSZ}$ turns negative at the largest apertures, producing an $f_{\rm gas}^{\rm obs}$ profile that falls below those of the other LRG bins and reaches negative values at the largest scales. To investigate whether this anomaly has a field-dependent origin, we repeat the kSZ stacking separately in the North Galactic Cap (NGC) and South Galactic Cap (SGC) portions of the DESI survey footprint and compare the resulting profiles. The lensing $\Delta\Sigma$ profiles are not split by field: no inconsistency is observed in the lensing signal for any redshift bin, the anomaly is kSZ-specific, and splitting the lensing sample would reduce the signal-to-noise without diagnostic benefit. This approach is also consistent with our general methodology of not requiring simultaneous ACT, DESI, and HSC sky coverage. The BGS sample is included in the kSZ consistency test but its $f_{\rm gas}^{\rm obs}$ profiles are not shown separately by subfield, as it shows no significant field-dependent anomaly.

\begin{figure}
    \centering
    \includegraphics[width=\columnwidth]{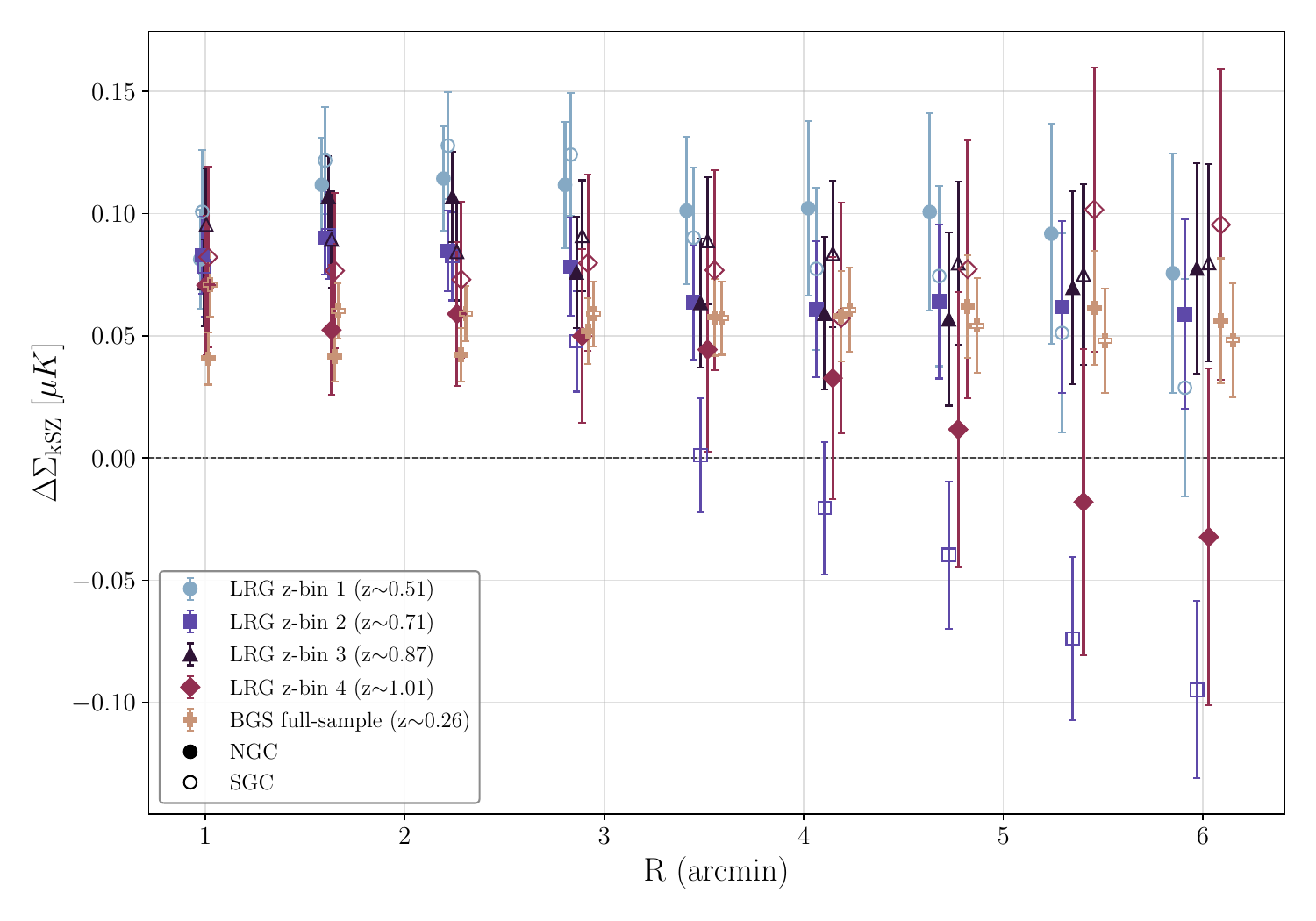}
    \caption{kSZ $\Delta\Sigma$ profiles measured independently in the NGC (filled markers) and SGC (hollow markers) fields. We use the same colour and marker convention as Figure~\ref{fig:ksz_profiles}. Points within each radial bin are offset slightly in the horizontal direction for readability. For all samples except LRG bin~2, NGC and SGC profiles are in good agreement. The LRG bin~2 SGC profile (hollow squares) turns negative at $R \gtrsim 4\,{\rm arcmin}$ while the corresponding NGC profile remains positive. Quantitative consistency statistics are listed in Table~\ref{tab:consistency_tests}.}
    \label{fig:ksz_consistency_profiles}
\end{figure}

\begin{table}[htbp!]
    \centering
    \setlength{\tabcolsep}{10pt}  
    \begin{tabular}{l | cccc }
        \hline \hline
        Sample     & $\chi^2_{\rm null}$ & dof & PTE & $\sigma_{\rm PTE}$ \\
        \hline
        BGS        & 12.03 & 9 & 0.211 & 1.25 \\
        LRG bin 1  &  4.87 & 9 & 0.845 & 0.20 \\
        LRG bin 2  & 15.08 & 9 & 0.089 & 1.70 \\
        LRG bin 3  & 10.34 & 9 & 0.324 & 0.99 \\
        LRG bin 4  & 10.55 & 9 & 0.308 & 1.02 \\
        \hline \hline
    \end{tabular}
    \caption{Consistency test results comparing kSZ $\Delta\Sigma$ profiles in the NGC and SGC fields. 
    For each sample, $\chi^2_{\rm null}$ is computed from the profile difference $\boldsymbol{\delta} = \boldsymbol{\Delta\Sigma}_{\rm NGC} - \boldsymbol{\Delta\Sigma}_{\rm SGC}$ with covariance $\mathbf{C}_{\rm diff} = \mathbf{C}_{\rm NGC} + \mathbf{C}_{\rm SGC}$ (each Hartlap-corrected), under the null hypothesis that the two fields are consistent. 
    All samples except LRG bin~2 are consistent between fields at $\leq 1.3\sigma$; LRG bin~2 shows a mild discrepancy at PTE~$= 0.089$ ($1.7\sigma$), though the two measurements are still statistically consistent. This discrepancy identifies the SGC as the origin of the anomalous signal discussed in Section~\ref{sec:combined_results}.}
    \label{tab:consistency_tests}
\end{table}

Figure~\ref{fig:ksz_consistency_profiles} shows the kSZ $\Delta\Sigma$ profiles measured independently in the two fields for all five galaxy samples. For LRG bins~1, 3, and~4 and the BGS sample, the NGC and SGC results are in good visual agreement at all apertures. For LRG bin~2, however, the SGC profile (hollow squares) turns negative at $R \gtrsim 4\,{\rm arcmin}$ while the NGC profile (filled squares) remains positive throughout. The consistency test statistics are listed in Table~\ref{tab:consistency_tests}: LRG bins~1, 3, and~4 and BGS all pass at $\leq 1.3\sigma$, while LRG bin~2 shows the largest discrepancy (PTE~$= 0.089$, $1.7\sigma$). Although this tension is not statistically significant at the conventional threshold, it clearly identifies the SGC kSZ measurement as the source of the anomaly.

\begin{figure*}[htbp!]
    \centering
    \begin{subfigure}{0.49\textwidth}
        \centering
        \includegraphics[width=\textwidth]{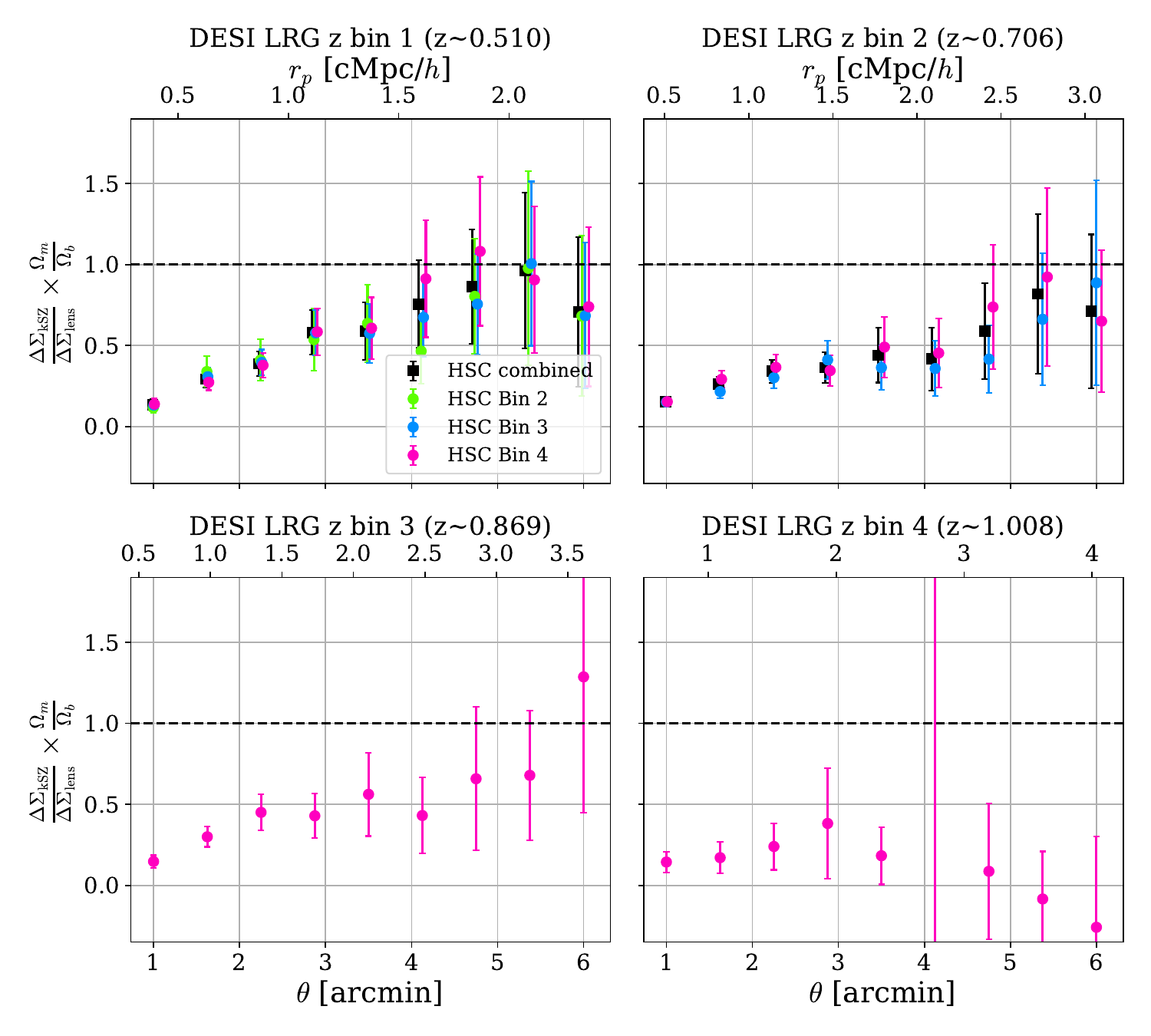}
        \caption{DESI LRG $f_{\rm gas}^{\rm obs}$ profiles for the NGC field.}
        \label{fig:NGC_ratios}
    \end{subfigure}
    \hfill
    \begin{subfigure}{0.49\textwidth}
        \centering
        \includegraphics[width=\textwidth]{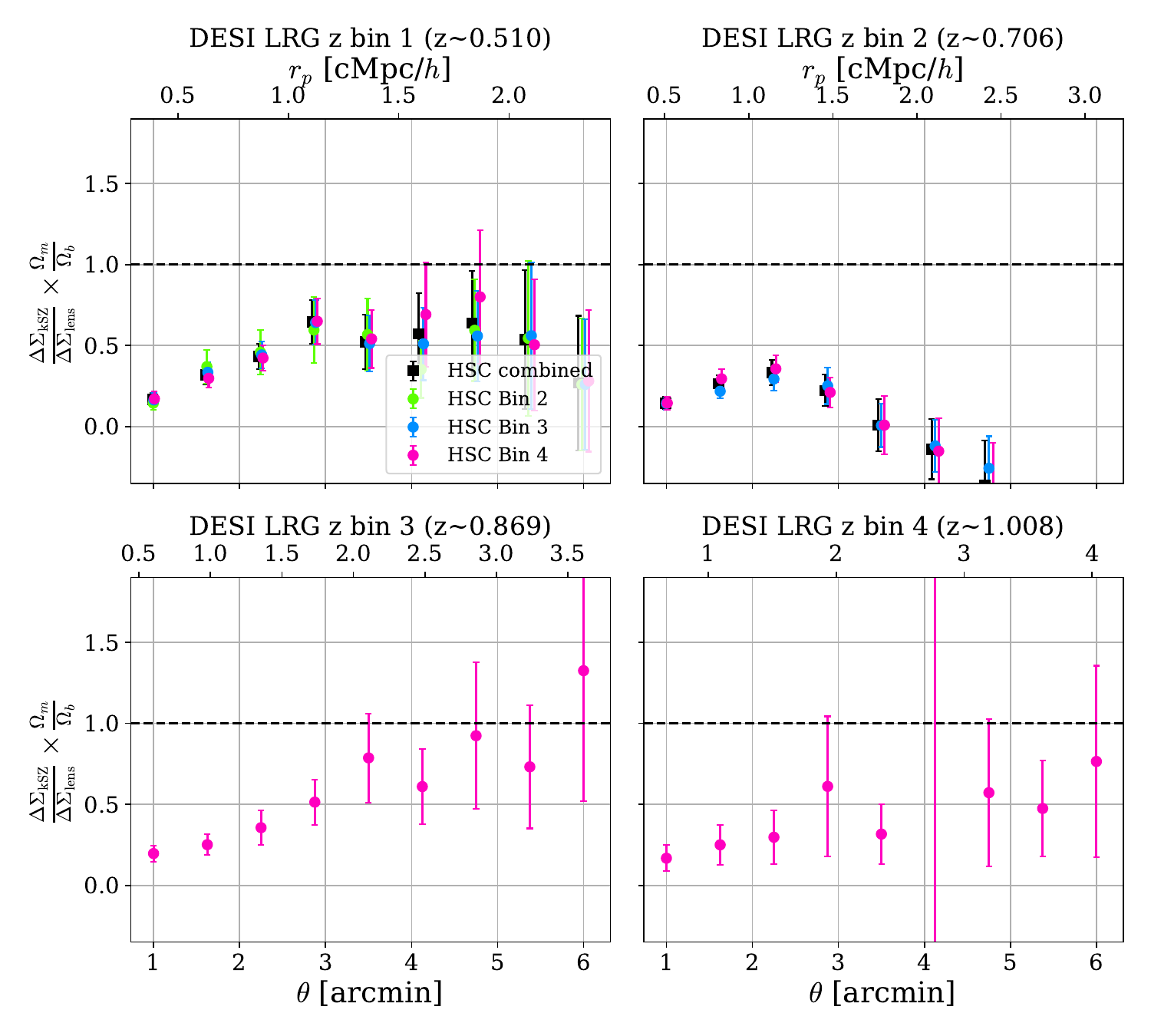}
        \caption{DESI LRG $f_{\rm gas}^{\rm obs}$ profiles for the SGC field.}
        \label{fig:SGC_ratios}
    \end{subfigure}
    \caption{Observed gas fraction $f_{\rm gas}^{\rm obs}(R)$ for the DESI LRG sample in four redshift bins, computed separately using the NGC (left) and SGC (right) kSZ $\Delta\Sigma$ profiles. The lensing $\Delta\Sigma$ profiles are taken from the full combined HSC sample in both panels, so all differences between the two panels are driven by the subfield kSZ measurements. In the NGC field, LRG bin~2 displays a rising $f_{\rm gas}^{\rm obs}$ profile consistent with the other LRG bins. In the SGC field, LRG bin~2 shows a suppressed profile with negative values at large $R$, confirming the field-specific origin of the anomaly. The large uncertainties on LRG bin~4 (pink, single HSC source bin) reflect its lower kSZ detection significance among all samples (Table~\ref{tab:SNR}).}
    \label{fig:NGC_SGC_ratios}
\end{figure*}

Figure~\ref{fig:NGC_SGC_ratios} shows $f_{\rm gas}^{\rm obs}(R)$ computed using the NGC and SGC kSZ measurements separately, with the lensing $\Delta\Sigma$ taken from the full combined sample in both panels. Since the lensing is identical in both panels, all differences between them are driven by the subfield kSZ measurements. In the NGC field, LRG bin~2 displays a rising $f_{\rm gas}^{\rm obs}$ profile broadly consistent with the other LRG bins. In the SGC field, the bin~2 profile is suppressed and turns negative at large apertures, confirming that the anomaly is localised to the SGC kSZ measurement. We also note that LRG bin~4 shows slightly negative $f_{\rm gas}^{\rm obs}$ values at the two largest apertures in the NGC field, but given \rev{its lower sample size (Table~\ref{tab:samples}) and} lower overall kSZ detection significance ($\mathrm{SNR} = 5.26$, the lowest among all bins; Table~\ref{tab:SNR}) and the large statistical uncertainties on individual radial bins, this anomaly is consistent with measurement noise.

One potential contributor to the SGC anomaly in LRG bin~2 is DESI target selection: within the SGC footprint, the Dark Energy Survey (DES) provides deeper photometric imaging than the DECam Legacy Survey (DECaLS) used elsewhere, which could introduce sample-variance differences in the selected galaxy population for this redshift bin. However, the SGC region outside the DES footprint also exhibits a mildly anomalous and slightly negative kSZ signal for LRG bin~2, indicating that target-selection depth alone does not fully account for the effect. Andrade et al.~\cite{Andrade2025} similarly find consistent DESI results in the DES region for related BAO consistency checks, \rev{and Saulder et al.~\cite{Saulder2025} found indications of regional variance, including mask-related issues in the southern SGC, in the same redshift range ($0.6<z<0.8$) of the photometric DESI Legacy Imaging Survey LRG sample.} Given the mild statistical significance of the field discrepancy ($1.7\sigma$) and the full consistency of the NGC-only result with the other LRG bins, we cannot distinguish between a statistical fluctuation and a residual systematic in the SGC kSZ measurement for this redshift bin, and adopt the combined NGC+SGC measurement throughout the main analysis.

\bibliographystyle{prsty.bst}
\bibliography{refs}

\end{document}